\documentclass{gGAF2e}
\usepackage[colorlinks,allcolors=blue]{hyperref}
\usepackage{mathrsfs}
\usepackage{lscape}
\renewcommand{\Omega}{{\varOmega}}
\renewcommand{\Gamma}{{\varGamma}}
\renewcommand{\Theta}{{\varTheta}}
\graphicspath{{./fig/}{./png/}}
\newcommand{\DD}{{\rm D}}

\newcommand{\bbb}{{\bm b}}

\newcommand{\jjj}{{\bm j}}

\newcommand{\uuu}{{\bm u}}

\newcommand{\ooo}{{\bm \omega}}

\newcommand{\FFF}{{\bm F}}

\newcommand{\UUU}{{\bm U}}
\newcommand{\mUUU}{\overline{\bm U}}

\newcommand{\eqs}[2]{equations~(\ref{#1}) and (\ref{#2})}
\newcommand{\EQ}{\begin{equation}}
\newcommand{\EN}{\end{equation}}
\newcommand{\EQA}{\begin{eqnarray}}
\newcommand{\ENA}{\end{eqnarray}}
\newcommand{\brac}[1]{\langle #1 \rangle}
\newcommand{\pd}{\upartial}

\newcommand{\mean}[1]{\overline{#1}}

\newcommand{\cP}{c_{\rm P}}
\newcommand{\cV}{c_{\rm V}}

\newcommand{\urms}{u_{\rm rms}}

\newcommand{\chiSGS}{\chi_{\rm SGS}}

\newcommand{\Co}{{\rm Co}}

\newcommand{\Hp}{H_{\rm p}}
\newcommand{\Nu}{{\rm Nu}}

\newcommand{\PraSGS}{{\rm Pr}_{\rm SGS}}

\newcommand{\PrM}{{\rm Pr}_{\rm M}}
\newcommand{\Pm}{{\rm Pm}}

\newcommand{\Ra}{{\rm Ra}}

\newcommand{\Rey}{{\rm Re}}

\newcommand{\ReM}{{\rm Re}_{\rm M}}
\newcommand{\Ro}{{\rm Ro}}

\newcommand{\Ta}{{\rm Ta}}

\newcommand{\nab}{\mbox{\boldmath $\nabla$} {}}

{}

\newcommand{\Fenthr}{\mean{F}^{\rm enth}_r}
\newcommand{\Fentht}{\mean{F}^{\rm enth}_\theta}

\newcommand{\Ftot}{F_{\rm tot}}

\newcommand{\mFenth}{\mean{F}_{\rm enth}}
\newcommand{\mFconv}{\mean{F}_{\rm conv}}

\newcommand{\mFkin}{\mean{F}_{\rm kin}}

\newcommand{\Lconvtp}{\brac{L_{\rm conv}}_{\theta \phi}}
\newcommand{\Lenthtp}{\brac{L_{\rm enth}}_{\theta \phi}}
\newcommand{\Lkintp}{\brac{L_{\rm kin}}_{\theta \phi}}
\newcommand{\Lenthdtp}{\brac{L_{\rm enth}^{\downarrow}}_{\theta \phi}}
\newcommand{\Lconvdtp}{\brac{L_{\rm conv}^{\downarrow}}_{\theta \phi}}
\newcommand{\Lkindtp}{\brac{L_{\rm kin}^{\downarrow}}_{\theta \phi}}
\newcommand{\Lenthutp}{\brac{L_{\rm enth}^{\uparrow}}_{\theta \phi}}
\newcommand{\Lconvutp}{\brac{L_{\rm conv}^{\uparrow}}_{\theta \phi}}
\newcommand{\Lkinutp}{\brac{L_{\rm kin}^{\uparrow}}_{\theta \phi}}
\newcommand{\nabad}{\nabla_{\rm ad}}
\newcommand{\nabrad}{\nabla_{\rm rad}}

\def\onethird{{\textstyle{1\over3}}}

\def\onehalf{{\textstyle{1\over2}}}
\def\threehalfs{{\textstyle{3\over2}}}

\newcommand{\figu}[1]{figure~\ref{#1}}

\newcommand{\sect}[1]{section~\ref{#1}} %for ApJ

\newcommand{\Tablel}[1]{table~\ref{#1}}

\newcommand{\apj}{\itshape Astrophys.\ J.}
\newcommand{\apjl}{\itshape Astrophys.\ J.\ Lett.}
\newcommand{\apjs}{\itshape Astrophys.\ J.\ Suppl.}
\newcommand{\apss}{\itshape Astrophys.\ and Space Sci.}
\newcommand{\aap}{\itshape Astron.\ Astrophys.}

\newcommand{\mnras}{\itshape Monthly Notices of the Roy.\ Astron.\ Soc.}
\newcommand{\pre}{\itshape Phys.\ Rev.\ E}

\newcommand{\memsai}{\itshape Mem.\ d.\ Soc.\ Astron.\ It.}
\newcommand{\solphys}{\itshape Solar Phys.}
\newcommand{\zap}{\itshape Z. Astrophys.}
\begin{document}
\jvol{00} \jnum{00} \jyear{2018} %\jmonth{February}
\markboth{\rm P.J.\ K\"APYL\"A et al.}{\rm GEOPHYSICAL AND ASTROPHYSICAL FLUID DYNAMICS}
%{Subadiabatic layer in convection-driven dynamos}

\title{Effects of a subadiabatic layer on convection and dynamos\\ in spherical wedge simulations}

\author{P.J.\ K\"APYL\"A$^{{\rm a,b,c,d,e}\,\ast}$\thanks{$^\ast$Corresponding author. Email: pkaepyl@uni-goettingen.de\vspace{6pt}},
M.\ VIVIANI$^{\rm d}$, M.J.\ K\"APYL\"A$^{\rm d,c}$,
A.\ BRANDENBURG$^{\rm e,f,g,h}$ and F.\ SPADA$^{\rm d}$
\\\vspace{6pt}
  $^{\rm a}$ Georg-August-Universit\"at G\"ottingen, Institut f\"ur
  Astrophysik, Friedrich-Hund-Platz 1, D-37077 G\"ottingen, Germany \\
  $^{\rm b}$Leibniz-Institut f\"ur Astrophysik, An der Sternwarte 16,
  D-14482 Potsdam, Germany \\
  $^{\rm c}$ReSoLVE Centre of Excellence, Department of Computer Science,
  P.O. Box 15400, FI-00076 Aalto, Finland \\
  $^{\rm d}$ Max-Planck-Institut f\"ur Sonnensystemforschung,
  Justus-von-Liebig-Weg 3, D-37077 G\"ottingen, Germany\\
  $^{\rm e}$NORDITA, KTH Royal Institute of Technology and Stockholm University,
  Roslagstullsbacken 23, SE-10691 Stockholm, Sweden\\
  $^{\rm f}$Department of Astronomy, AlbaNova University Center,
  Stockholm University, SE-10691 Stockholm, Sweden\\
  $^{\rm g}$JILA and Department of Astrophysical and Planetary Sciences,
  Box 440, University of Colorado, Boulder, CO 80303, USA\\
  $^{\rm h}$Laboratory for Atmospheric and Space Physics,
  3665 Discovery Drive, Boulder, CO 80303, USA\\
  \vspace{6pt}\received{\it \today,~ $ $Revision: 1.1 $ $} }

\maketitle

\begin{abstract}
We consider the effect of a subadiabatic layer at the base of the
convection zone on convection itself and the associated large-scale
dynamos in spherical
wedge geometry.
We use a heat conduction prescription based on the Kramers opacity law
which allows the depth of the convection zone to dynamically adapt to
changes in the physical characteristics such as rotation rate and magnetic
fields. We find that the convective heat transport is strongly
concentrated toward the equatorial and polar regions in the cases without
a substantial radiative layer below the convection zone.
The presence of a stable layer below the convection zone significantly
reduces the anisotropy of radial enthalpy transport.
Furthermore,
the dynamo solutions are sensitive to subtle changes in the convection
zone structure.
We find that the kinetic helicity changes sign in the deeper parts of
the convection zone at high latitudes in all runs. This region expands
progressively toward the equator in runs with a thicker stably
stratified layer.
\end{abstract}

\begin{keywords}convection, turbulence, dynamos, magnetohydrodynamics
\end{keywords}

\section{Introduction}

Both differential rotation and dynamo action in late-type stars such as the Sun
are driven by the interaction of turbulent convection and global
rotation of the stars \citep[e.g.][]{MT09,2017LRSP...14....4B}. While
a popular class of mean-field dynamos, known as the flux
transport dynamos \citep[e.g.][]{DC99}, rely on processes in the
boundary layers at the base and near the surface of the convection
zone (CZ), large-eddy simulations of stellar convection have
demonstrated that solar-like magnetic activity can be obtained without
the inclusion of such layers
\citep[e.g.][]{GCS10,KMB12,WKKB14,PC14,ABMT15,KKOBWKP16}.
However, this does not necessarily imply that the solar dynamo works
like the simulations suggest, because they face problems of their own.
For example, numerical simulations appear to produce much higher
velocity
amplitudes at large horizontal scales in comparison to what is found
with helioseismic inversions \citep{HDS12,2012PNAS..10911896G}.

There is another piece of evidence that also suggests that
the velocities are too high in simulations. This evidence
comes from simulations that adopt the solar luminosity and rotation rate:
instead of a solar-like differential rotation profile with fast
equator and slow poles, an anti-solar one with slow equator and fast
poles is obtained. This is indicative of a lower
rotational influence on the flow in simulations in comparison to the Sun
\citep[e.g.][]{GYMRW14,KKB14,HRY15a}. This discrepancy between
observations and simulations is called the `convective
conundrum' \citep{2016AdSpR..58.1475O}.
Furthermore, the simulated rotation profiles
are nearly in Taylor-Proudman balance, corresponding to cylindrical
isocontours of constant angular velocity
\citep[e.g.][]{2002ApJ...570..865B,MBT06,KMB11} in comparison to more
spoke-like isocontours inferred for the Sun \citep{Schouea98}.

A possible remedy to the Taylor-Proudman dilemma is to assume that the
lower part of the CZ is slightly subadiabatic \citep{Re05}, in which
case a thermal wind produced by the negative entropy fluctuations
leads to a more conical angular velocity profile \citep{MBT06}. A
related idea has
been invoked to crack the convective conundrum: if convection is
driven only in the near-surface layers by radiative cooling
\citep{Sp97,Br16}, the larger-scale convective modes such as giant
cells are not excited, leading to a reduction of power at large
horizontal scales \citep[e.g][]{CR16}. In this scenario the bulk of
the revised CZ is being mixed due to overshooting by downflow plumes
originating near the surface.

Recent numerical simulations indeed suggest that convection is driven
by cooling near the surface \citep{CR16,2017ApJ...845L..23K} and that the
lower part of the convection zone is weakly subadiabatic
\citep[e.g.][]{2015ApJ...799..142T,2017ApJ...845L..23K,2017ApJ...843...52H,2017ApJ...851...74B,2018PhFl...30d6602K,2018ApJ...859..117N}.
Evidence of a changing structure of convection from a tree-like
(decreasing number of downflow plumes with increasing depth) to a
forest-like structure (constant number of plumes) has also been reported
\citep{2017ApJ...845L..23K}. In the simulations of
\cite{2017ApJ...843...52H}, the extent of the subadiabatic
region has been reported to encompass at most roughly 40 per
cent of the combined depth of the
convection and overshoot zones. In a
subsequent study, \cite{2018PhFl...30d6602K} found a similar effect in
non-rotating hydrodynamic convection simulations at thermal Prandtl
numbers above unity. However, the effect was significantly weaker in
simulations including rotation. The main difference of the present
study compared to that of \cite{2018PhFl...30d6602K} is that we
also include setups where overshoot and radiative layers are present,
and investigate cases where dynamo action occurs.

Large-scale dynamos in stellar convective envelopes can also be affected
by a subadiabatic layer at the base of the convection zone: such a layer can
store magnetic flux \cite[e.g.][]{BMBT06} and it can possibly
contribute to inverting the sign of kinetic helicity of the flow in the deep
parts of the CZ \citep{DWBG16}. Such inversion is a possible way out
of the `modern dynamo dilemma' that plagues current simulators: the
equatorward migrating dynamo waves are most likely due to a region of
negative radial shear within the CZ \citep{WKKB14}, which is not
present in the Sun, except for the near-surface shear layer (NSSL);
see \cite{Br05}. The problem of the observed equatorward migration
of the sunspot belts is a variation of Parker's dynamo dilemma
\citep{Pa87a} where the observed differential rotation profile and
theoretically expected sign of kinetic helicity lead to poleward
migration of activity belts \citep[see also][]{1986GApFD..37...85D}.

In the current study we present first results from convection-driven
dynamo simulations in spherical wedges where stably stratified layers
are present with a setup that is similar to that of the hydrodynamic
Cartesian runs of \cite{2017ApJ...845L..23K}, where a physics-based
rather than a prescribed formulation for the heat conduction was used.

\section{Model}

Our simulation setup is similar to that used earlier
\citep{KMCWB13,KKOBWKP16,2017A&A...599A...4K}. However, the current
models differ in a few key aspects from the previous studies. We solve
the equations of fully compressible magnetohydrodynamics
\begin{align}
\frac{\pd {\bm A}}{\pd t}\, =\,&\, {\bm U} \times {\bm B} - \eta \mu_0 {\bm J}, \\
\frac{\DD \ln\rho}{\DD t}\, =\,&\,  -\, \bm\nabla\bm\cdot{\bm U}, \\
\frac{\DD {\bm U}}{\DD t}\, =\,&\,  {\bm g}  - 2\bm\Omega_0\times{\bm U} - \frac{1}{\rho}(\bm\nabla p + {\bm J} \times {\bm B} + \bm\nabla \bm\cdot 2\nu\rho\bm{\mathsf{S}}),\\
T\frac{\DD s}{\DD t}\, = \,&\,  \frac{1}{\rho} \left[ \eta\mu_0 {\bm J}^2 - \bm\nabla\bm\cdot( {\bm F}^{\rm rad} + {\bm F}^{\rm SGS} ) - \Gamma_{\rm cool} \right] + 2\nu \bm{\mathsf{S}}^2, \label{equ:ss}
\end{align}
where ${\bm A}$ is the magnetic vector potential, ${\bm U}$ is the
velocity, ${\bm B} = \bm\nabla\times{\bm A}$ is the magnetic field,
$\eta$ is the magnetic diffusivity, $\mu_0$ is the permeability of
vacuum, ${\bm J}=\bm\nabla\times{\bm B}/\mu_0$ is the current density,
$\DD/\DD t = \pd/\pd t + {\bm U}\bm\cdot\bm\nabla$ is the advective time
derivative, $\rho$ is the density, ${\bm g}=-GM_\odot\hat{\bm r}/r^2$
is the acceleration due to gravity, where
$G=6.67\cdot10^{-11}$~N~m$^2$~kg$^{-2}$ is the universal gravitational
constant, and $M_\odot=2.0\cdot10^{30}$~kg is the solar mass,
$\bm\Omega_0=(\cos\theta,-\sin\theta,0)\Omega_0$ is the
angular velocity vector, where $\Omega_0$ is the rotation rate of the
frame of reference, $\nu$ is the kinematic viscosity, $p$ is the
pressure, and $s$ is the specific entropy with $Ds=\cV D\ln p-\cP
D\ln\rho$, where $\cV$ and $\cP$ are the specific heats in constant
volume and pressure, respectively. The gas is assumed to obey the
ideal gas law, $p=\mathcal{R}\rho T$, where $\mathcal{R}=\cP-\cV$ is
the gas constant. The rate of strain tensor is given by
\begin{eqnarray}
\mathsf{S}_{ij}\, = \,\onehalf (U_{i;j} + U_{j;i}) - \onethird \delta_{ij} \bm\nabla\bm\cdot {\bm U},
\end{eqnarray}
where the semicolons refer to covariant derivatives
\citep{MTBM09}. The radiative flux is given by
\begin{eqnarray}
{\bm F}^{\rm rad} \,=\, -\,K\bm\nabla T,
\label{equ:Frad}
\end{eqnarray}
where $K$ is the heat conductivity, which is allowed to
vary in a dynamic and
local fashion. We use two heat
conduction schemes, where $K$ is either a fixed function of height
$K=K(r)$ or it depends on density and temperature $K=K(\rho,T)$. In
the former case we use the same profile as defined in
\cite{KMCWB13}. In the latter case $K$ is computed from
\begin{eqnarray}
K \,= \,\frac{16 \sigma_{\rm SB} T^3}{3 \kappa \rho}\,,
\label{equ:Krad1}
\end{eqnarray}
where $\sigma_{\rm SB}$ is the Stefan-Boltzmann constant and $\kappa$
is the opacity. The latter is assumed to obey a power law
\begin{eqnarray}
\kappa\, = \,\kappa_0 (\rho/\rho_0)^a (T/T_0)^b,
\label{equ:kappa}
\end{eqnarray}
where $\rho_0$ and $T_0$ are reference values of density and
temperature. Combining (\ref{equ:Krad1}) and (\ref{equ:kappa})
gives \citep{BB14}
\begin{eqnarray}
K(\rho,T)\, =\, K_0 (\rho/\rho_0)^{-(a+1)} (T/T_0)^{3-b}.
\label{equ:Krad2}
\end{eqnarray}
In the current study we use the combination $a=1$ and $b=-7/2$,
which corresponds to the
Kramers opacity law for free-free and bound-free transitions
\citep{WHTR04}. This scheme has been used both in
Cartesian \citep{2000gac..conf...85B,2017ApJ...845L..23K}
and in spherical wedge \citep{2018arXiv180709309K}
simulations of convection.

The subgrid scale (SGS) flux is given by
\begin{eqnarray}
{\bm F}^{\rm SGS}\, = \,-\,\chiSGS \rho T \bm\nabla s',
\label{equ:FSGS}
\end{eqnarray}
where $\chiSGS$ is the (constant) SGS diffusion coefficient for the
entropy fluctuation $s'(r,\theta,\phi)=s-\brac{s}_{\theta\phi}$, where
$\brac{s}_{\theta\phi}$ is the horizontally averaged or spherically
symmetric
part of the specific entropy.

The last term on the right-hand side of (\ref{equ:ss}) models the cooling
near the surface of the star:
\begin{eqnarray}
  \Gamma_{\rm cool}\, =\, \,-\Gamma_0 f(r) (T_{\rm cool}-\brac{T}_{\theta \phi}),
\label{equ:Gamma}
\end{eqnarray}
where $\Gamma_0$ is a cooling luminosity, $\brac{T}_{\theta \phi}$
is the horizontally averaged temperature, and $T_{\rm cool}=T_{\rm
  cool}(r)$ is a radius-dependent cooling temperature coinciding with
the initial isentropic stratification.
In our previous studies \citep{KKBMT10,KMGBC11}.
we cooled the near-surface layers toward an isothermal state. The
main effect of the changed cooling profile
is that no strongly subadiabatic isothermal layer
forms near the surface.

The simulations were performed using the {\sc Pencil Code}\footnote{\url{https://github.com/pencil-code/}}.
The code employs a high-order finite difference method for solving the compressible equations of MHD.

\subsection{System parameters and diagnostics quantities}

The wedges used in the current simulations span $r_{\rm in}<r<R_\odot$ in
radius, where $r_{\rm in}=0.7R_\odot$ and $R_\odot=7\cdot10^8$~m is
the solar radius,
$\theta_0<\theta<180^\circ-\theta_0$ in colatitude, where
$\theta_0=15^\circ$, and $0<\phi<90^\circ$ in longitude.
Our simulations are defined by the energy flux imposed at the bottom
boundary, $F_{\rm b}=-(K \pd T/\pd r)|_{r=r_{\rm in}}$, the values of
$K_0$, $a$, $b$, $\rho_0$, $T_0$, $\Omega_0$, $\nu$, $\eta$,
$\chiSGS$, and the fixed profile of $K$ in cases where the Kramers
opacity law is not used. Furthermore, the radial profile of $f(r)$ is
piecewise constant with $f(r)=0$ in $r_{\rm in} < r <0.98R_\odot$,
and smoothly connecting to $f(r)=1$ above $r=0.98R_\odot$. We use a
significantly higher luminosity and thus a higher Mach number
than what is estimated for the Sun to avoid the time step
being too severely limited by sound waves. This also necessitates a
  correspondingly higher rotation rate to capture the same rotational
  influence on the flow in the simulations in comparison to the Sun;
  see appendix~A of \cite{2018arXiv180709309K} for a thorough description of this
  procedure. This study also indicates that the results depend only
  weakly on the Mach number. The ratio $L_{\rm ratio}=L_0/L_\odot$,
where $L_0$ is the luminosity in the simulations and
$L_\odot=3.83\cdot10^{26}$~W is the corresponding solar value,
quantifies the luminosity.
The non-dimensional luminosity is given by
\begin{eqnarray}
\mathcal{L}\,=\,\frac{L_0}{\rho_0 (GM_\odot)^{3/2}R_\odot^{1/2}}\,.
\end{eqnarray}
The initial stratification is determined by the non-dimensional
pressure scale height at the surface
\begin{eqnarray}
\xi_0\,=\,\frac{\mathcal{R}T_1}{GM_\odot/R_\odot},
\end{eqnarray}
where $T_1$ is the temperature at the surface ($r=R_\odot$)\,.

The relations between viscosity, magnetic diffusivity, and SGS
diffusion are given by the Prandtl numbers
\begin{eqnarray}
\PraSGS\, =\, {\nu}\big/{\chiSGS}\,, \hskip 10mm  \Pm\,=\,{\nu}\big/{\eta}\,.
\end{eqnarray}
We use $\PraSGS=\Pm=1$ in all of our runs. The Prandtl number is related
to the radiative conductivity,
\begin{eqnarray}
\Pr \,= \,{\nu}\big/{\chi}\,,
\end{eqnarray}
where $\chi=K/\cP\rho$ is the radiative diffusivity, which varies
as a function of radius, latitude, and time.
The efficiency of convection is quantified by the Rayleigh number
\begin{eqnarray}
\Ra \,= \,\frac{GM_\odot (\Delta r)^4}{\nu \chiSGS R_\odot^2}\left(- \frac{1}{c_{\rm P}} \frac{{\rm d}s_{\rm hs}}{{\rm d}r}\right)_{r_{\rm m}},
\end{eqnarray}
where
$\Delta r=0.3R_\odot$ is the depth of the layer, $s_{\rm hs}$ is
the entropy in a one-dimensional non-convecting
hydrostatic model, evaluated at the middle of the domain at $r_{\rm  m}=0.85R_\odot$.
We note that in the cases with a Kramers-based heat conduction
prescription, only a very thin surface layer is convectively unstable
\citep[see, e.g., Figure~7 of][]{Br16}, such that $\Ra<0$ at $r=r_{\rm m}$.
We additionally quote the Nusselt number, which describes the
efficiency of convection in comparison to radiative diffusion
\citep[e.g.][]{HTM84,Br16}:
\begin{eqnarray}
\Nu\, = \,{\nabrad}\big/{\nabad}\,,\label{equ:Nus}
\end{eqnarray}
just below the cooling layer at $r=0.98R_\odot$, where
\begin{eqnarray}
\nabrad\,=\,\frac{\mathcal{R}}{Kg}\Ftot  \hskip 10mm  \mbox{and} \hskip 10mm  \nabad=1-\frac{1}{\gamma}\,,
\end{eqnarray}
are the radiative and adiabatic temperature gradients,
$g=|{\bm g}|$, $\Ftot=L_0/(4\pi r^2)$, and $\gamma=\cP/\cV$.
For runs with a fixed $K$, $\Nu$ remains constant throughout the
  duration of the simulation whereas in the cases with Kramers
  conductivity the saturated value, $\Nu_{\rm sat}$ differs from the
  initial value $\Nu$.
The effect of rotation is controlled by the Taylor number
\begin{eqnarray}
\Ta\, = \,(2\Omega_0 \Delta r^2/\nu^2)^2.
\end{eqnarray}

The fluid and magnetic Reynolds numbers are
\begin{eqnarray}
\Rey\,=\,\frac{\urms}{\nu k_1}\hskip 10mm \mbox{and} \hskip 10mm
\ReM\,=\,\frac{\urms}{\eta k_1}\,,
\label{equ:Rey}
\end{eqnarray}
respectively, where $\urms = \sqrt{\threehalfs (U_r^2+U_\theta^2)}$
is the volume averaged rms velocity and $U_\phi^2$ has been
replaced by $(U_r^2+ U_\theta^2)/2$ to avoid contributions from
differential rotation \citep[cf.][]{KMGBC11}.
The inverse of the basic wavenumber $k_1=2\pi/\Delta r\approx21/R_\odot$
is used to characterise the radial extent of convection cells.

The rotational influence on the flow is quantified by the Coriolis number
\begin{eqnarray}
\Co\,=\,\frac{2\,\Omega_0}{\urms k_1}\,.
\label{equ:Cori}
\end{eqnarray}
Mean quantities refer either to
azimuthal (denoted by an overbar) or horizontal averages (denoted by
angle brackets with subscript $\theta\phi$).
Additional time averaging is also performed
unless stated otherwise.

\subsection{Initial and boundary conditions}

The initial stratification is
polytropic with index $n=1.5$ corresponding to an isentropic
stratification. We use $\xi_0=0.01$, which results in an initial
density contrast of roughly 77. In cases with a fixed heat
conductivity profile, the
value of $K$ at $r=r_{\rm in}$ is set such that the flux through the
lower boundary is $L_0/4\pi r_{\rm in}^2$. The luminosity $L_0$ is
based on the total horizontal area of the star, although the simulations
  cover only a fraction of the full $4\pi$ area. The flux at the
outer radius, however, is initially much lower and the
convective instability
arises from the fact that the system is not in thermodynamic
equilibrium driven by the efficient surface cooling \citep[see
  e.g.][]{KMCWB13}. In the cases with Kramers
heat conductivity, the value of $K$
at the bottom of the domain is varied by changing the value of $K_0$
in (\ref{equ:Krad2}) to probe the influence it has on the depth of the
convection zone.
In the fiducial case, a nominal value $K_0^{\rm nom}$ is chosen such that
$F_{\rm rad}=F_{\rm tot}$ at the bottom of the domain. We probe a set of
runs where the value of $\tilde{K}_0=K_0/K_0^{\rm nom}$ is
increased. These runs correspond to more efficient radiative diffusion
for a given thermal stratification. The expectation is that
an increasing value of $\tilde{K}_0$ leads
to the formation of a stably stratified radiative layer at the
bottom of the domain.

The radial and latitudinal boundaries are assumed impenetrable and
stress-free for the flow. On the bottom boundary, a fixed heat
flux is prescribed and the temperature is fixed on the outer
  boundary. On the latitudinal
boundaries, the gradients of thermodynamic quantities are set to
zero; see \cite{KMCWB13}. For the magnetic field we apply a vertical
field condition at the upper, and a perfect conductor condition at the
lower boundary. On the latitudinal boundaries the field is assumed to be
tangential to the boundary. These conditions are given by:
%\begin{eqnarray}
%  \frac{\pd A_r}{\pd r} &=& 0,\ \ \frac{\pd^2 A_\theta}{\pd r^2} = -\frac{2}{r}\frac{\pd A_\theta}{\pd r}, \ \ \frac{\pd^2 A_\phi}{\pd r^2}= -\frac{2}{r}\frac{\pd A_\phi}{\pd r} \ \ (r=r_{\rm in}), \label{equ:BCBbot} \\
%A_r &=& 0, \ \ \frac{\pd A_\theta}{\pd r} = -\frac{A_\theta}{r}, \ \ \frac{\pd A_\phi}{\pd r} = -\frac{A_\phi}{r} \hspace{1.4cm} (r=R_\odot), \\
%A_r &=& \frac{\pd A_\theta}{\pd \theta} = A_\phi = 0 \hspace{3.2cm} (\theta=\theta_0,\pi-\theta_0).
%\end{eqnarray}
\begin{align}
  \qquad
  \frac{\pd A_r}{\pd r}\,=\,&\, 0\,, & \frac{\pd^2 A_\theta}{\pd r^2}\, = \,&\,-\,\frac{2}{r}\frac{\pd A_\theta}{\pd r}\,, & \frac{\pd^2 A_\phi}{\pd r^2}\,=\,&\, -\,\frac{2}{r}\frac{\pd A_\phi}{\pd r} &  (r=r_{\rm in})\,,
  \qquad\label{equ:BCBbot} \\
  \qquad
  A_r\,=\,&\, 0\,, & \frac{\pd A_\theta}{\pd r}\, =\,&\,  -\,\frac{A_\theta}{r}\,, & \frac{\pd A_\phi}{\pd r} \,=\, &\, -\,\frac{A_\phi}{r} & (r=R_\odot)\,,
  \qquad\\
  \qquad
  A_r\, =\,& \,\frac{\pd A_\theta}{\pd \theta} = A_\phi = 0\,.
\!\!\!\!\!\!\!\!\!\!\!\!\!\!\!\!\!\!\!\!\!\!\!\!\!\!\!\!\!\!\!\!\!\!\!\!\!\!\!\!\!\!\!\!\!
&&&&& \!\!\!\!\!\!\!\!\!\!\!\!\!\!\!(\theta=\theta_0,\pi-\theta_0).
\qquad
\end{align}

Equation~(\ref{equ:BCBbot}) differs from previously used conditions
\citep[see, e.g., equation~(10) of][]{KMCWB13}, where instead the tangential electric
field was assumed to vanish on the boundary.
We show in \cite{2018arXiv180709309K} that the differences
  between the current boundary conditions and those used in
  \cite{KMCWB13} are minor.
The azimuthal direction is periodic for all quantities.
The velocity and magnetic fields are initialised with random Gaussian
noise fluctuations with amplitudes on the order of $0.1$~m~s$^{-1}$ and
$0.1$~Gauss, respectively.

%\begin{landscape}
\begin{table}
  \tbl{Summary of the runs. All runs have $L_{\rm ratio}=2.1\cdot10^5$, $\Omega_0=3\Omega_\odot$,
    $\PraSGS=\PrM=1$, $\nu=1.46\cdot10^8$~m$^2$~s$^{-1}$, $\Ta=2.33\cdot10^7$,
    $\xi_0=0.01$, and grid resolution $144\times288\times144$.}
{\begin{tabular}{@{}lcccccc ccccc ccc}\toprule
    Run & $\Ra\ [10^7]$ & $\Nu$ & $\Nu_{\rm sat}$ & $\Rey$ & $\Co$ & $r_{\rm BZ}$ & $r_{\rm DZ}$ & $r_{\rm OZ}$ &$d_{\rm BZ}$ & $d_{\rm DZ}$ & $d_{\rm OZ}$ & $\Delta t$ [yr] & $K$ & $\tilde{K}_0$ \\ \colrule
    HDp & 3.0 &   156 &  156 & 36 (36) &     --     &  (0.76 &  0.70 &  0.70 &  0.24 &  0.06 &  0.00) &  35 & profile &  -- \\
    HD1 &  -  &  3167 & 2599 & 34 (33) &     --     &  (0.76 &  0.71 &  0.70 &  0.24 &  0.06 &  0.01) &  10 & Kramers & 1.0 \\
    HD2 &  -  &  1843 & 1524 & 31 (29) &     --     &  (0.79 &  0.73 &  0.70 &  0.21 &  0.07 &  0.03) &  10 & Kramers & 1.7 \\
    HD3 &  -  &   972 &  786 & 28 (25) &     --     &   0.82 &  0.77 &  0.71 &  0.18 &  0.06 &  0.06  &  12 & Kramers & 3.2 \\
    HD4 &  -  &   590 &  440 & 26 (22) &     --     &   0.85 &  0.80 &  0.73 &  0.15 &  0.05 &  0.07  &  11 & Kramers & 5.4 \\
    \hline
    RHDp & 3.0 &  156 &  156 & 27 (27) &  4.6 (4.6) &  (0.75 &  0.70 &  0.70 &  0.25 &  0.05 &  0.00) &  49 & profile &  -- \\
    RHD1 &  -  & 3167 & 3034 & 30 (30) &  4.1 (4.1) &  (0.75 &  0.70 &  0.70 &  0.25 &  0.05 &  0.00) &  29 & Kramers & 1.0 \\
    RHD2 &  -  & 1843 & 1772 & 28 (26) &  4.3 (3.5) &  (0.78 &  0.74 &  0.71 &  0.22 &  0.04 &  0.03) &  27 & Kramers & 1.7 \\
    RHD3 &  -  &  972 &  882 & 25 (22) &  4.8 (3.0) &   0.79 &  0.78 &  0.72 &  0.21 &  0.01 &  0.06  &  29 & Kramers & 3.2 \\
    RHD4 &  -  &  590 &  479 & 23 (19) &  5.3 (2.5) &   0.82 &  0.81 &  0.76 &  0.18 &  0.01 &  0.05  &  22 & Kramers & 5.4 \\
    \hline
    MHDp & 3.0 &  156 &  156 & 27 (27) &  4.5 (4.5) &  (0.76 &  0.70 &  0.70 &  0.24 &  0.06 &  0.00) &  44 & profile &  -- \\
    MHD1 &  -  & 3167 & 3004 & 30 (30) &  4.1 (4.1) &  (0.76 &  0.70 &  0.70 &  0.24 &  0.06 &  0.00) &  63 & Kramers & 1.0 \\
    MHD2 &  -  & 1843 & 1743 & 27 (25) &  4.5 (3.6) &  (0.78 &  0.74 &  0.71 &  0.22 &  0.05 &  0.03) &  74 & Kramers & 1.7 \\
    MHD3 &  -  &  972 &  868 & 23 (20) &  5.3 (3.2) &   0.80 &  0.78 &  0.72 &  0.20 &  0.02 &  0.06  &  64 & Kramers & 3.2 \\
    MHD4 &  -  &  590 &  473 & 21 (18) &  5.8 (2.7) &   0.82 &  0.81 &  0.77 &  0.18 &  0.01 &  0.04  &  72 & Kramers & 5.4 \\
    \botrule
  \end{tabular}}
\tabnote{The values of $r_{\rm BZ}$, $r_{\rm DZ}$, $r_{\rm OZ}$,
    $d_{\rm BZ}$, $d_{\rm DZ}$, and $d_{\rm OZ}$ for Runs~RHDp, RHD1,
  RHD2, MHDp, MHD1, and MHD2,
where strong latitudinal variations are seen,
are listed in parentheses to indicate uncertainty.
The value of $\Nu$ refers to the initial state and $\Nu_{\rm sat}$ to
the saturated convective state, both computed from
  (\ref{equ:Nus}). The values in brackets for $\Rey$ and $\Co$ are
  calculated taking the volume averaged $\urms$ from the revised
  convection zone ($r_{\rm DZ} < r < R_\odot$) using
  $k_1=2\pi/(R_\odot-r_{\rm DZ})$ as the wavenumber.}
\label{tab:runs}
\end{table}
%\end{landscape}

\section{Results}

We perform three sets of simulations denoted as HD, RHD, and
MHD. In Set~HD, we model non-rotating convection, where $\tilde{K_0}$ is varied
to control the depth of the convection zone. The effect of increasing
$\tilde{K_0}$ is to make radiative diffusion more efficient. This is
particularly important in the deep parts of the domain where the
temperature is high due to the strong temperature dependency of the
heat conduction ($K\propto T^{6.5}$, see (\ref{equ:Krad2})). Thus the
expectation is that with higher values of $\tilde{K_0}$, a radiative
layer develops at the bottom of the domain. In the RHD runs, we take
the HD runs and add rotation with $\Omega_0=3\Omega_\odot$, where
$\Omega_\odot = 2.7\cdot 10^{-6}$~s$^{-1}$ is the mean solar rotation
rate. In the MHD
set, magnetic fields are added to the RHD setup to study the effects of
stably stratified layers on the dynamo. Each set consists of four
  runs, denoted by a suffix running from 1 to 4, where the value of
  $K_0$ is systematically increased. A run with a fixed profile of
$K$, denoted by a suffix `p', is used as a reference in each set with
the same variation of physical
ingredients. The runs are listed in \Tablel{tab:runs}.

\begin{figure}
\begin{center}
%PJK: for arXiv-submission
    \includegraphics[width=\textwidth]{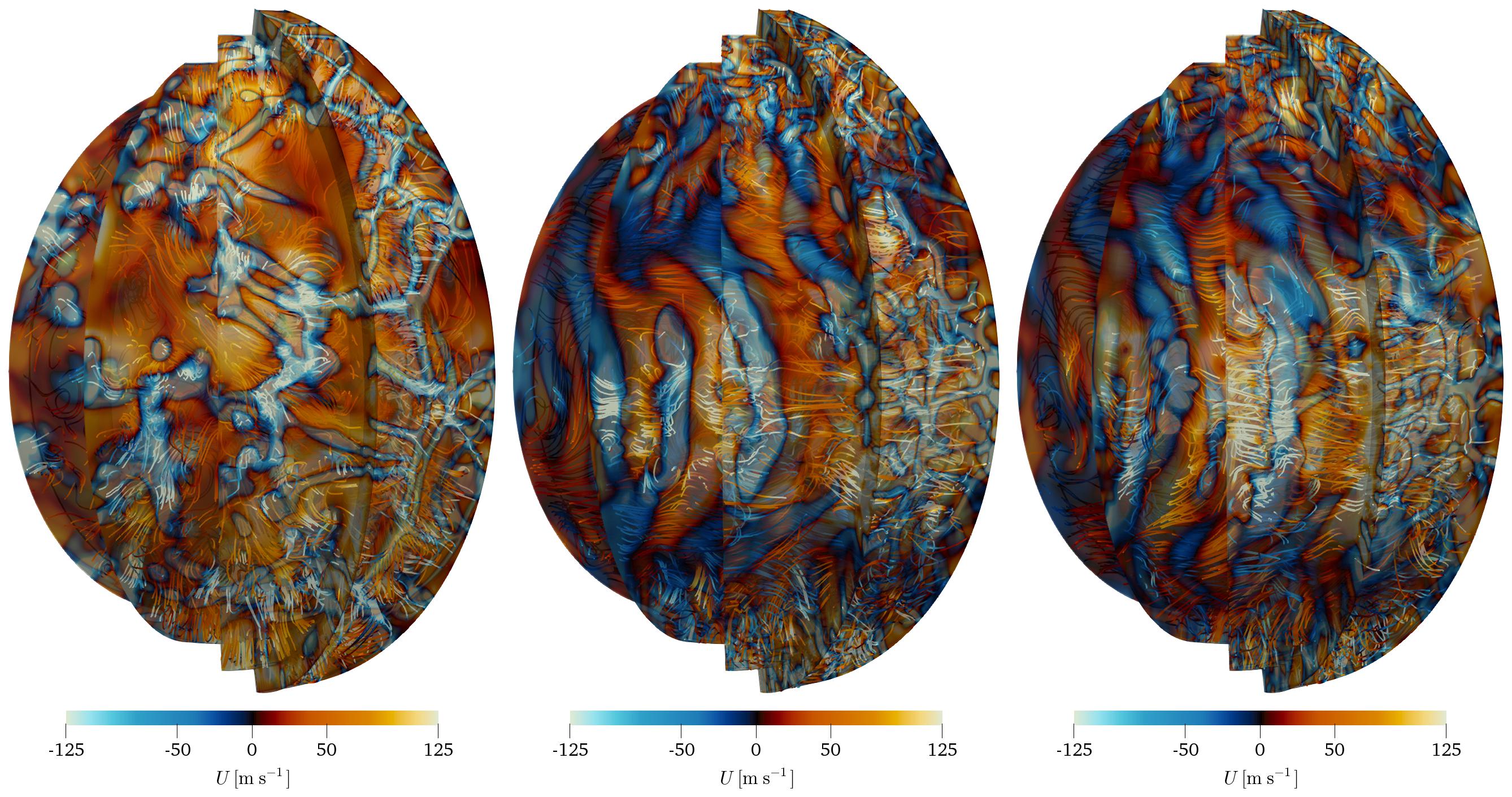}
    \caption{Streamlines of the total velocity and contours of
      vertical velocity at the periphery in snapshots of Runs~HD2
        (left), RHD2 (middle), and MHD2 (right). The colour-coding of
      both is indicated
      at the bottom of each panel. The horizontal
      cuts from left to right are shown from depths
      $r/R_\odot=0.78$, $0.85$, $0.92$, and $0.99$.
Animated visualisations of Runs~HD2 and RHD2 are available in the
online material (colour online).}
\label{fig:wedges_Ur}
\end{center}
\end{figure}

The value of $\Omega_0$ in the rotating simulations is chosen such
that a solar-like differential rotation is obtained. The current
setups with a Kramers-based heat conduction still tend to produce
anti-solar differential rotation at solar luminosity and rotation
rate. Visualisations of the flow fields realized in representative
hydrodynamic runs without (HD2) and with rotation (RHD2), and a
  corresponding MHD run (MHD2) are shown in \figu{fig:wedges_Ur}. The
non-rotating cases qualitatively resemble mixing length ideas in that
the horizontal scale of the convective eddies increases as a function
of depth. The rotating cases are dominated by banana cells
\citep[e.g.][]{1970ApJ...159..629B,1986ApJS...61..585G} in the
equatorial regions and by small scale convection at high
latitudes, and this carries over also to the magnetic cases.
The flow structure in the current MHD runs is typically very
  similar to the corresponding RHD runs.
The convective scales show significantly less variation in
depth in comparison to non-rotating convection.

\subsection{Convective energy transport and structure of the convection zone}

In an earlier study, \cite{2017ApJ...845L..23K} found that a stably
stratified layer, where the enthalpy flux is nevertheless directed
outward, develops at the bottom of the convection zone if a smoothly
varying profile for the heat conduction is used. Furthermore, when the
Kramers opacity law is applied, the depth of the convection zone is a
result of the simulation rather than a priori fixed.
Here we extend these studies to more realistic spherical geometry and
take into account global rotation and dynamo-generated magnetic
fields.

\begin{figure}
\begin{center}
    \includegraphics[width=0.5\textwidth]{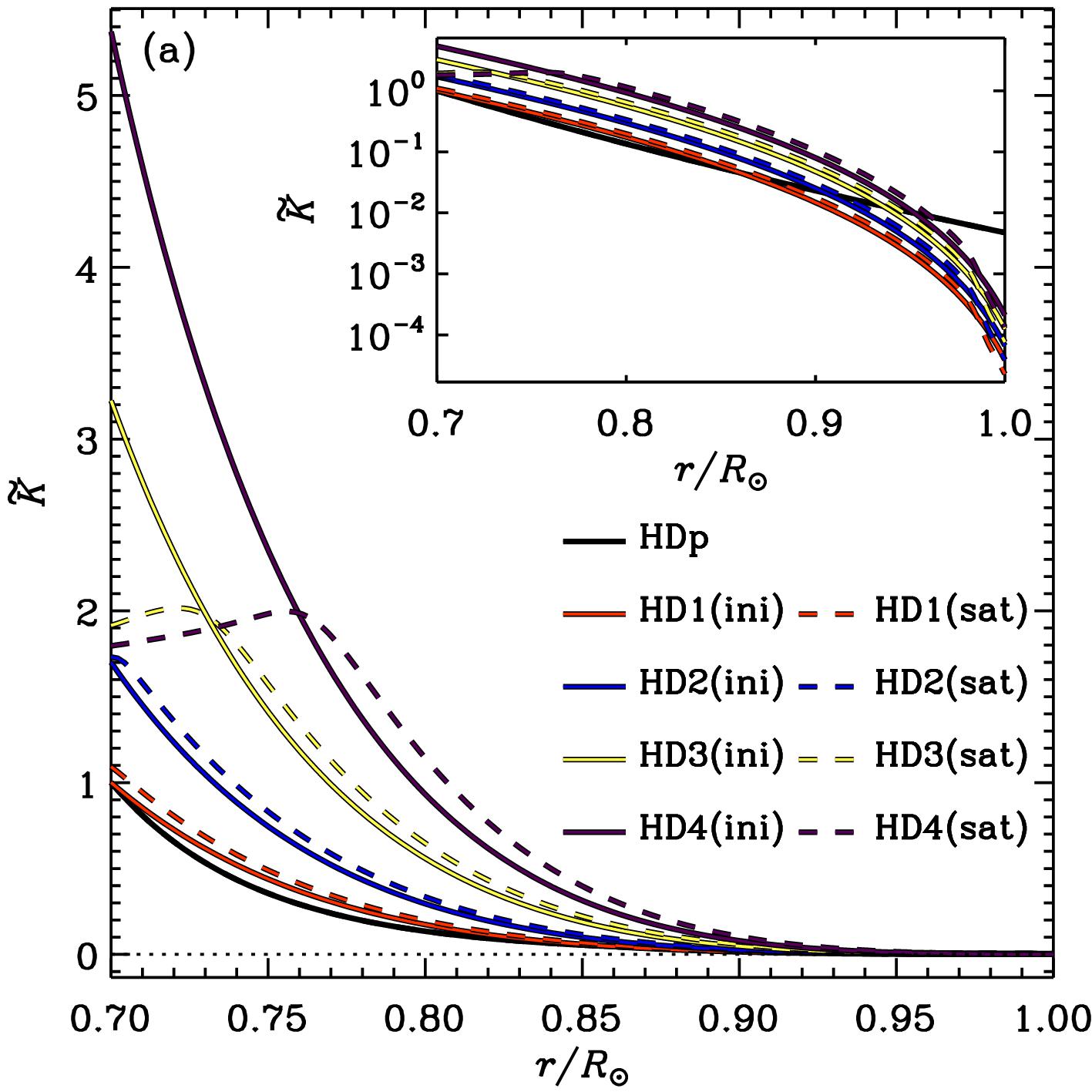}\includegraphics[width=0.5\textwidth]{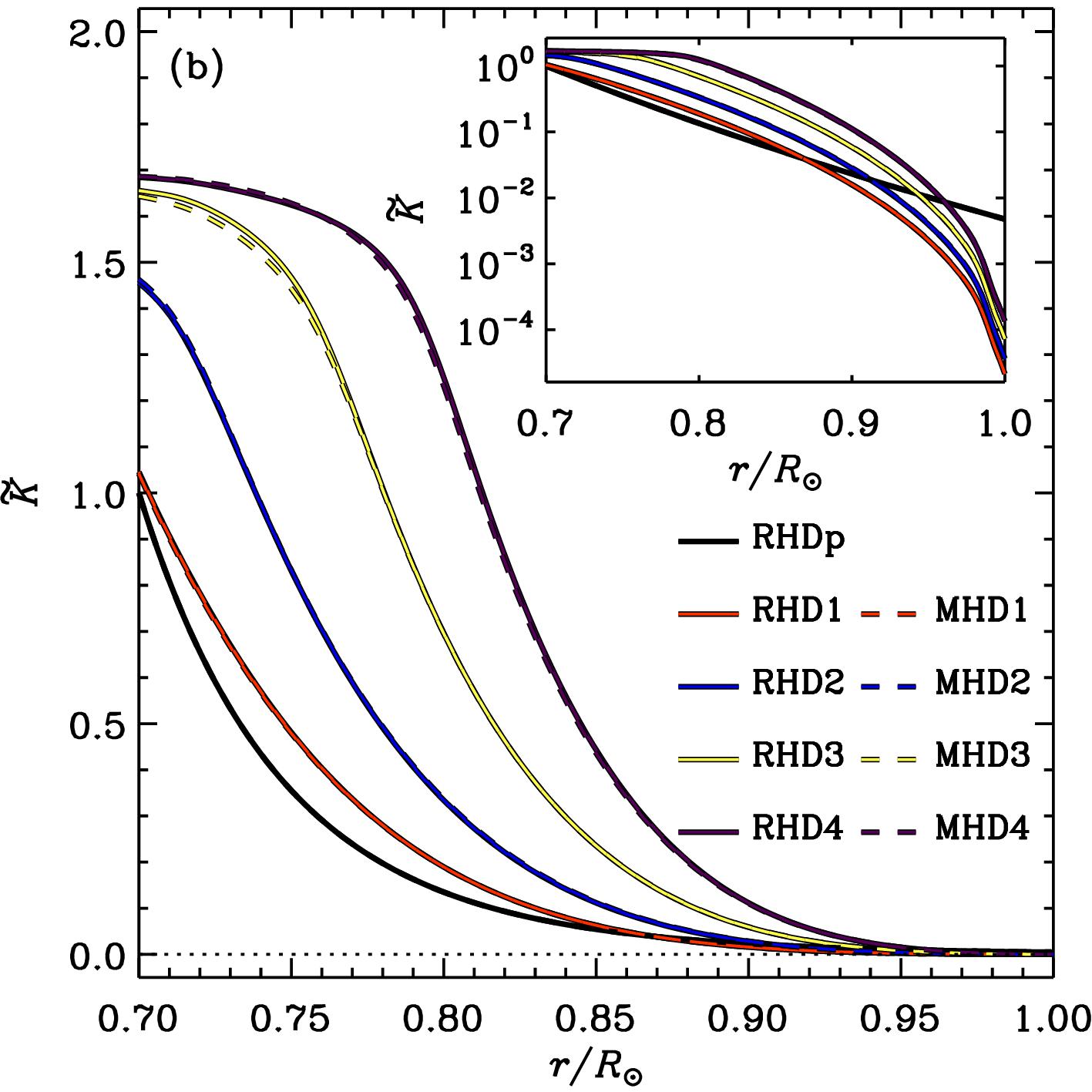}
    \caption{(a) Profiles of the initial (solid lines) and saturated (dashed)
        mean heat conductivity profiles $\tilde{K}=\brac{K}_{\theta\phi}/K_0^{\rm
        nom}$. (b) Same as (a) but for sets RHD (solid lines) and MHD
        (dashed lines) from the saturated states (colour online).}
\label{fig:pkappa}
\end{center}
\end{figure}

We begin by inspecting horizontally averaged diagnostic
  quantities from our simulations. The profiles of
  $\brac{K}_{\theta\phi}$ from the HD set in the initial and thermally
  saturated states are shown in \figu{fig:pkappa}(a). The mean
  $K$-profiles in
  Runs~HD1 and HD2 remain almost unaffected in the thermally relaxed
  regime. In Runs~HD3 and HD4, the lower parts of the domain
  ($r\lesssim0.73R_\odot$ and $r\lesssim0.76R_\odot$, respectively),
  where $\brac{K}_{\theta\phi}$ is the largest, become convectively
  stable and a
  lower temperature gradient is sufficient to carry the luminosity
  through these layers. In the RHD and MHD runs the temperature
  gradient is steeper throughout and the values of
  $\brac{K}_{\theta\phi}$ are
  reduced overall; see \figu{fig:pkappa}(b). Furthermore, the MHD runs
  differ only marginally from their RHD counterparts. This suggests a
  relatively weak influence of magnetic fields in the current
  simulations. However, we note that the magnetic Reynolds number in
  the current simulations is relatively moderate and clearly below the
  excitation threshold for small-scale dynamo action.

\begin{figure}
\begin{center}
    \includegraphics[width=0.5\textwidth]{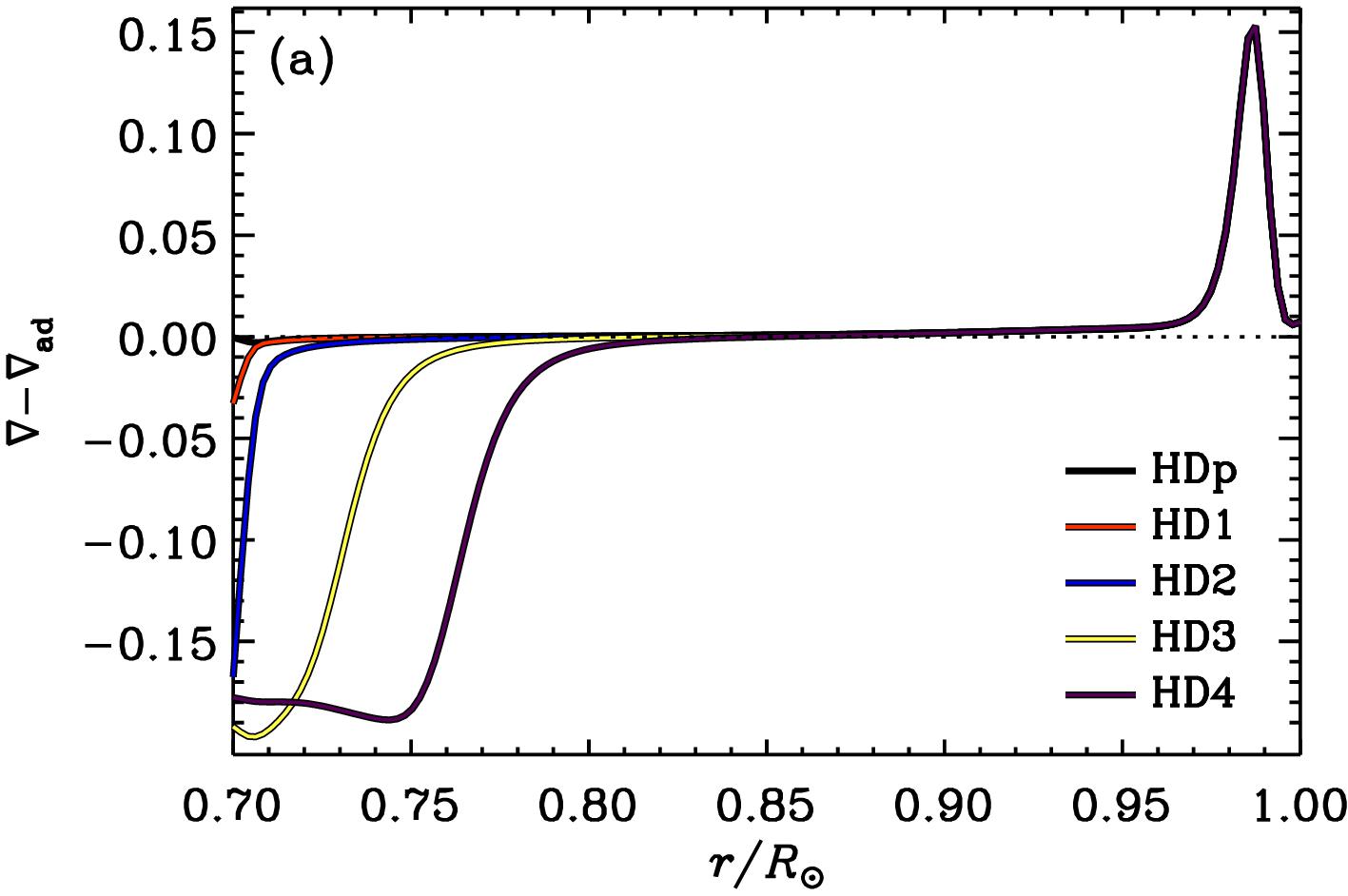}\includegraphics[width=0.5\textwidth]{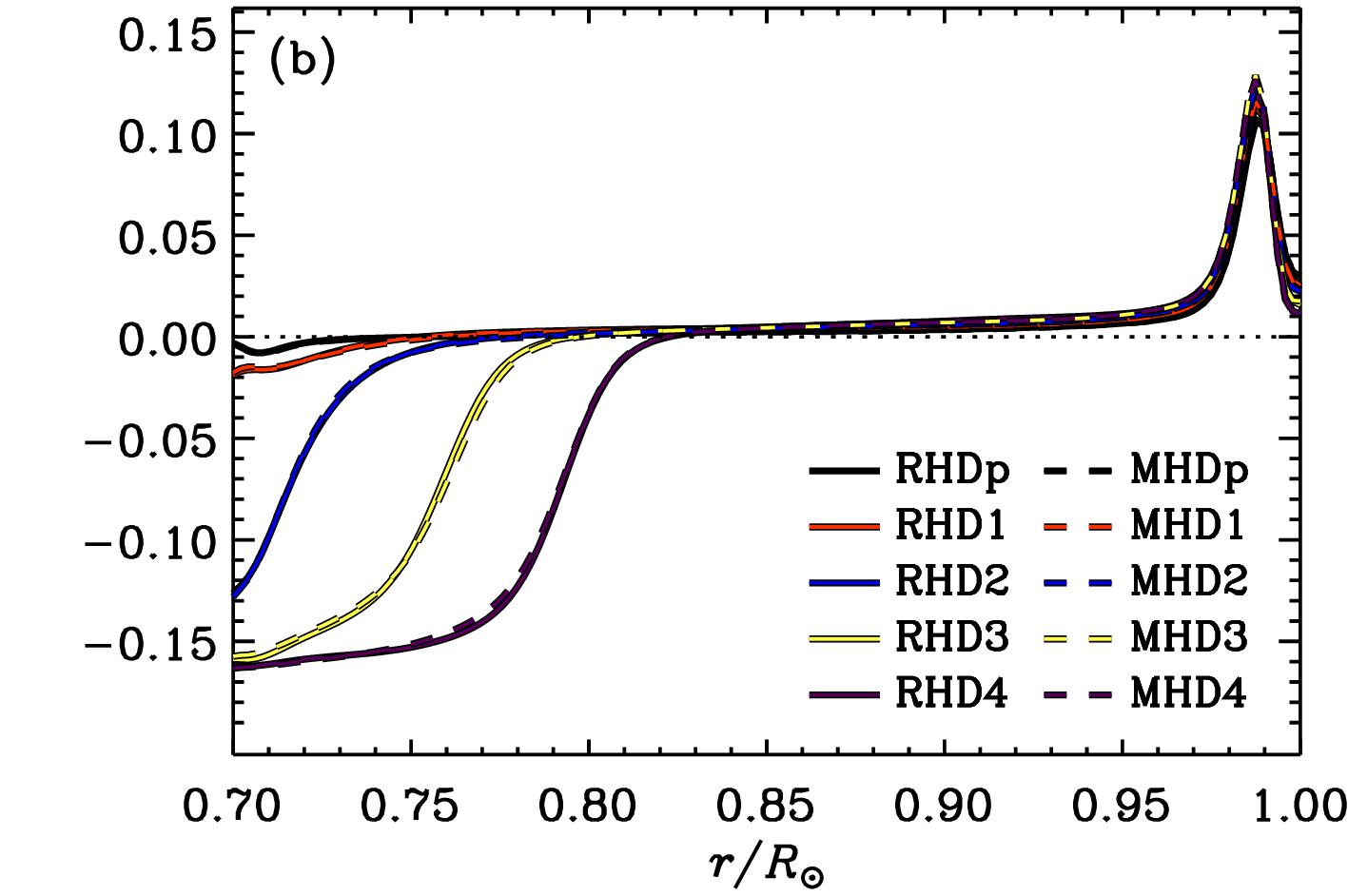}
    \includegraphics[width=0.5\textwidth]{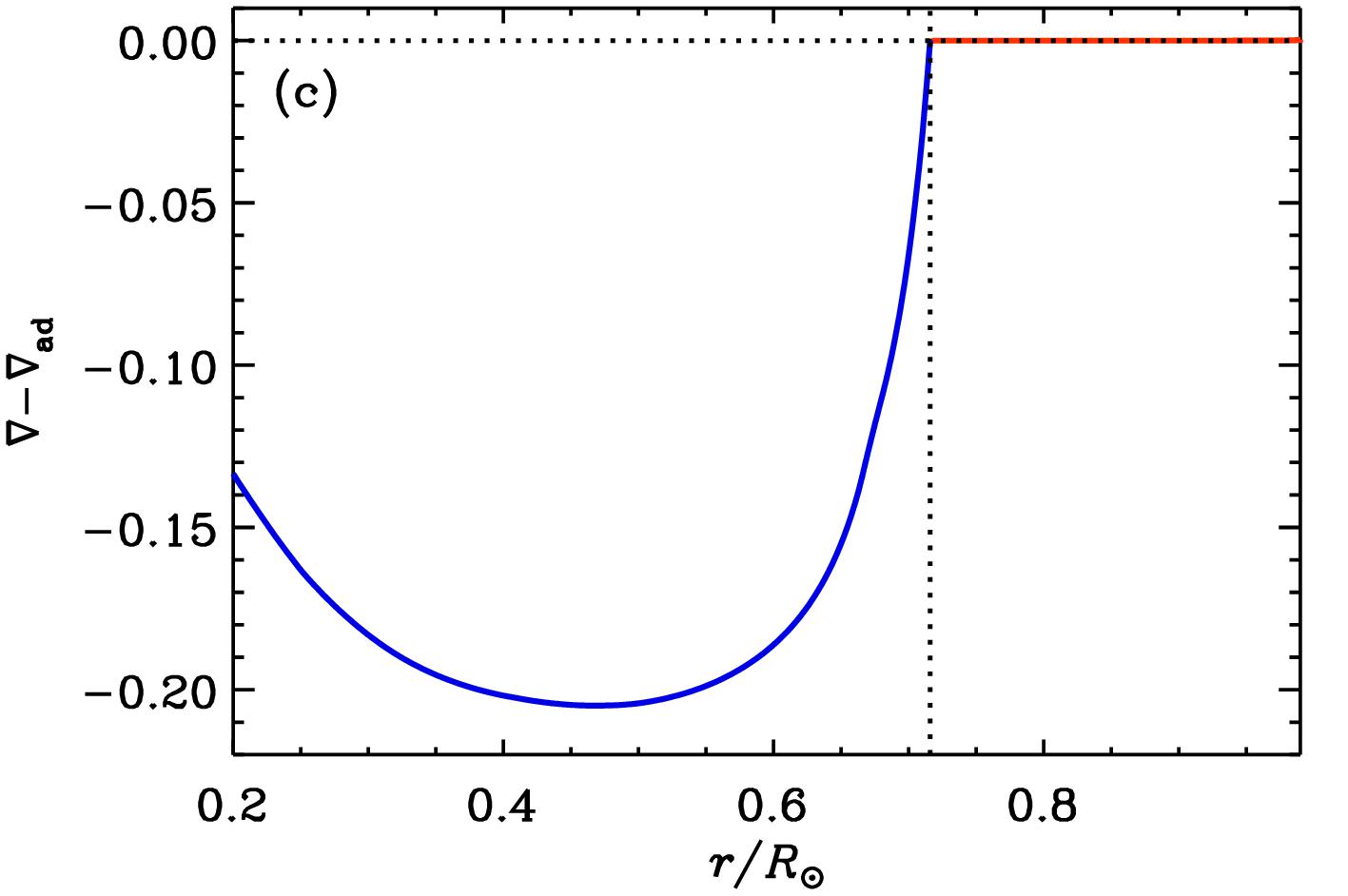}
    \caption{(a) Profiles of the superadiabatic temperature
        gradient $\nabla-\nabad$ from non-rotating HD runs. (b) Same
        as (a) but for sets RHD (solid lines) and MHD (dashed). Panel
        (c) shows $\nabla-\nabad$ from a standard solar model in the
        range $r/R_\odot=0.2\ldots0.99$. The blue (red) curve
      corresponds to the radiative (convection) zone with the
      interface marked by the dotted vertical line at
      $r/R_\odot=0.716$ (colour online).}
\label{fig:pnabla_all}
\end{center}
\end{figure}

The horizontally and temporally averaged superadiabatic
  temperature gradient $\nabla-\nabad$ is shown for all of our runs in
  \figu{fig:pnabla_all}(a)--(b). We find that in Runs~HDp and HD1, as well as
  their rotating and MHD counterparts, $\nabla-\nabad$ is close to
  zero in the bulk of the domain, with a mildly subadiabatic layer near
  the base. Furthermore, with increasing $K_0$, a gradually deeper
  subadiabatic layer forms in the lowermost parts of the domain. The
  values of $\nabla-\nabad$ in the `3' and `4' runs of all sets are on
  the order of $-0.19\ldots-0.16$. This is close to that of the
  hydrostatic case which approaches a polytropic state with index
  $n=3.25$ \citep{BB14}. Panel (c) of \figu{fig:pnabla_all} shows the
  superadiabatic temperature gradient from a standard solar model
  produced with the Yale Rotating Stellar Evolution Code (YREC)
  \citep[][]{2008Ap&SS.316...31D,2017ApJ...838..161S}. The minimum
  values of $\nabla-\nabad$ in the radiative layer in this
  model and our simulations are comparable and about $-0.2$.

The enthalpy flux is defined as
\begin{equation}
\mean{F}_i^{\rm enth}\, = \,\cP \mean{(\rho u_i)' T'}\,,
\end{equation}
where the primes denote fluctuation from the azimuthal
mean denoted by an overbar.
We use the same nomenclature as in \cite{2017ApJ...845L..23K} to
distinguish the different layers in the domain
\citep[see also][]{2000gac..conf...85B,2015ApJ...799..142T}. This entails
classifying the layers by the signs of the radial enthalpy flux
$\Fenthr$ and the radial gradient of entropy, $\nabla_r \mean{s} = \pd
\mean{s}/\pd r$, see \Tablel{tab:class}. The bottom of the buoyancy
zone (BZ) is where $\nabla_r \mean{s}$ changes from negative to
positive, whereas the bottom of the Deardorff zone (DZ) is where
$\Fenthr$ changes from positive to negative; see \cite{Br16} for an
explanation of a non-gradient contribution to $\Fenthr$ by \cite{De66}.
Finally, the bottom of
the overshoot zone (OZ) is where the $|\Fenthr|$ falls below a
threshold value, here chosen to be 2.5 per cent of the total
  flux corresponding to luminosity $L_0$.
In the commonly accepted view, the convection zone consists of the
Schwarzschild-unstable layer without Deardorff zone
\citep[e.g.][]{Za91}. This coincides with the predictions from standard
mixing length theory \citep[e.g.][]{Vi53}. In our revised view, the
convection zone (CZ) is considered to encompass both the BZ and the
DZ.

\begin{figure}
\begin{center}
\begin{minipage}{150mm}
\subfigure[HDp]{
\resizebox*{4.75cm}{!}{\includegraphics{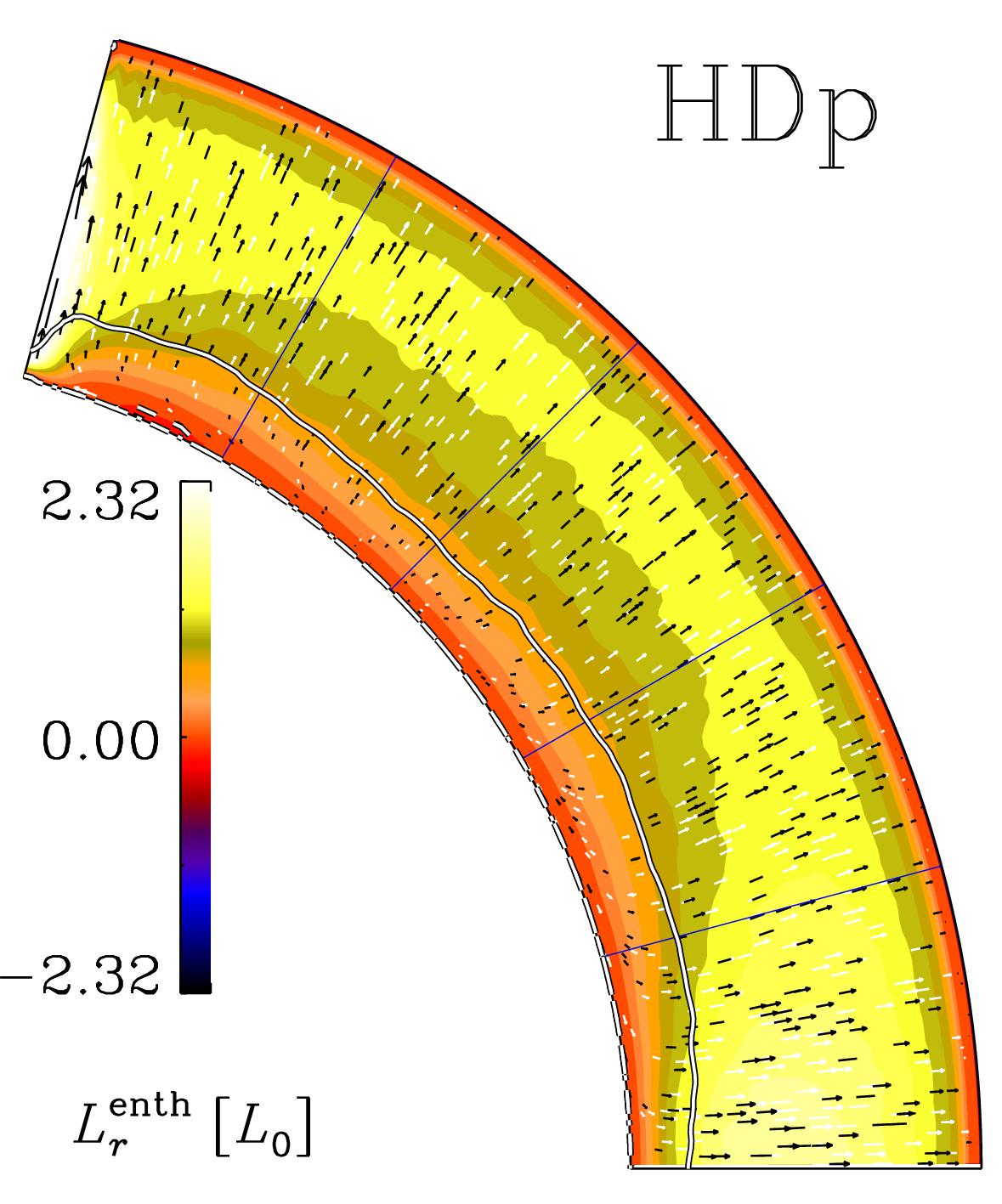}}}%
\subfigure[HD1]{
\resizebox*{4.75cm}{!}{\includegraphics{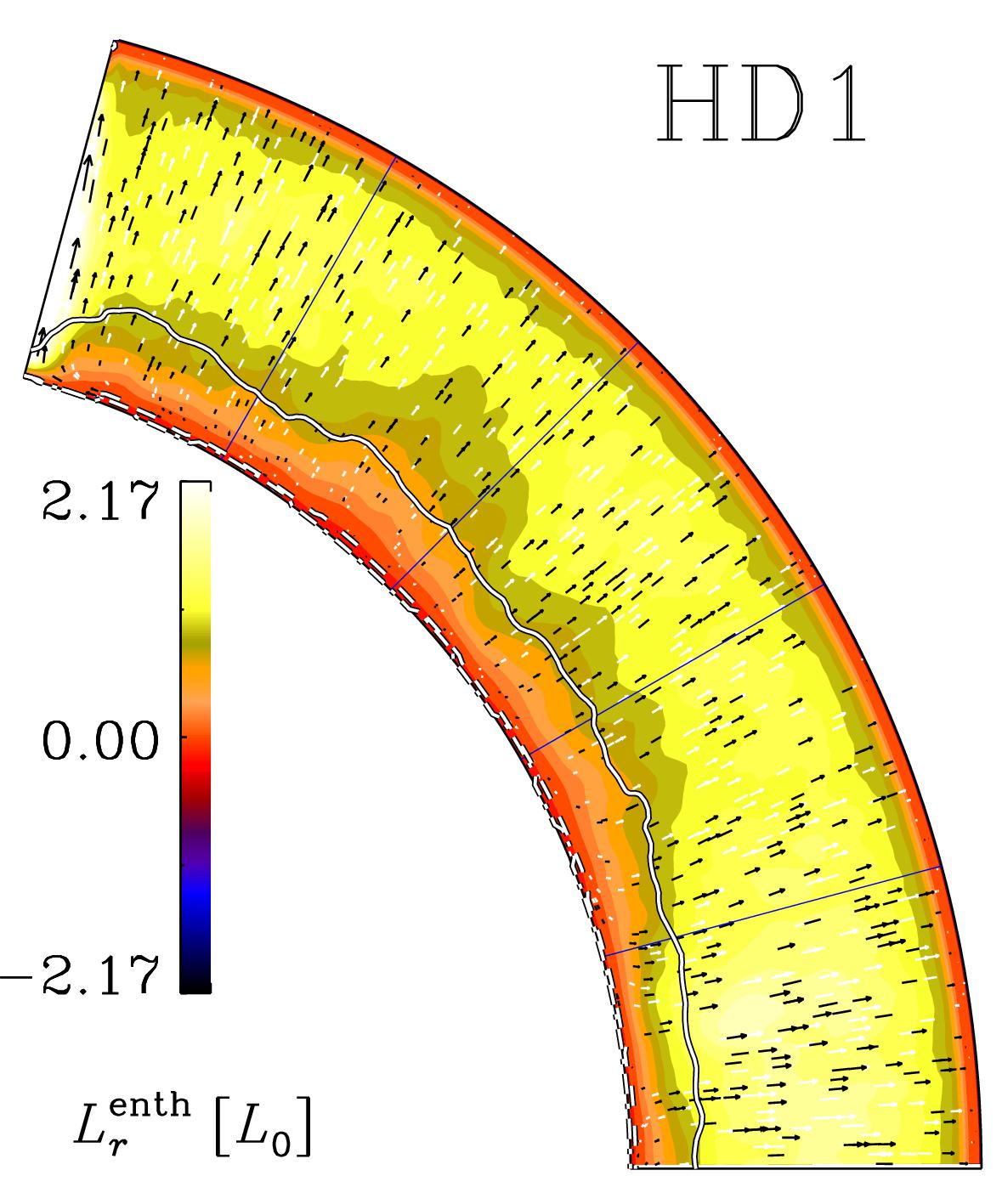}}}%
\subfigure[HD4]{
\resizebox*{4.75cm}{!}{\includegraphics{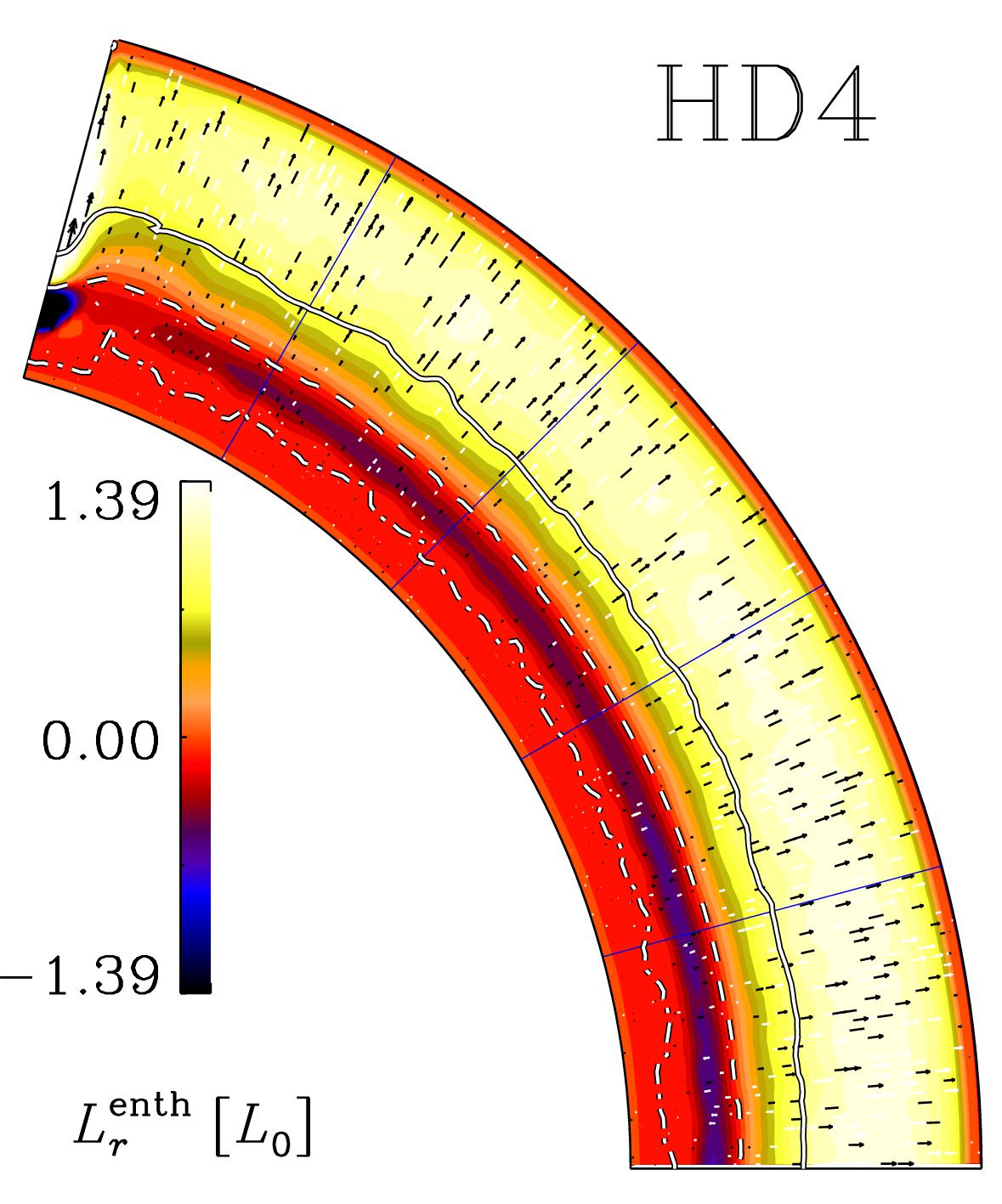}}}%
%\\
\begin{center}
\subfigure[RHDp]{
\resizebox*{4.75cm}{!}{\includegraphics{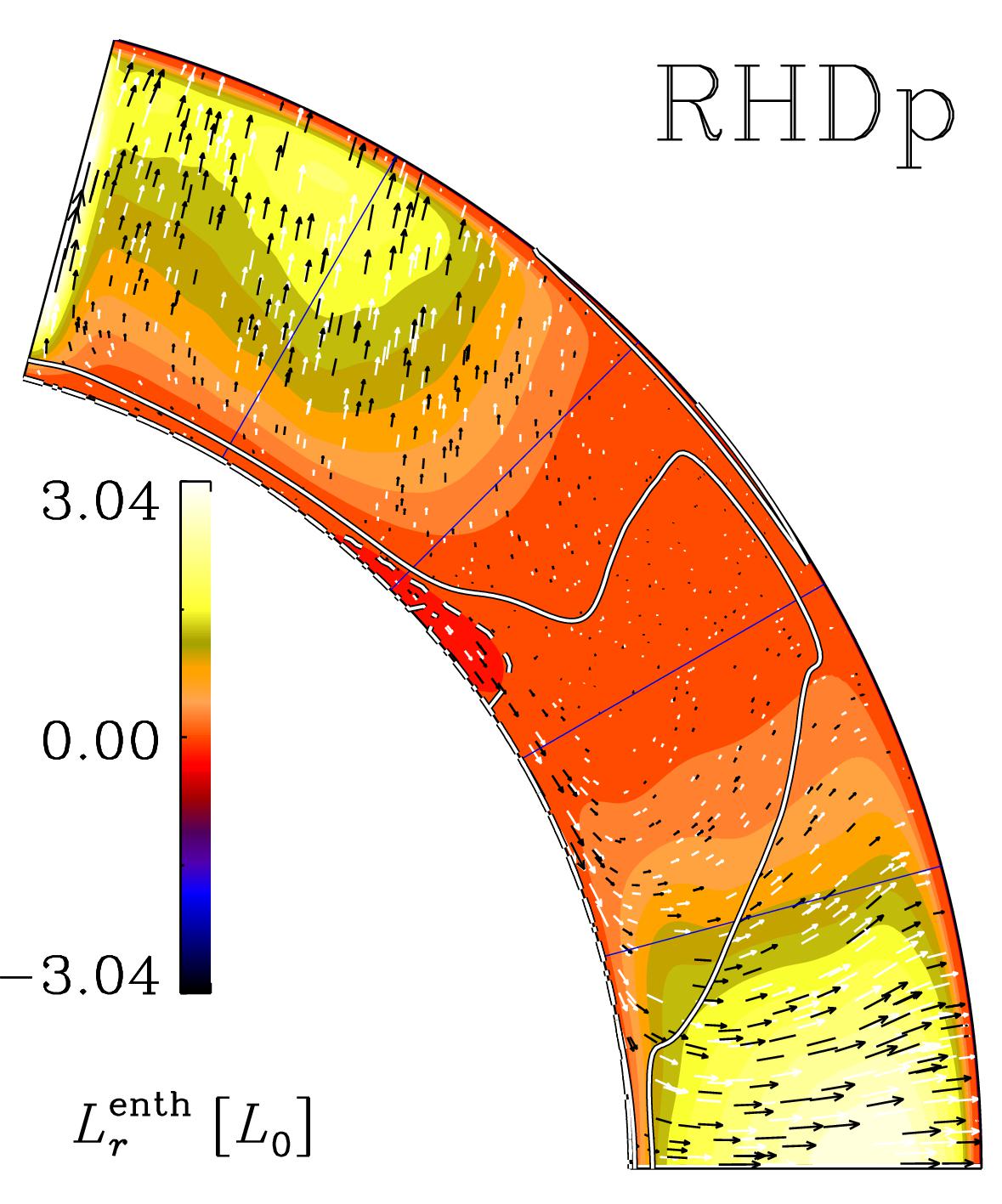}}}%
\subfigure[RHD1]{
\resizebox*{4.75cm}{!}{\includegraphics{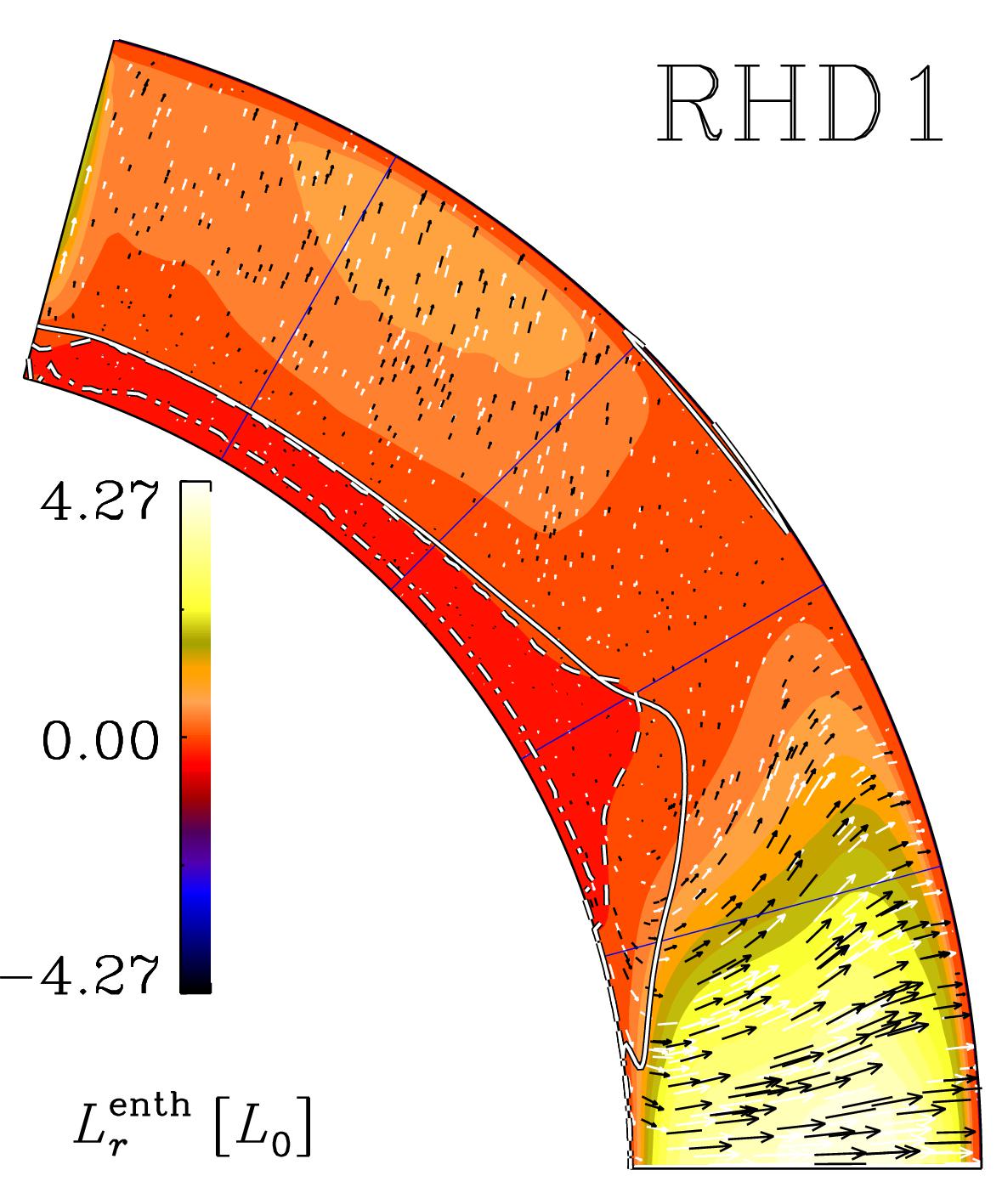}}}%
\subfigure[RHD4]{
\resizebox*{4.75cm}{!}{\includegraphics{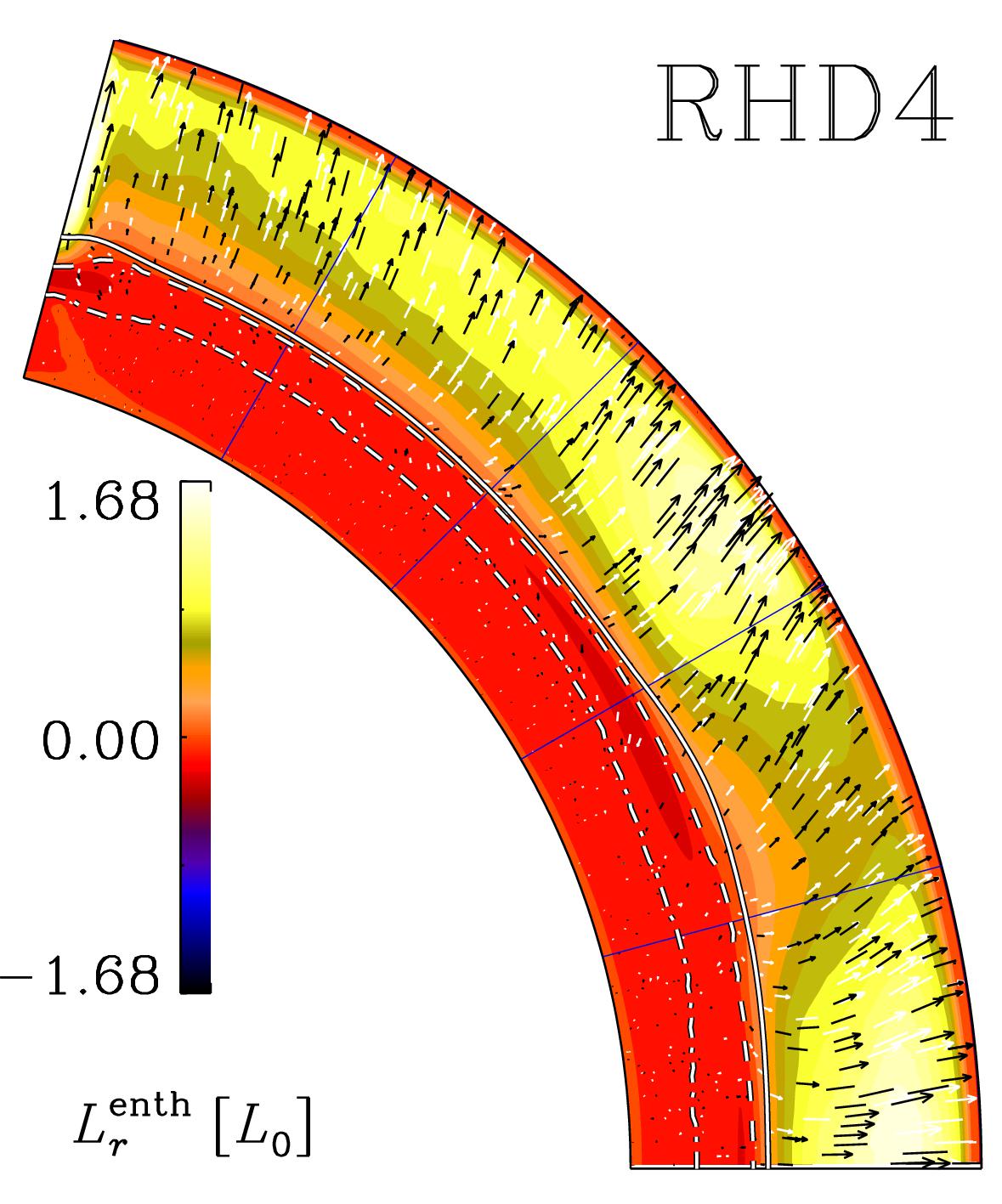}}}%
\end{center}
\begin{center}
\subfigure[MHDp]{
\resizebox*{4.75cm}{!}{\includegraphics{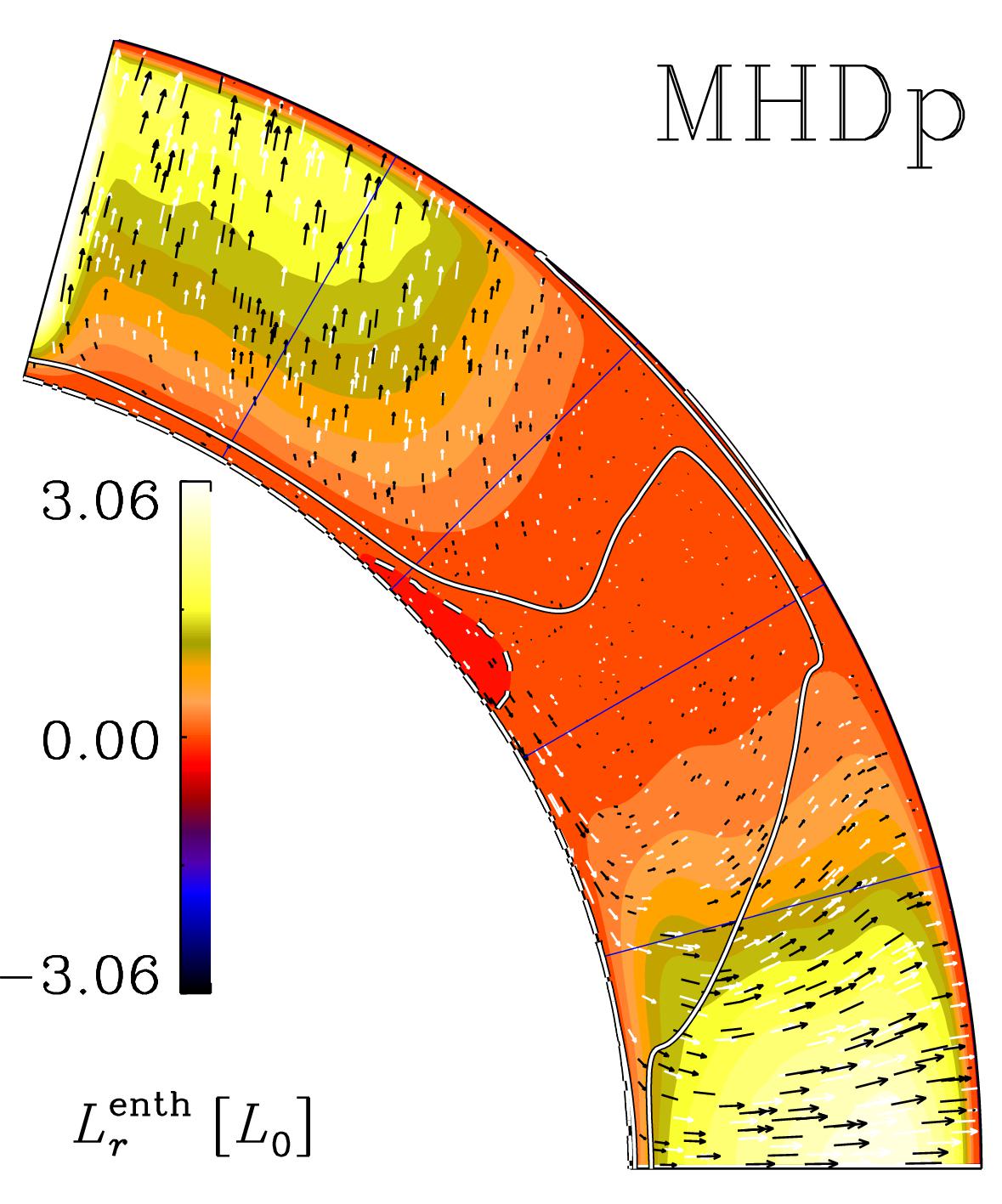}}}%
\subfigure[MHD1]{
\resizebox*{4.75cm}{!}{\includegraphics{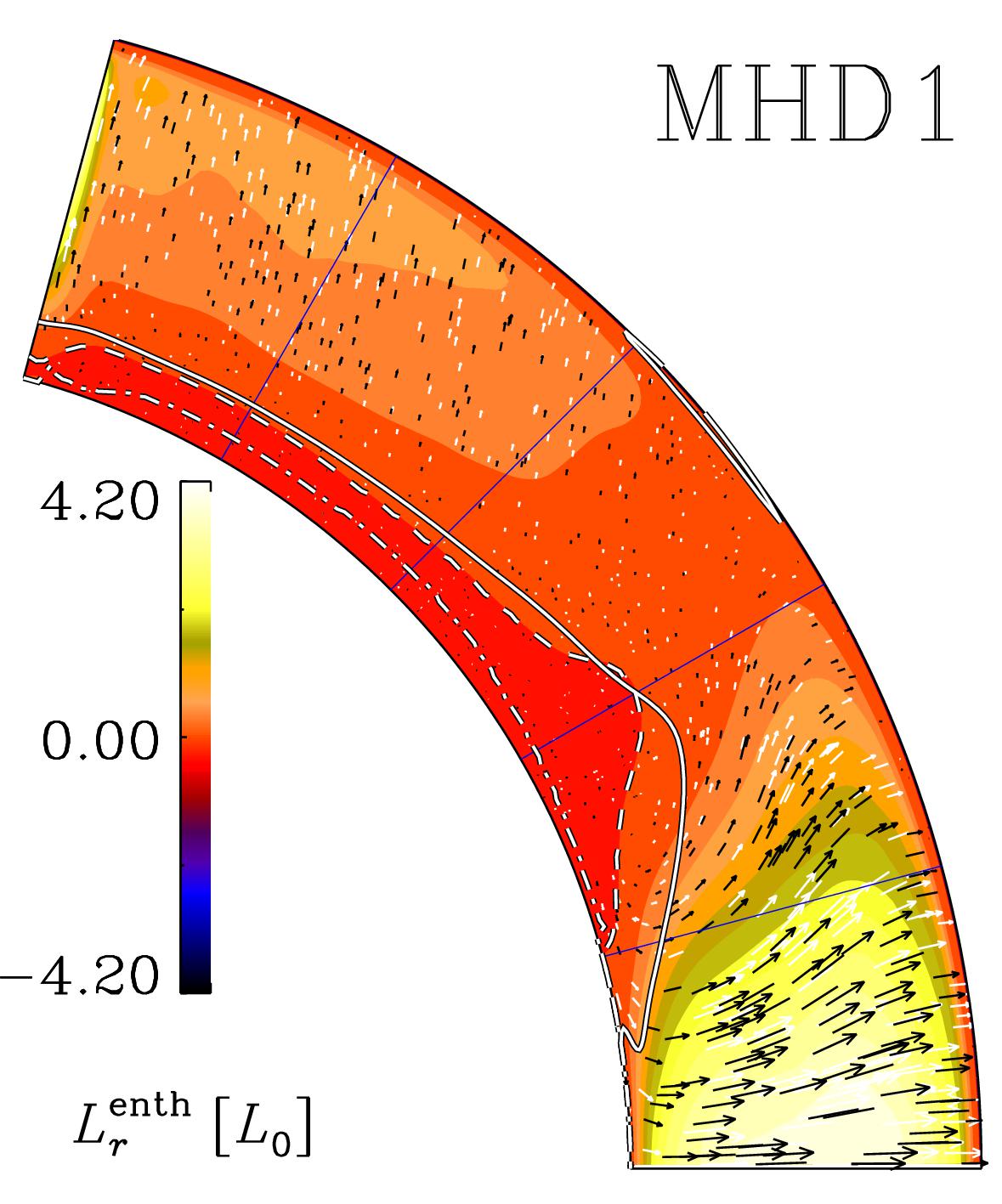}}}%
\subfigure[MHD4]{
\resizebox*{4.75cm}{!}{\includegraphics{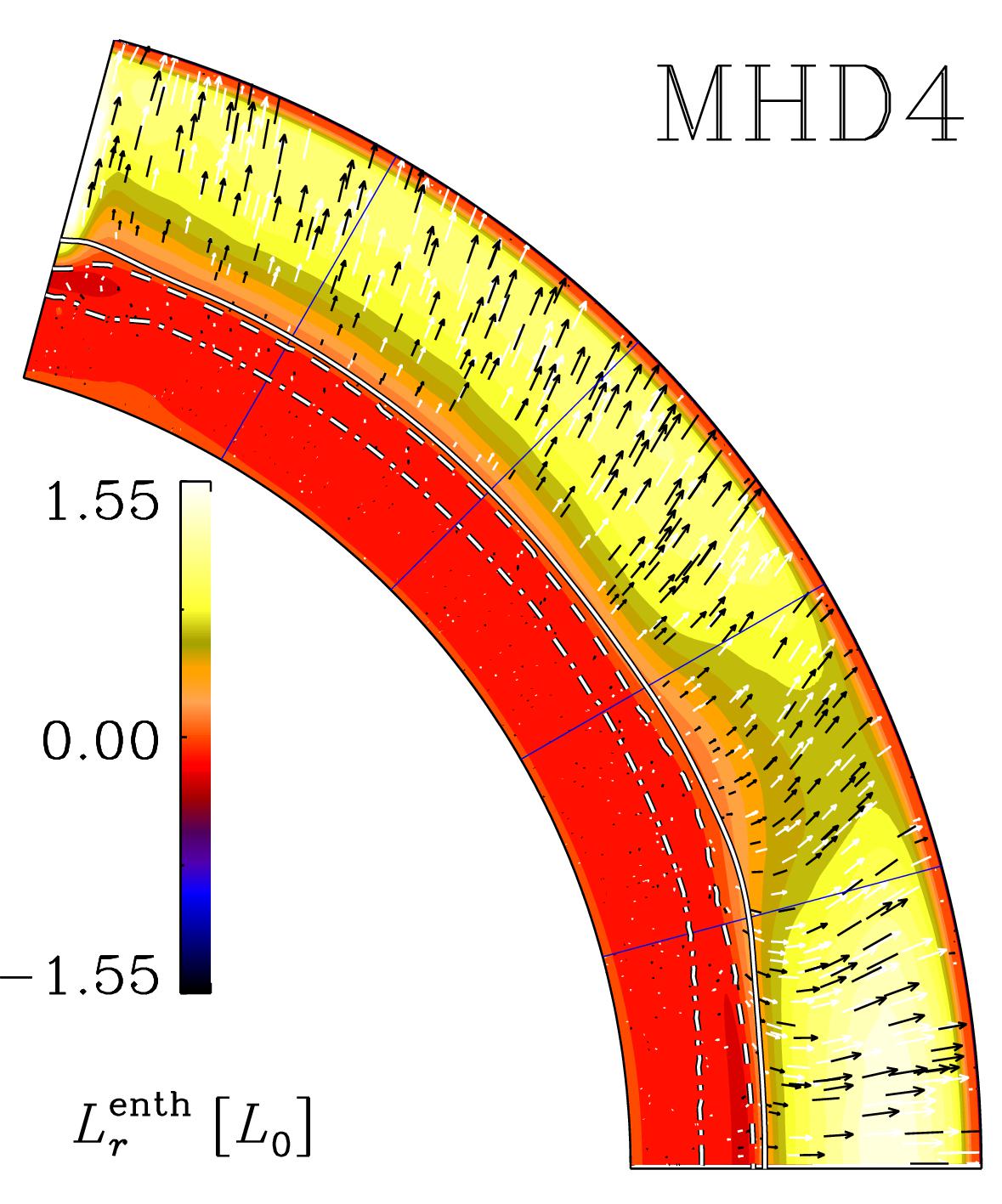}}}%
\end{center}
\caption{Colour contours: time-averaged luminosity of the radial enthalpy
  flux normalised by the total luminosity from non-rotating,
  hydrodynamic runs HDp, HD1, and HD4 (top row), rotating runs
    RHDp, RHD1, and RHD4 (middle row), and dynamo runs MHDp, MHD1,
  and MHD4 (lower row). The arrows indicate the
  magnitude and direction of the
  vectorial enthalpy flux, $\mean{\bm F}^{\rm enth} =
  (\Fenthr,\Fentht,0)$ in the meridional plane. The black and
  white solid, dashed, and dash-dotted lines indicate the bottoms of
  the buoyancy, Deardorff, and overshoot zones, respectively. The thin
  blue lines indicate latitudes $15^\circ$, $30^\circ$, $45^\circ$, and $60^\circ$
  (colour online).}
\label{fig:pFenth}
\end{minipage}
\end{center}
\end{figure}

We show the time-averaged luminosity of the radial enthalpy flux,
$L^{\rm enth}_r=4\pi r^2 \Fenthr$ and the direction of vectorial
enthalpy flux, $\mean{\bm F}^{\rm enth} = (\Fenthr,\Fentht,0)$ in the
meridional plane for a selection of runs in \figu{fig:pFenth}. In the
non-rotating, hydrodynamic run HD1, the enthalpy flux is directed
radially outward and approximately uniformly distributed in latitude with the
exception of regions in the immediate vicinity of the latitudinal
boundaries where the enthalpy flux is enhanced. The differences between
Runs~HDp and HD1 are very minor in that both develop a DZ, covering
roughly 20 per cent of the depth of the domain, at the base of the
CZ. No appreciable overshoot region develops in either run. This is
unsurprising because the heat conductivity in these models is chosen
such that it just delivers the input flux through the lower
boundary and
decreases rapidly in the upper layers, necessitating convection to
transport some fraction of the energy there. Increasing the value of
$\tilde{K}_0$ (from Runs~HD2 to HD4) enhances the radiative diffusion---in
particular in the deep parts where the temperature is high. This leads
gradually to the formation of a radiative zone (RZ) at the base of the
domain; see \figu{fig:pFenth}(c). In the non-rotating case it is
meaningful to average over latitude and to obtain estimates of the
depths of the different layers. These are listed as $d_{\rm BZ}$,
$d_{\rm DZ}$, and $d_{\rm OZ}$ in \Tablel{tab:runs}.
We note that only in Runs~HD3 and HD4 the
domain is deep enough to allow the formation of an RZ and that the
depths of the DZ and/or OZ are thus underestimated for Runs~HDp, HD1,
and HD2. A similar argument applies to the runs presented by
\cite{2017ApJ...851...74B} and \cite{2018PhFl...30d6602K}. In Runs~HD3
and HD4, the subadiabatic but mixed layers (DZ and OZ) cover 38 and 44
per cent of the total depth of the mixed zone. This is in good
agreement with the results from local simulations
\citep[e.g.][]{2017ApJ...845L..23K,2017ApJ...843...52H}.

\begin{table}
  \tbl{Classification of zones.}
{\begin{tabular}{@{}lcccc}\toprule
    Quantity/zone              & Buoyancy (BZ) & Deardorff (DZ) & Overshoot (OZ) & Radiation (RZ) \\ \colrule
    $\Fenthr$         &  $> 0$  &  $> 0$ & $< 0$ & $\approx 0$ \\ \colrule
    $\nabla_r \mean{s}$ & $< 0$ & $> 0$ & $> 0$  &  $> 0$  \\ \colrule
    \botrule
  \end{tabular}}
\label{tab:class}
\end{table}

This picture is radically altered in the rotating cases; see the
middle and lower
panels of \figu{fig:pFenth}. The most prominent new feature is the
strong latitude dependence of the radial enthalpy flux: the energy
transport is strongly concentrated toward high latitudes
and near the equator.
Comparing figures~\ref{fig:pFenth}(d)--(f) and \ref{fig:pFenth}(g)--(i)
  shows that the qualitative differences between the RHD and MHD runs are
  small. This is also the case for most other diagnostics and thus we
  will mostly discuss the MHD cases in what follows.
A major difference between Runs~MHDp and MHD1 is that in the former,
the enthalpy flux is roughly equally efficient at high and low
latitudes ($\Theta\gtrsim55^\circ$ and $\Theta\lesssim30^\circ$, where
$\Theta=90^\circ-\theta$ is the latitude), whereas in the latter the
high latitude flux is suppressed. Another difference is that in
Run~MHDp the Deardorff layer at mid-latitudes ($20^\circ\lesssim
\Theta \lesssim 35^\circ$) covers almost the entire depth of the domain
whereas in Run~MHD1 the latitude variation is less extreme although
still substantial; see the solid black and white lines in
\figu{fig:pFenth}(g) and (h). Near the equator
($\Theta\lesssim10^\circ$), the Deardorff layer is either very thin
(MHDp) or missing completely (MHD1). A possible explanation to the
very deep mid-latitude Deardorff layer in Run~MHDp is that the current
simulations are only moderately supercritical in terms of the Rayleigh
number and that convection is dominated by polar and equatorial
  modes in such parameter regimes \citep[e.g.][]{Gi77}. This is
exacerbated by
the rigid combination of a fixed heat conductivity profile and a
constant temperature boundary condition applied at the radial top
boundary. The reason why the Deardorff zone is significantly shallower
in Run~MHD1 is because, unlike in Run~MHDp with a fixed
  $K$-profile, the heat conductivity adapts in response to
changes in the thermal structure. Convectively stable mid-latitudes
have been reported from similar simulation setups with fixed
$K$-profile and surface temperature by \cite{KMGBC11}. However, in
cases where, for example, a black body radiation condition is applied
at the surface, the mid-latitudes remain convectively unstable and
allow for significant latitudinal variation of the surface temperature
\citep{WKKB16,2018arXiv180709309K}. This is likely due to the enhanced luminosity used in
the current simulations. We also note that in the rotating cases the
heat flux
is mostly radial near the equator, but more inclined with the
rotation vector at high latitudes. This is a manifestation of
latitudinal turbulent heat flux, which is often invoked to break the
Taylor-Proudman balance in the Sun \citep[e.g.][]{BMT92,KR95}.
A poleward enthalpy flux has been reported in numerous earlier studies
\citep[e.g.][]{PTBNS93,2005A&A...431..345R,KMGBC11,2017ApJ...836..192B}.

The strong latitudinal variation of the depths of the various layers
render latitudinal averaging of $r_{\rm BZ}$, $r_{\rm DZ}$, and
$r_{\rm OZ}$ useless in these cases. In
runs where an RZ develops (`3' and `4' runs in each set), the
latitudinal variation
of the depths of the different zones and of the enthalpy flux are
significantly weaker, see \figu{fig:pFenth}(f) and (i). However, for
completeness, we list the latitudinally averaged coordinates of the
bottoms of BZ, DZ, and OZ and the depths of the corresponding layers
for all runs in \Tablel{tab:runs}. The values for runs RHDp, RHD1,
RHD2, MHDp, MHD1, and MHD2, where strong latitudinal variations are seen,
are listed in parentheses and should be considered as uncertain. We
found that the DZ diminishes substantially in Runs~MHD3 and MHD4
in comparison to the non-rotating case HD3 and HD4, whereas the depth of the
OZ is influenced less. It is also noteworthy that in the rotating
Kramers-based Runs~RHD1 and MHD1, the overall velocities, measured by
the Reynolds numbers, are higher than in the fixed profile runs RHDp and
MHDp; see the fifth column of \Tablel{tab:runs}.

\begin{figure}
\begin{center}
   \includegraphics[width=0.65\textwidth]{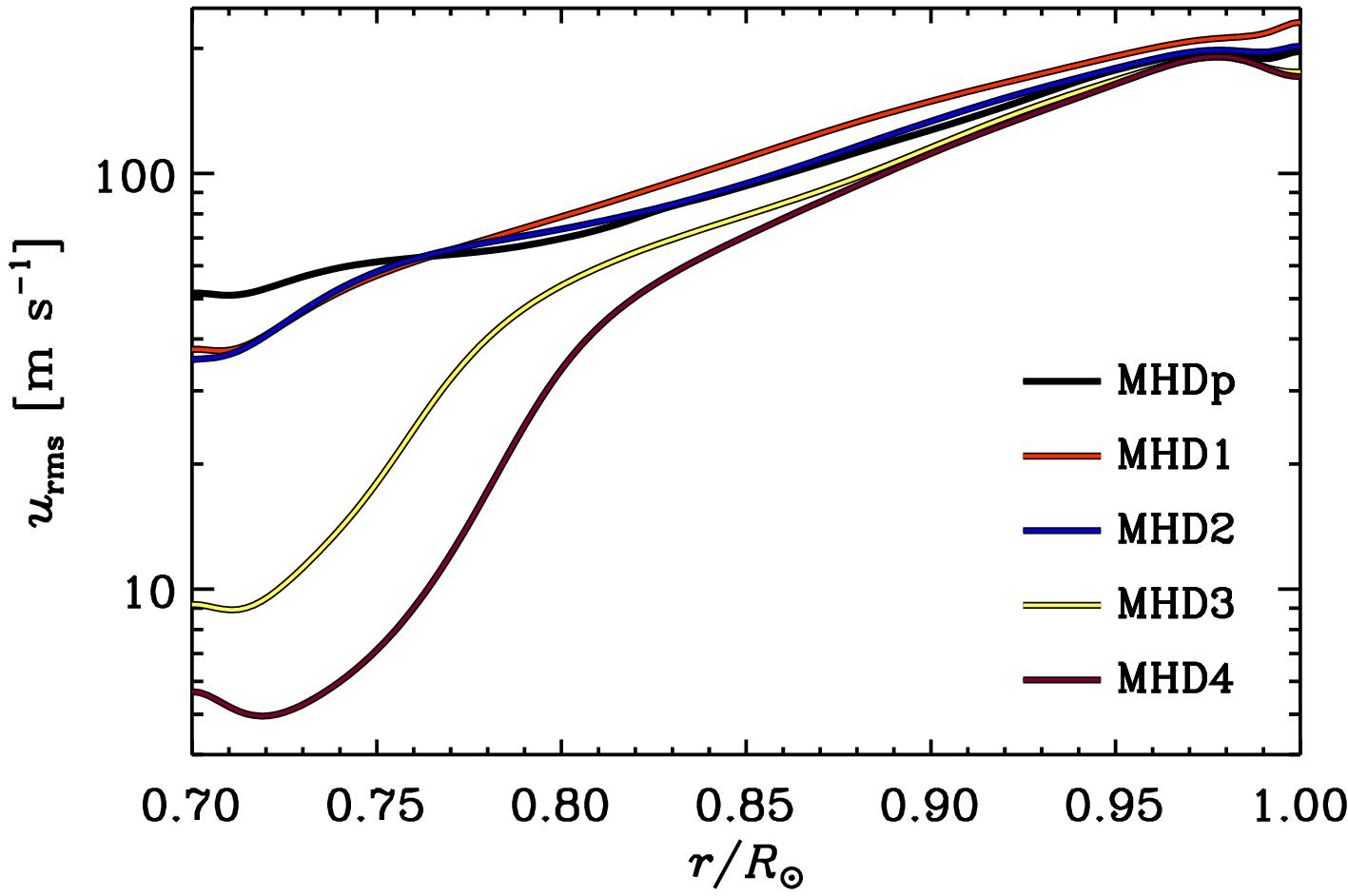}
    \caption{Horizontally averaged rms velocity $\urms$ from the
        runs in the MHD set (colour online).}
\label{fig:purms}
\end{center}
\end{figure}

Figure~\ref{fig:purms} shows the temporally and
  horizontally averaged rms velocity from the runs in the MHD set with
  the same definition of $\urms$ as used in (\ref{equ:Rey}). The
  velocities
  near the surface are not much affected by the appearance of a
  radiative layer in the deep parts. A significant decrease of $\urms$
  in the latter occurs only for the `3' and `4' runs in each set of
  simulations. In such cases the definitions of the Reynolds
  and Coriolis numbers in \eqs{equ:Rey}{equ:Cori}, respectively, become
  inaccurate. We thus provide these diagnostics computed using the rms
  velocity and the depth of the revised convection zone $r_{\rm DZ} <
  r < R_\odot$ in
  brackets in the fifth and sixth columns of \Tablel{tab:runs}.

Our earlier Cartesian study indicated that the downflows are mostly
responsible for the enthalpy flux in non-rotating overshooting
convection \citep[cf.\ Figure~2(c) of][]{2017ApJ...845L..23K}. However,
convective flows produce also a substantial (but downward) kinetic energy flux
\begin{eqnarray}
\mFkin\,=\,\onehalf\, \mean{\rho \bm{u}^2 u_r}\,,
\end{eqnarray}
where $\uuu= \UUU - \mUUU$.
Thus the total convected flux
\begin{eqnarray}
\mFconv\, =\, \mFenth + \mFkin\,,
\label{equ:cfluxes}
\end{eqnarray}
can be substantially different from the enthalpy flux. This is
particularly true for the downflows, where the signs of the enthalpy
and kinetic energy fluxes are opposite \citep[cf.\ Figure~1
  of][]{2017ApJ...845L..23K}. An early study \citep{CBTMH91}
suggested that the two contributions nearly cancel for the
downflows. However, later studies of \cite{CG92} and \cite{BCT02}
confirmed that partial cancellation occurs, but that the
downflows still contribute approximately equally much as the
  upflows to the total energy transport. The main
difference between the study of \cite{CBTMH91} and those of
\cite{CG92} and \cite{BCT02} is that the latter include a stably
stratified overshoot layer below the CZ, whereas in the former the
whole domain is convectively unstable.

\begin{figure}
\begin{center}
\begin{minipage}{150mm}
\subfigure[HDp]{
\resizebox*{7.5cm}{!}{\includegraphics{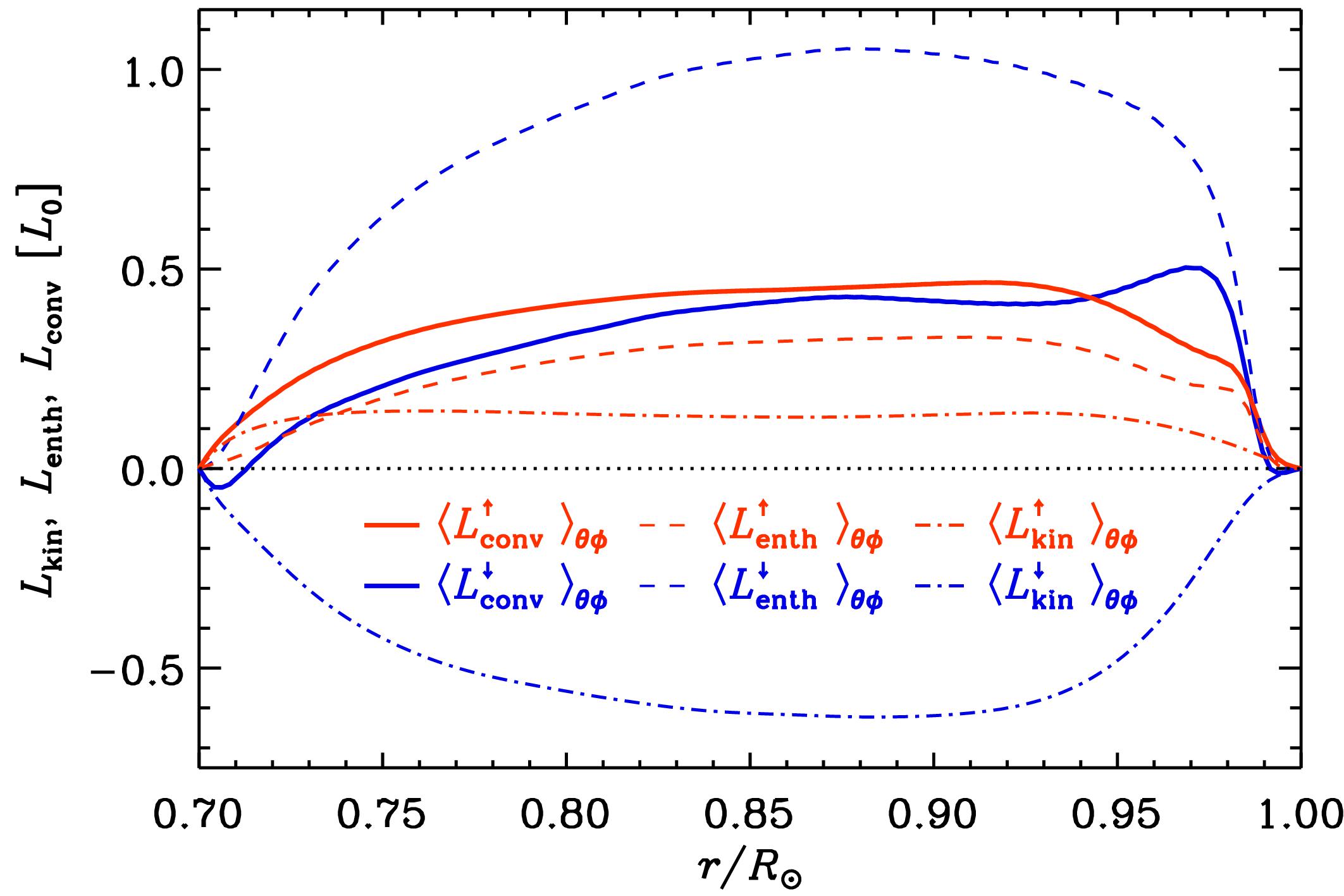}}}%
\subfigure[HD4]{
\resizebox*{7.5cm}{!}{\includegraphics{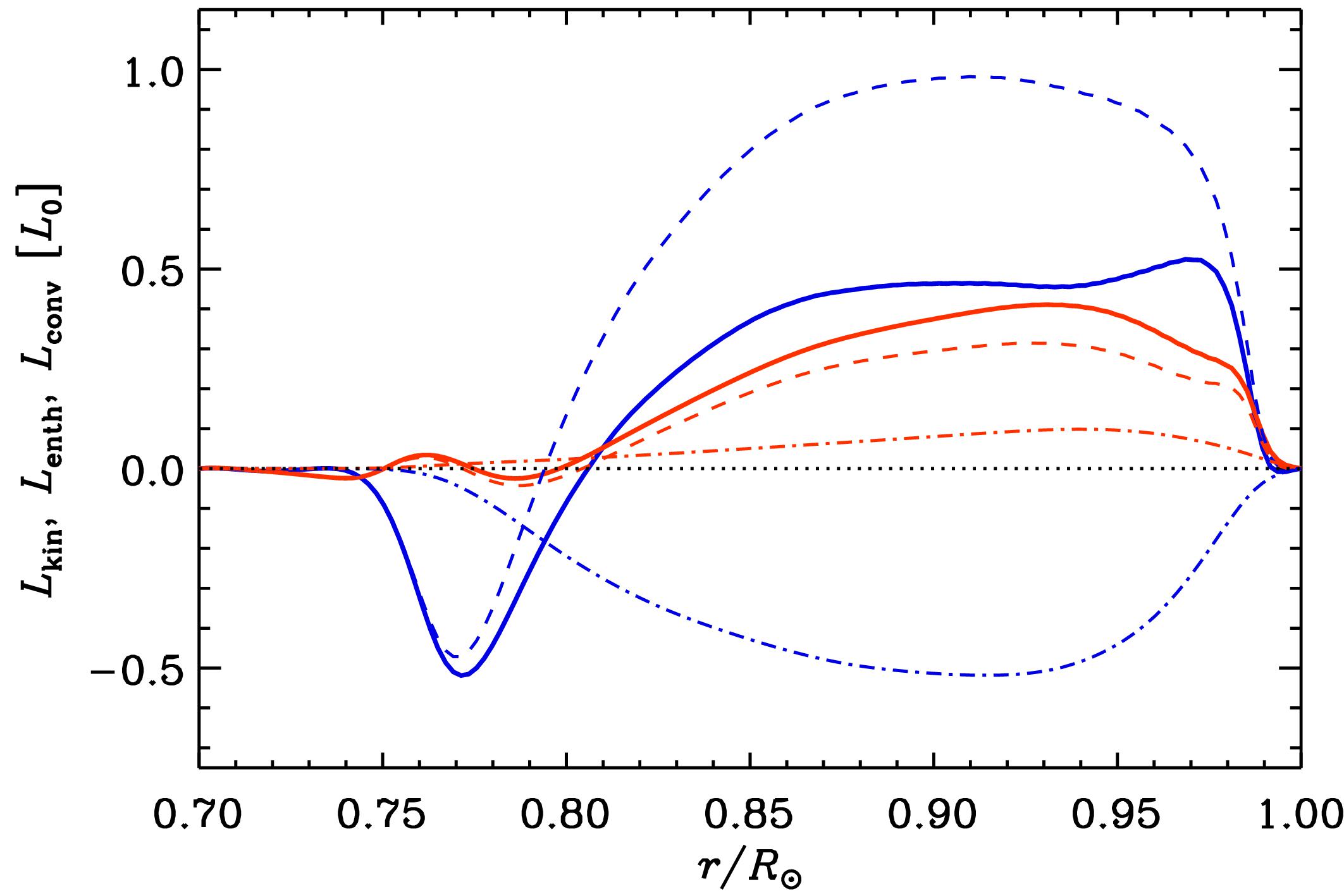}}}%
\\
\begin{center}
\subfigure[MHD1]{
\resizebox*{7.5cm}{!}{\includegraphics{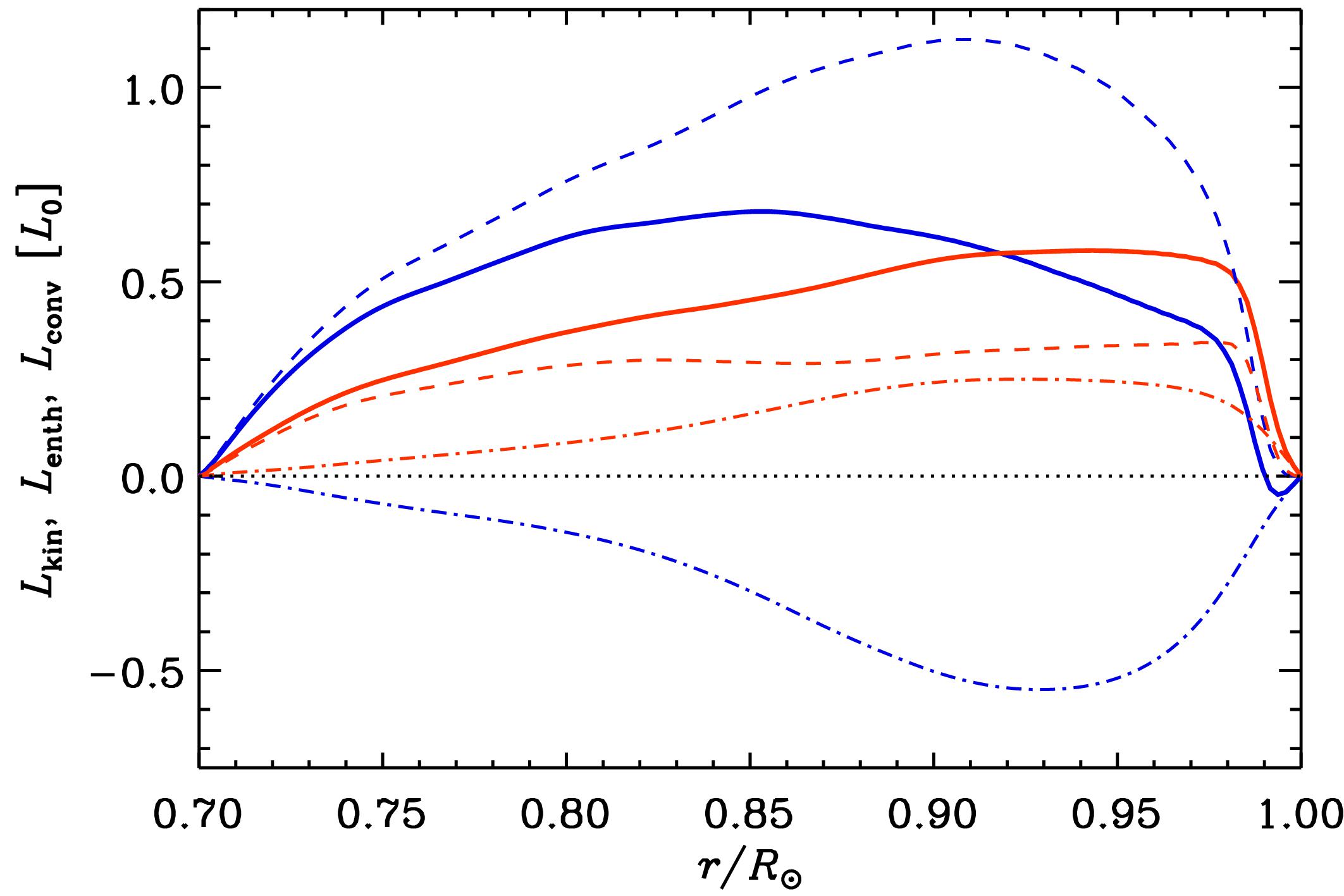}}}%
\subfigure[MHD4]{
\resizebox*{7.5cm}{!}{\includegraphics{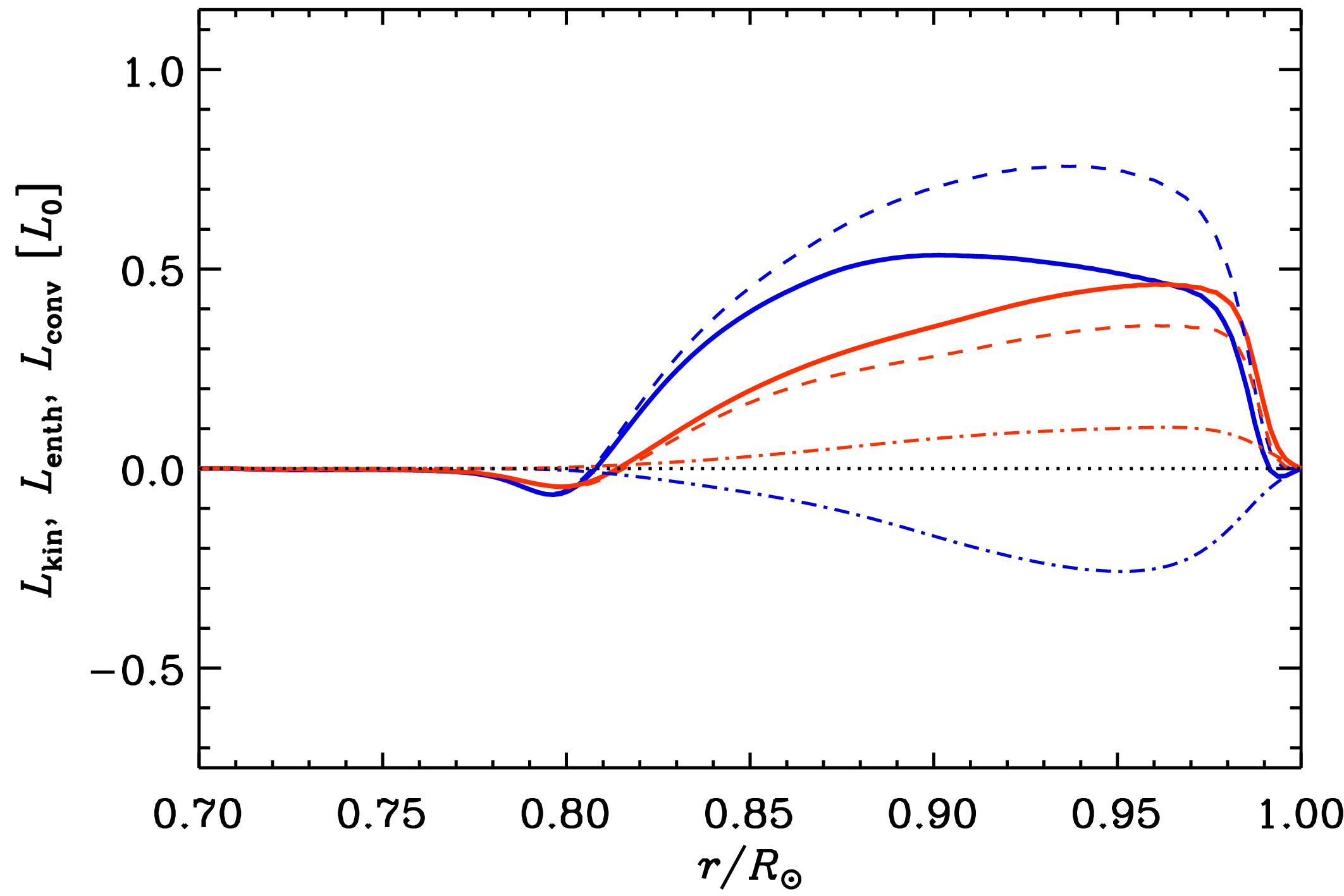}}}%
\end{center}
\caption{Convective (thick solid), enthalpy (dashed), and kinetic
  energy (dash-dotted) fluxes for upflows (red) and downflows (blue)
  from Runs~HDp, HD4, MHD1, and MHD4 (colour online).}
\label{fig:pfluxes}
\end{minipage}
\end{center}
\end{figure}

We study the detailed flux dynamics by separating the convective flux
into kinetic and
enthalpy fluxes from up- and downflows and represent them in terms of
the corresponding luminosities:
\begin{eqnarray}
\Lconvtp\, &=&\, \Lenthtp + \Lkintp\,,  \\
\Lenthtp\, &=&\, \Lenthutp+ \Lenthdtp\,,\\
\Lkintp \, &=&\, \Lkinutp + \Lkindtp\,.
\end{eqnarray}
Here $\uparrow$ and $\downarrow$ refer to contributions from up- and
downflows, respectively, and $L_i=4\pi r^2 F_i$ are the corresponding
luminosities. Representative
results are shown in \figu{fig:pfluxes} from Runs~HDp, HD4, MHD1, and
MHD4.
We find that both $\Lenthdtp$ and $\Lkindtp$ are large and of
  opposite sign, leading to a net positive $\Lconvdtp$ that is much
  smaller than either of its constituents. However, $\Lconvdtp$ contributes
  equally, or more, than the upflows ($\Lconvutp$) to the total
  convected flux ($\Lconvtp$) in all cases. This agrees with the
Cartesian simulations of \cite{CG92}, \cite{BCT02} and
\cite{2017ApJ...845L..23K}. No qualitative difference is seen between
setups without and with stably stratified overshoot and radiative
layers.
These results are contrasted with those of
  \cite{2017PhRvE..96c3104K} from Boussinesq convection, where the
  upflows contribute only to downward transport of thermal energy.

\subsection{Force balance}

Recently, \cite{2017ApJ...845L..23K} and \cite{2017ApJ...843...52H}
studied the force balance on up- and downflows in non-rotating
Cartesian convection. A remarkable result from these studies is that
the downflows appear to feel the Schwarzschild criterion, such that
they are accelerated in unstable and decelerated in stable regions,
while the upflows do not appear to do so. Here we study whether this
result holds also in astrophysically more realistic setups that
include rotation and magnetic fields in spherical coordinates.

\begin{figure}
\begin{center}
\begin{minipage}{150mm}
\subfigure[HDp]{
\resizebox*{3.75cm}{!}{\includegraphics{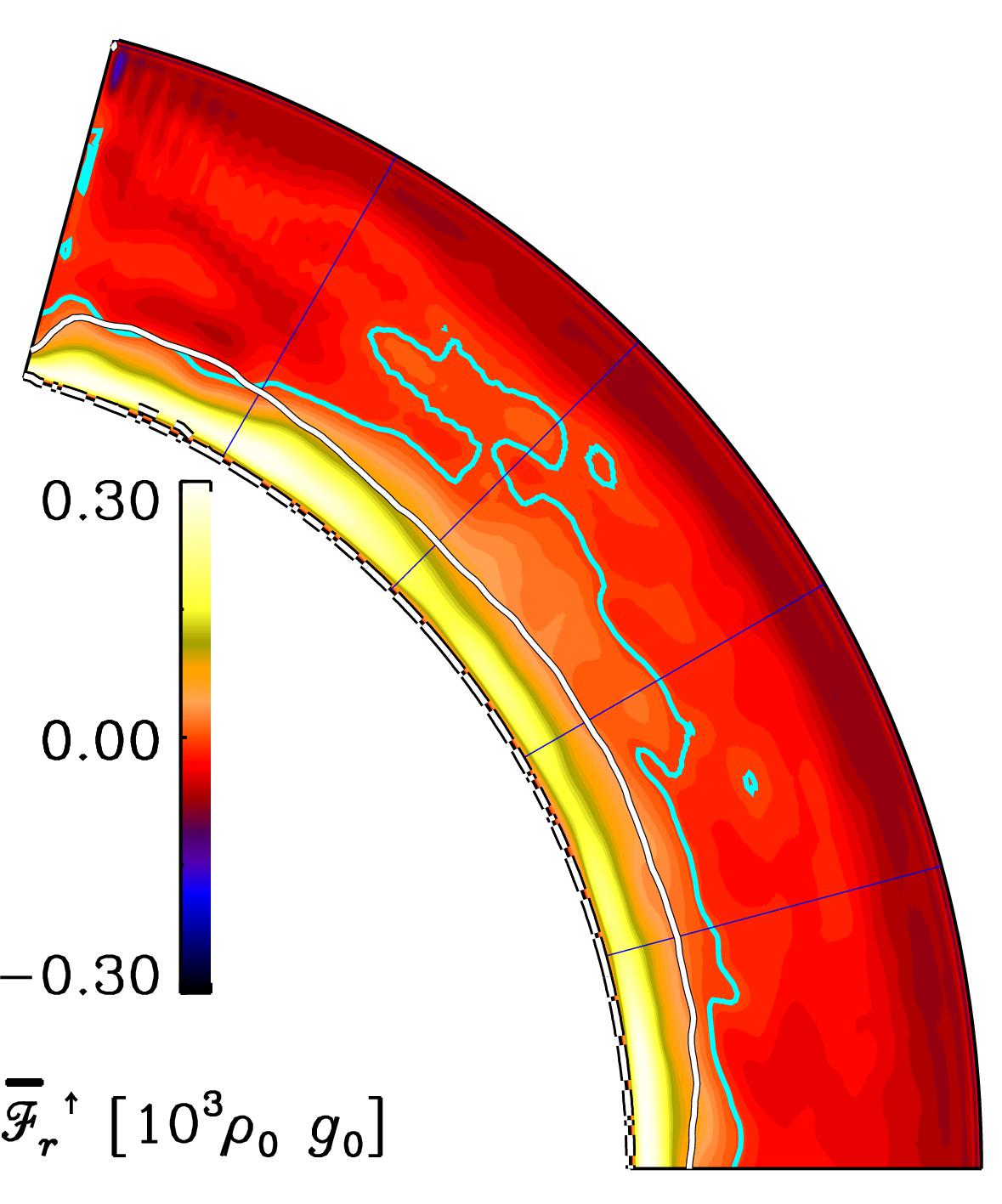}}}%
\subfigure[HD4]{
\resizebox*{3.75cm}{!}{\includegraphics{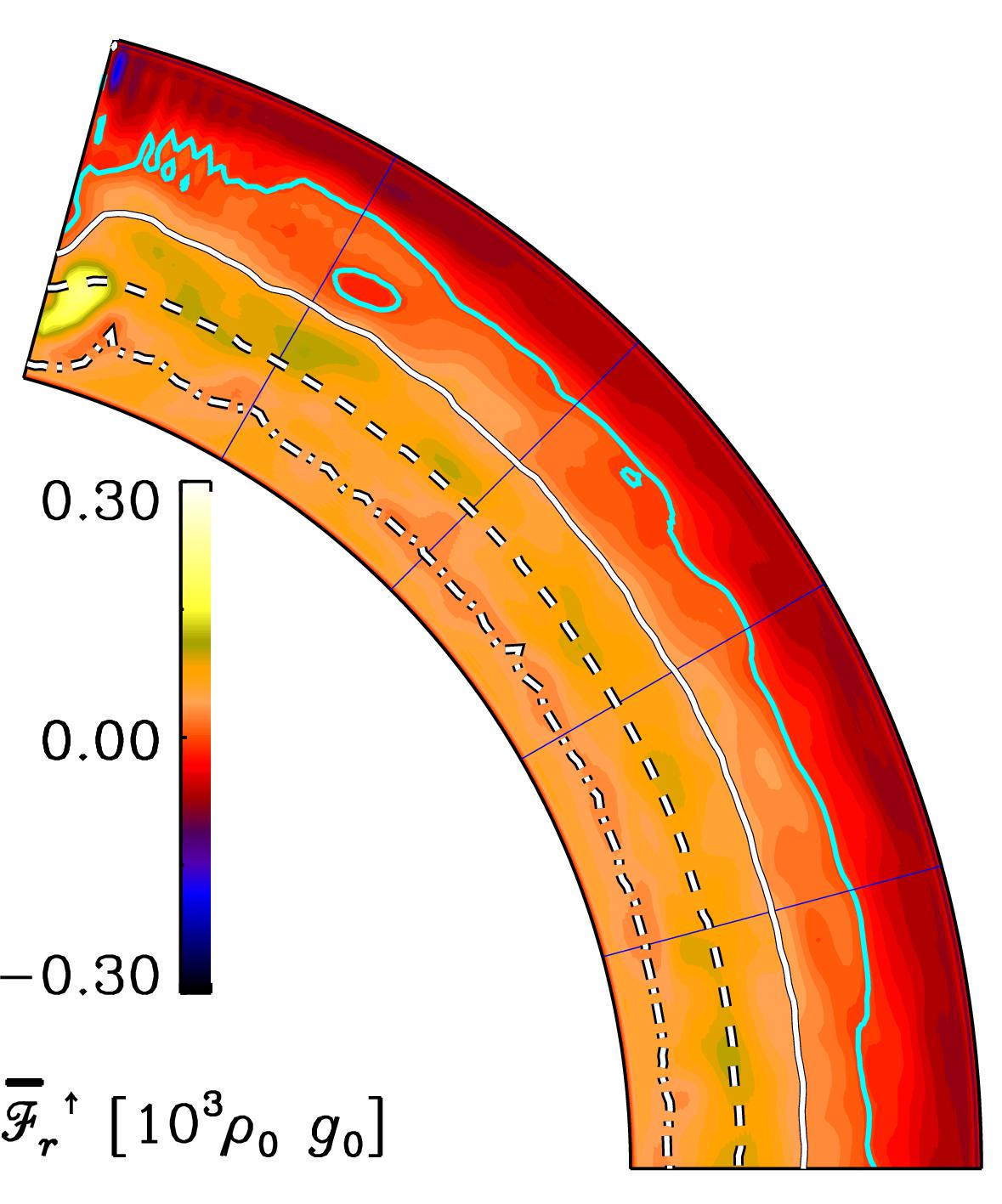}}}%
\subfigure[MHD1]{
\resizebox*{3.75cm}{!}{\includegraphics{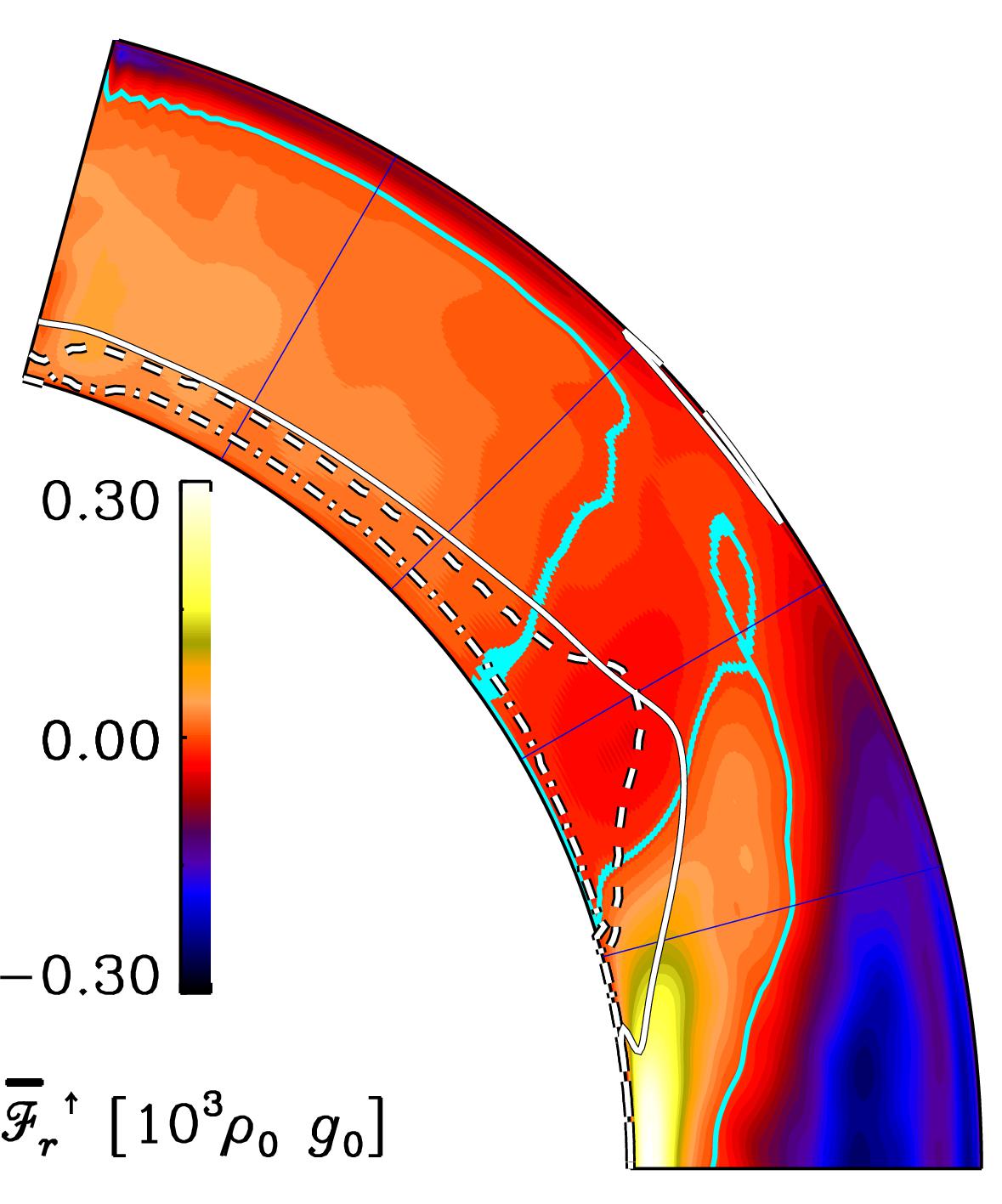}}}%
\subfigure[MHD4]{
\resizebox*{3.75cm}{!}{\includegraphics{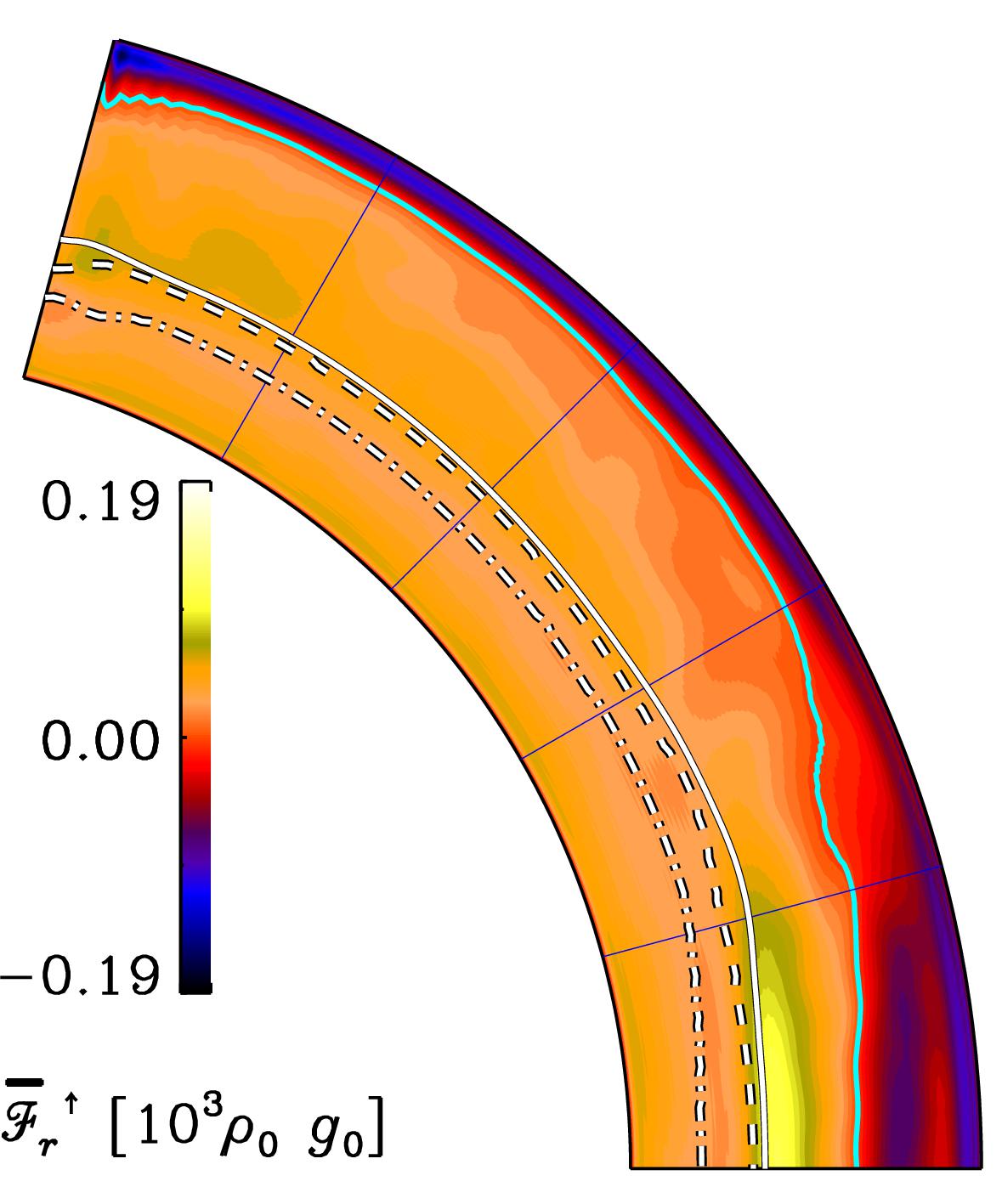}}}%
\\
\begin{center}
\subfigure[HDp]{
\resizebox*{3.75cm}{!}{\includegraphics{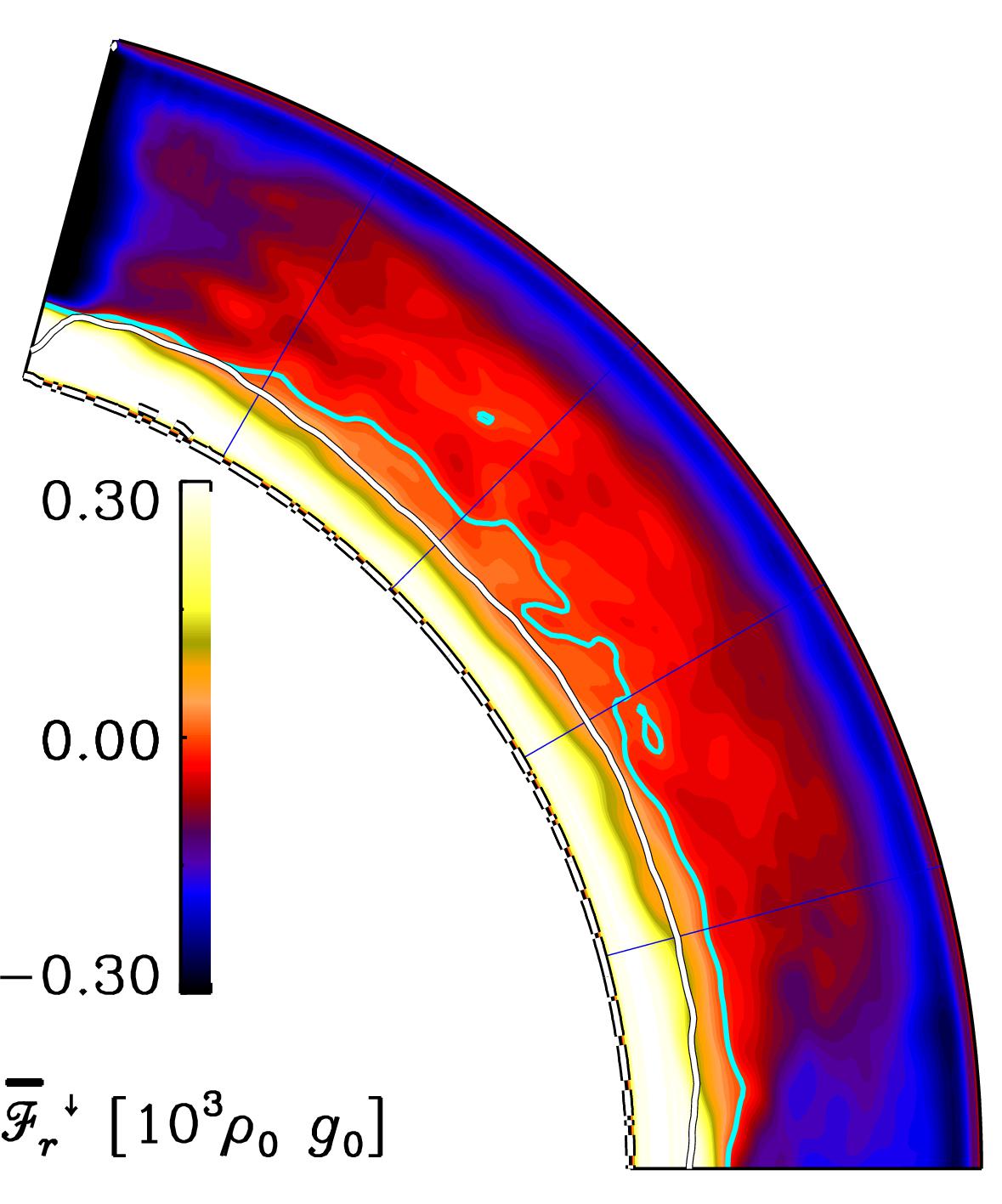}}}%
\subfigure[HD4]{
\resizebox*{3.75cm}{!}{\includegraphics{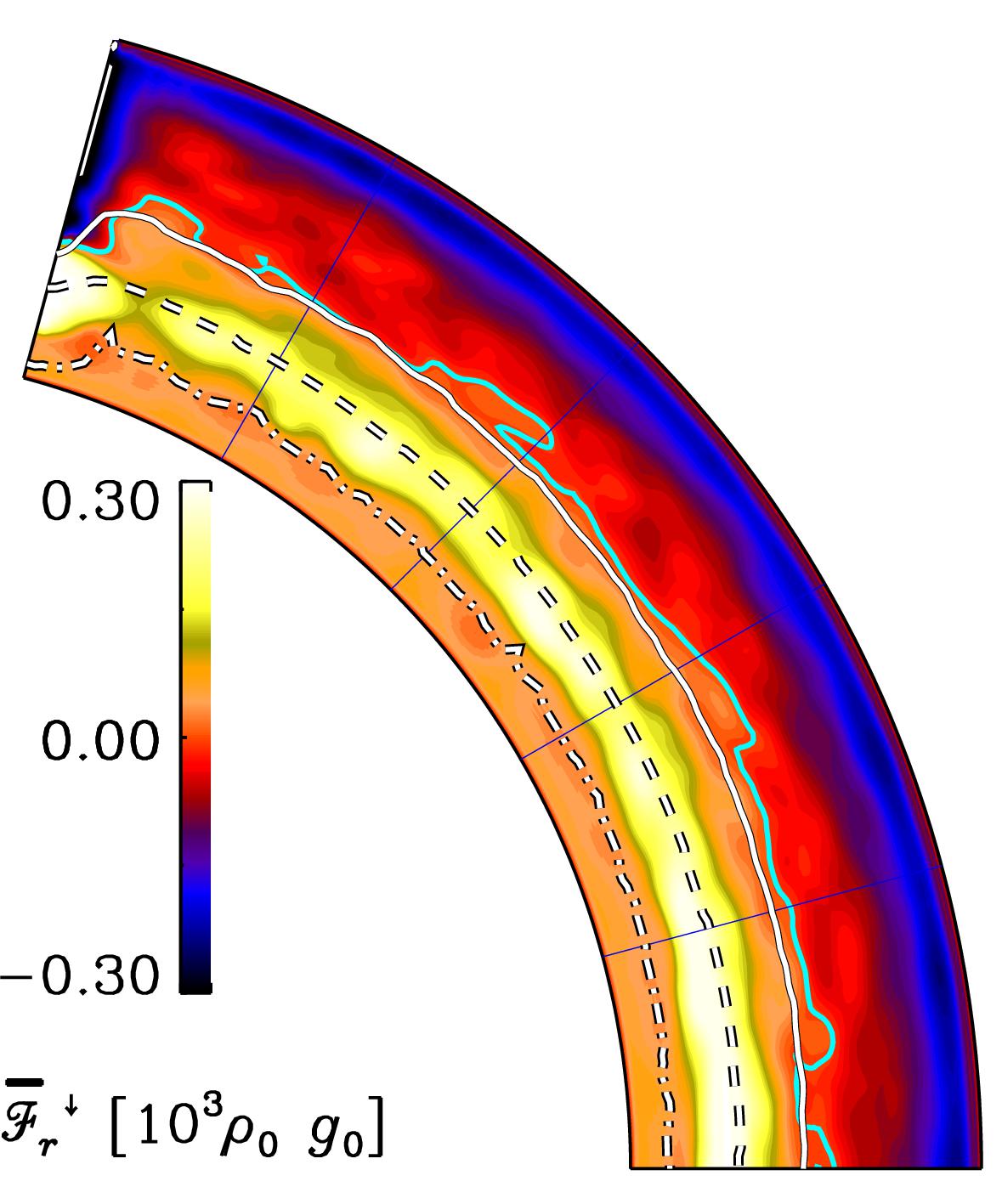}}}%
\subfigure[MHD1]{
\resizebox*{3.75cm}{!}{\includegraphics{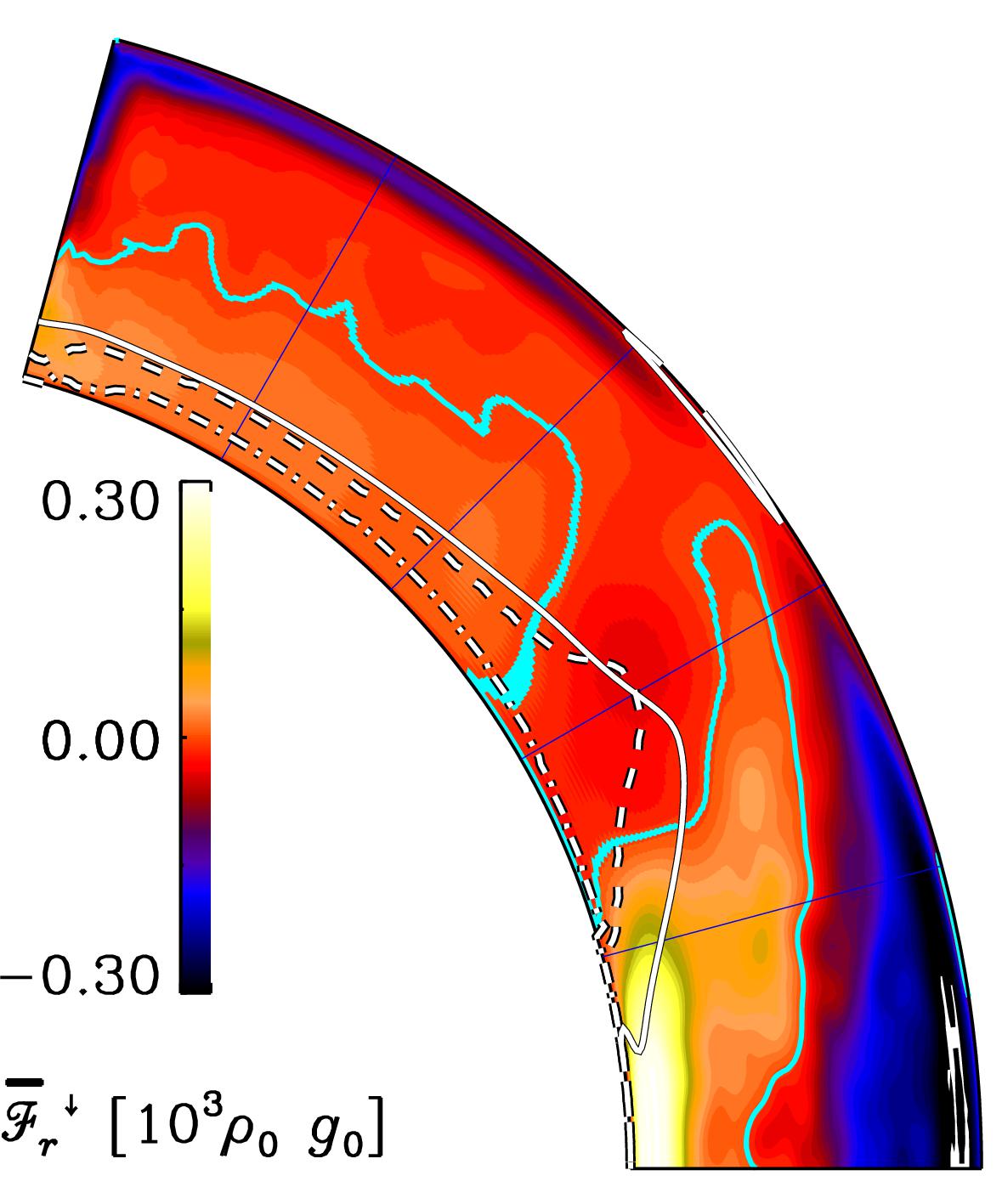}}}%
\subfigure[MHD4]{
\resizebox*{3.75cm}{!}{\includegraphics{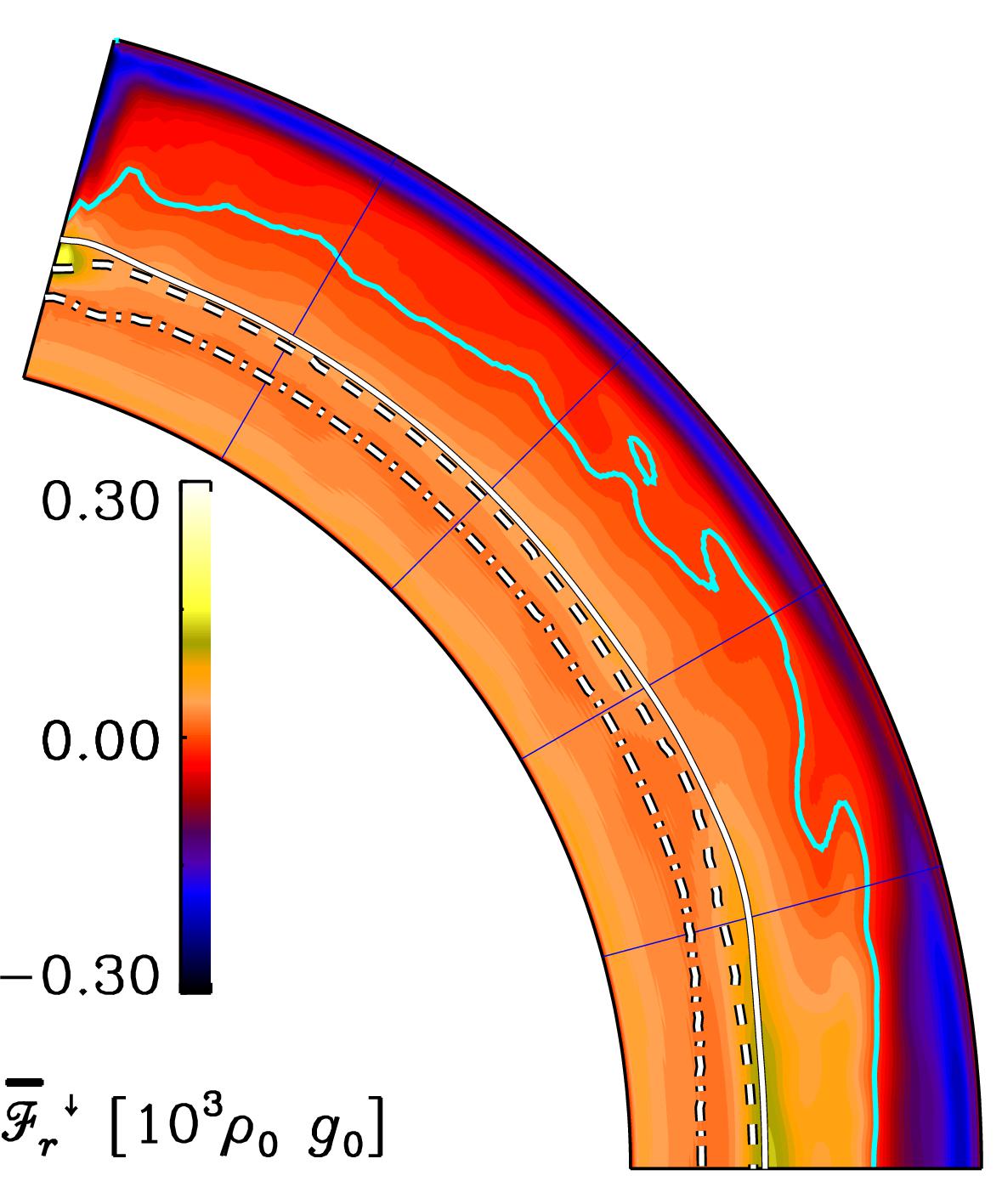}}}%
\end{center}
\caption{Total azimuthally averaged radial force
  $\mean{\mathscr{F}}_r=\mean{\rho \DD u_r/\DD t}$ on the upflows (upper
  row) and downflows (lower) for Runs~HDp, HD4, MHD1, and MHD4. The
  thick cyan line indicates the zero level of the
  force. The black and white solid, dashed, and dash-dotted lines
    indicate the bottoms of the buoyancy, Deardorff, and overshoot
    zones, respectively (colour online).}
\label{fig:pforces}
\end{minipage}
\end{center}
\end{figure}

We study this by measuring the total force on the fluid
\begin{eqnarray}
\mathscr{F}_r\,=\, \mean{\rho\, \frac{\DD u_r}{\DD t}}\,,
\end{eqnarray}
separately for the up- and downflows which are denoted by $\uparrow$
and $\downarrow$, respectively. A positive (negative) force
  accelerates upflows (downflows). Representative results are shown
in \figu{fig:pforces} for the same set of runs as in \figu{fig:pfluxes}.
Comparing the thick black-and-white and cyan curves in \figu{fig:pforces}(a) and
(e), it is seen that for Run~HDp, the sign change of
$\mathscr{F}_r^\downarrow$ occurs roughly at the same average position
as that of the radial entropy gradient (solid line black and
  white line). This appears to be the case
also for $\mathscr{F}_r^\uparrow$ at high latitudes and near the
equator, whereas at mid-latitudes, $\mathscr{F}_r^\uparrow$ is positive
until roughly $r\approx 0.85R_\odot$. These results indicate that the
downflows are accelerated in the Schwarzschild-unstable layer whereas
the upflows accelerate mainly in the Schwarzschild-stable
layer. This is clearly deviating from the behaviour of the Cartesian
simulations with proper OZ and RZ, see \cite{2017ApJ...845L..23K} and
\cite{2017ApJ...843...52H}. However, as seen in \figu{fig:pforces}(b)
and (f) for Run~HD4, the results of the Cartesian simulations are again
restored: the downflows appear to adhere to the Schwarzschild
criterion, while the upflows are accelerated in the stably
stratified OZ, in the DZ, and in the lower part of the BZ.

The situation is significantly more complex in runs where rotation and
magnetic fields are
included. This is particularly clear in cases where the stably
stratified layers are absent or very thin. This is seen, for example,
in Run~MHD1 in \figu{fig:pforces}(c) and (g): at high latitudes, the
upflows are accelerated everywhere except in a thin layer
($r\gtrsim0.95R_\odot$) near the surface, whereas the downflows are
accelerated roughly above $r\gtrsim0.8R_\odot$. The upflows are,
however, driven upward also in the stably stratified OZ and DZ. No
clear relation to the Schwarzschild criterion can be identified. At
mid-latitudes around the tangent cylinder, the total force is downward
for both, up- and downflows. Outside the tangent cylinder, the force is
very roughly following a radially decreasing trend as a function of
cylindrical radius. For runs with more substantial OZ, such as MHD4 in
\figu{fig:pforces}(d) and (h), the latitudinal variation is clearly
weaker. The force on the upflows does not follow the Schwarzschild
criterion in the deep parts in that the upflows are accelerated in the
BZ as well as in the stably stratified DZ and OZ. The layer near
the surface, where deceleration of the upflows occurs, is deeper near
the equator also in these cases. The downflows, on the other hand, are
decelerated in the lower part of the BZ well above the level
where the Schwarzschild criterion indicates stability.

Although a detailed interpretation of the results is non-trivial, we can
conclude that the presence of a substantial OZ has a significant
influence on the large-scale dynamics of the system.
Whether this is also the case in the Sun depends on the extent of
  overshooting at the base of the solar CZ. The recent results of
\cite{2017ApJ...843...52H}, who used Cartesian simulations to conclude
that convective overshooting below the CZ of the Sun is only 0.4 per
cent or 250~km, were used to argue that the interface between RZ and CZ could be well
modelled by imposing suitable boundary conditions. However, these
  results are at odds with earlier models
  \citep[e.g.][]{1982A&A...113...99V,1984ApJ...282..316S,1986A&A...157..338P}
  and helioseismic constraints \citep[e.g.][]{2016LRSP...13....2B},
  which suggest an overshooting depth of the order of $0.05$--$0.1
  \Hp$ or $2500$--$5000\,$km.
Furthermore, global
rotation also changes the behaviour of
the system qualitatively. The current study explores only a
single rotation rate, leading to a rotationally constrained flow, at a
modest supercriticality of convection. Studying the effects of
rotation and higher Rayleigh numbers in more detail will be presented
elsewhere.

\subsection{Differential rotation}
\label{sec:diffrot}

\begin{figure}
\begin{center}
\begin{minipage}{150mm}
\subfigure[MHDp]{
\resizebox*{5cm}{!}{\includegraphics{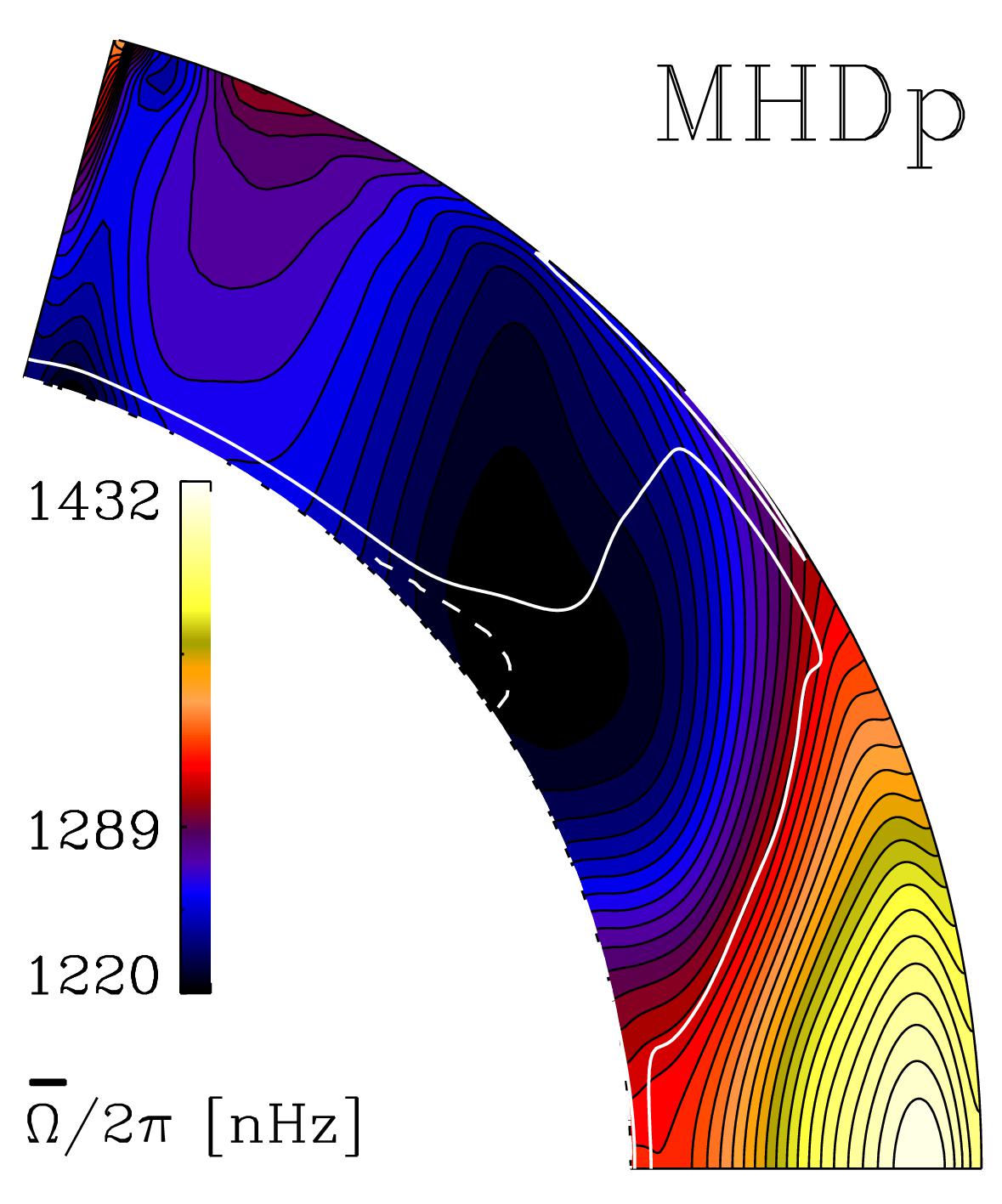}}}%
\subfigure[MHD1]{
\resizebox*{5cm}{!}{\includegraphics{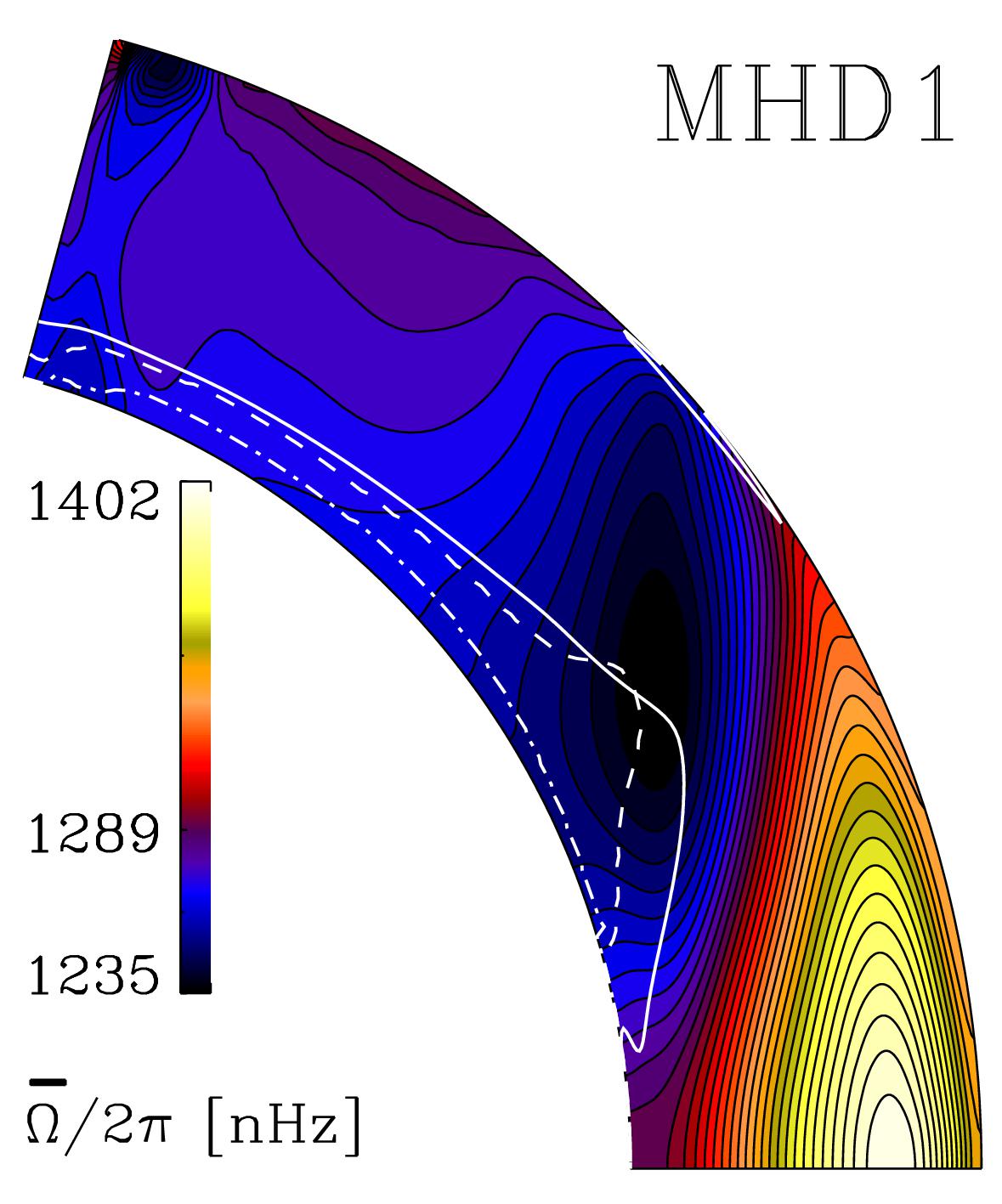}}}%
\subfigure[MHD2]{
\resizebox*{5cm}{!}{\includegraphics{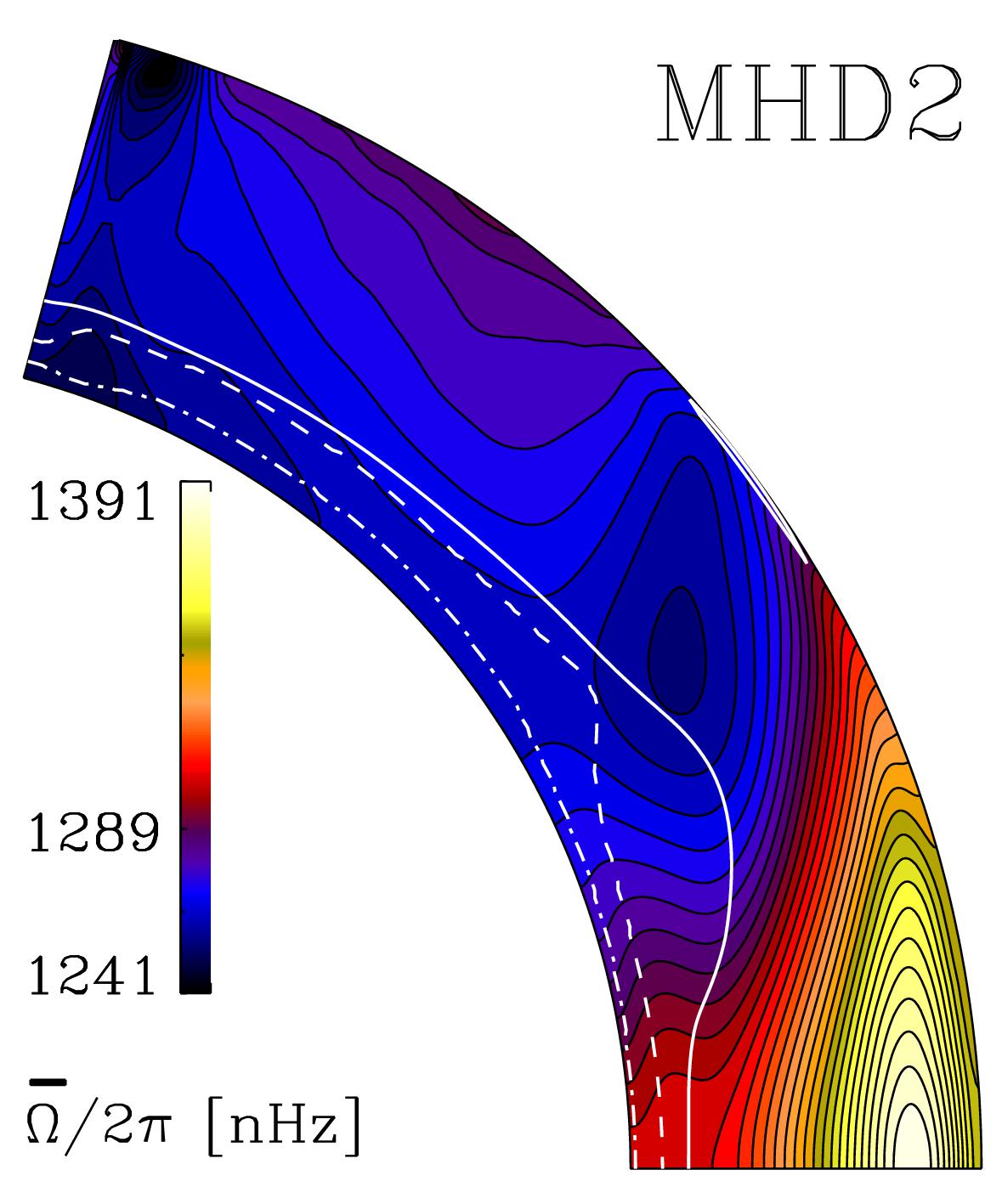}}}%
\\
\begin{center}
\subfigure[MHD3]{
\resizebox*{5cm}{!}{\includegraphics{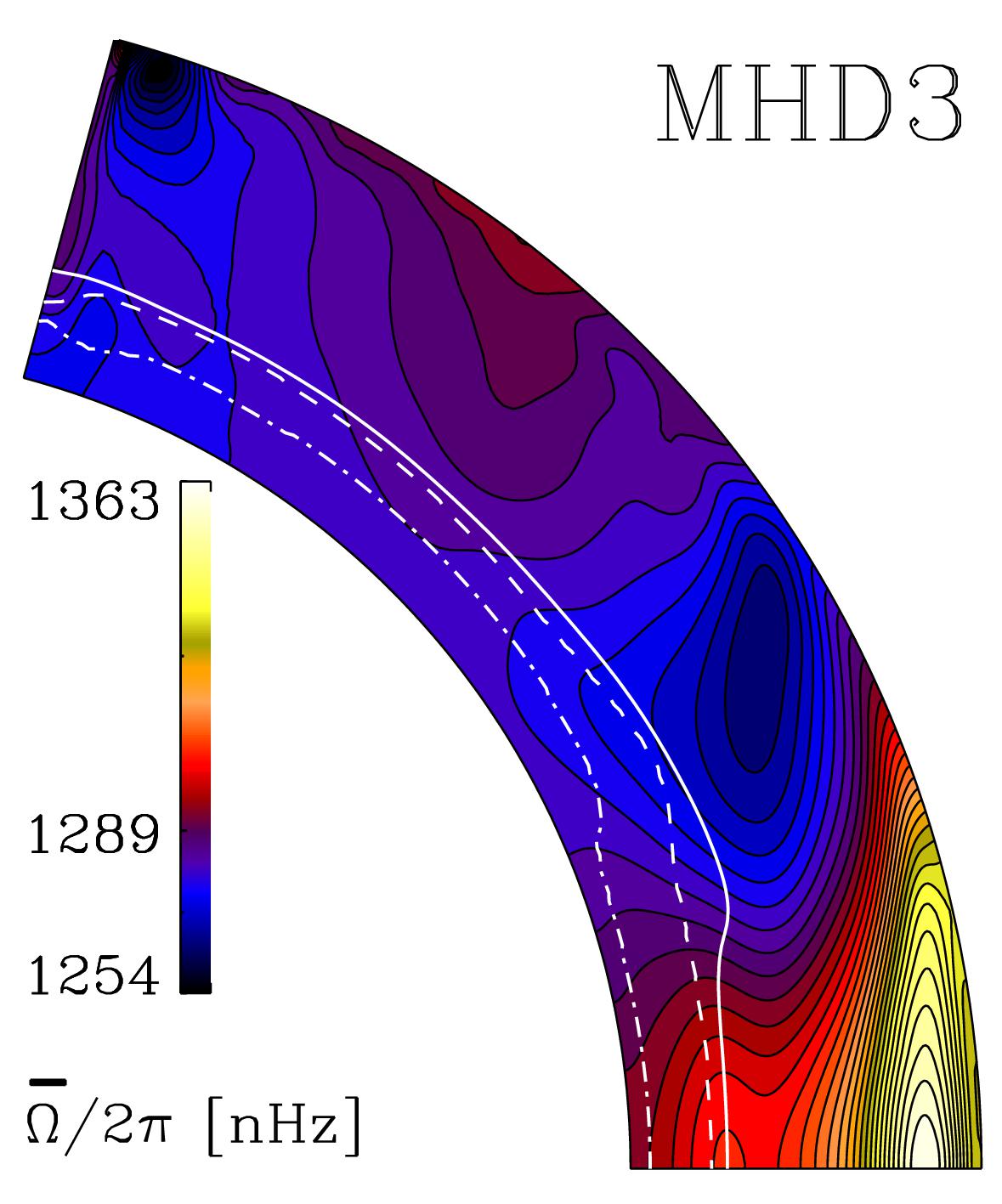}}}%
\subfigure[MHD4]{
\resizebox*{5cm}{!}{\includegraphics{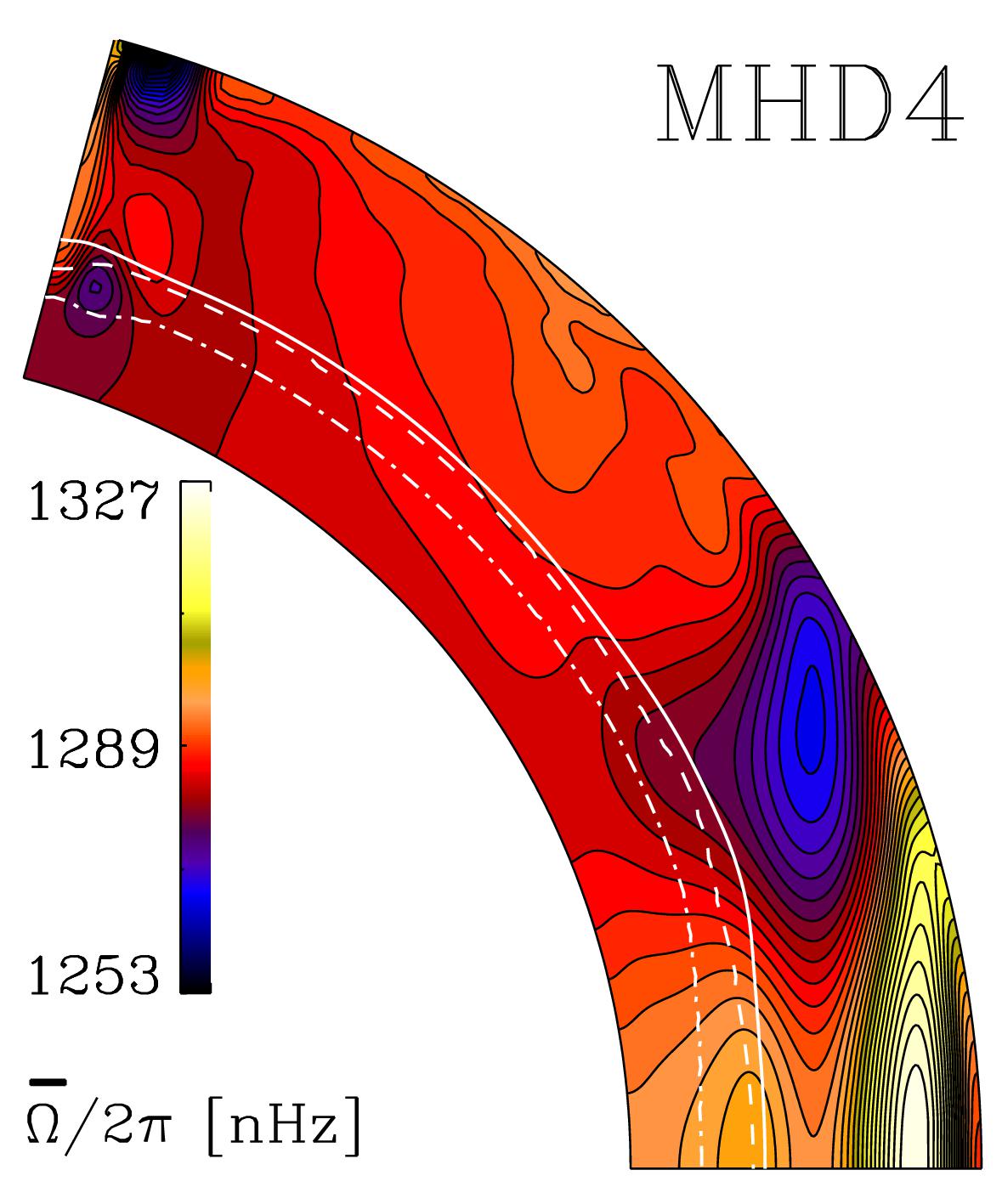}}}%
\subfigure[RHD2]{
\resizebox*{5cm}{!}{\includegraphics{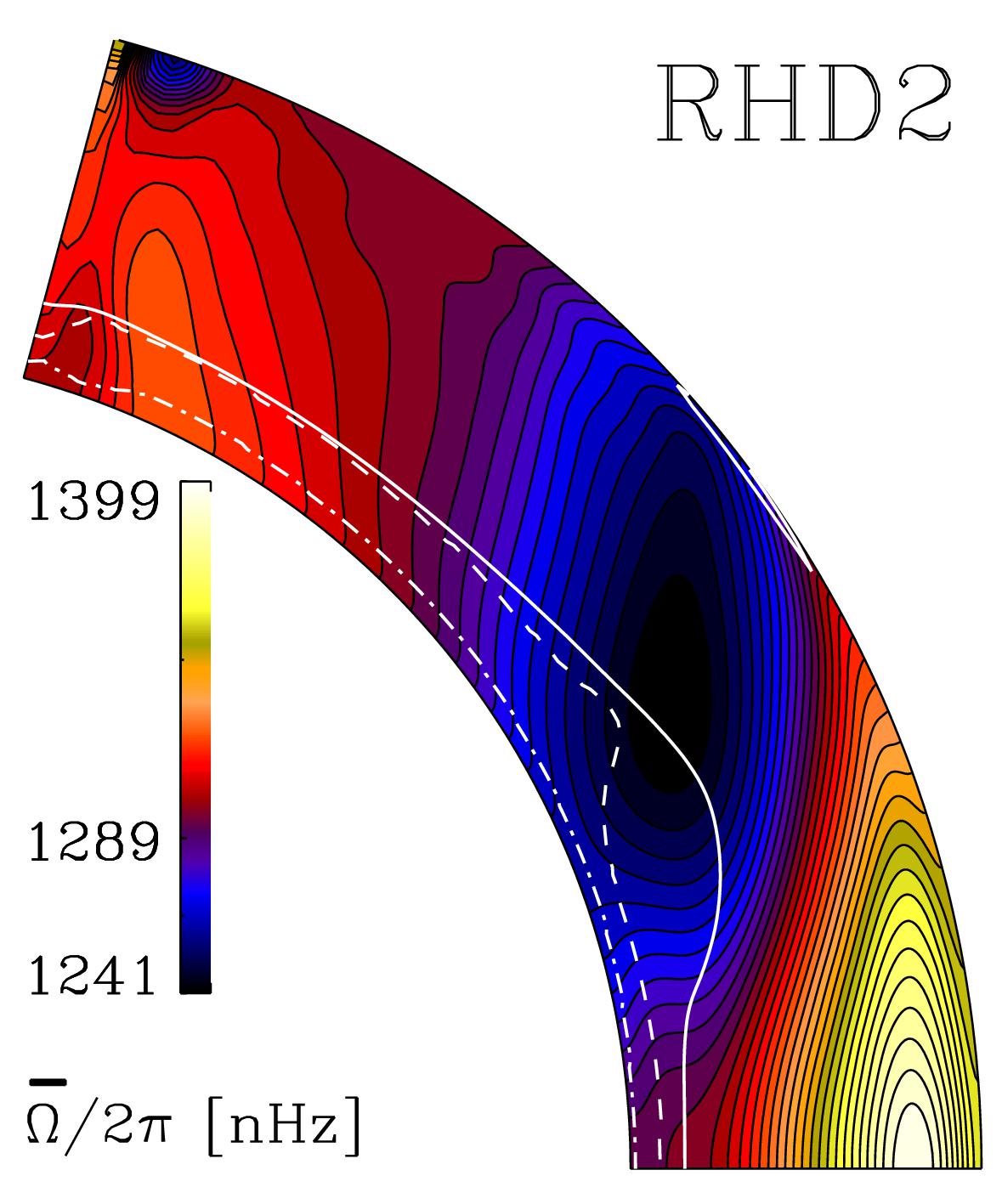}}}%
\end{center}
\caption{Time-averaged rotation profiles from the runs in the MHD
  set. The lower right panel shows Run~RHD2 for comparison.
  The white solid, dashed, and dash-dotted lines indicate the
  bottoms of the BZ, DZ, and OZ,
  respectively (colour online).}
\label{fig:pOm}
\end{minipage}
\end{center}
\end{figure}

Some mean-field models of solar differential rotation \citep[][]{Re05}
have invoked a subadiabatic lower part of the CZ to break the
Taylor-Proudman constraint which, in turn, manifests itself through cylindrical
isocontours of constant angular velocity. Given the
subadiabatic layers in the current simulations, it is of interest to
study the rotation profiles in comparison to earlier studies.
We show in \figu{fig:pOm} the time-averaged rotation profiles from the
MHD runs along with a hydrodynamic run, RHD2.
The rotation rate is here chosen to be $\Omega_0=3\Omega_\odot$
  in order to reach a parameter regime where solar-like (fast equator,
  slow poles) differential rotation appears. Corresponding simulations
  with the solar rotation rate would lead to anti-solar differential
  rotation \citep[e.g.][]{KKB14}. Obtaining solar-like differential rotation with
  $\Omega_0=\Omega_\odot$ is challenging and can be achieved only if
  the radiative diffusion is unrealistically large
  \citep[e.g.][]{FF14,KKB14,HRY16} or the luminosity is artificially
  reduced \citep[e.g.][]{HRY15a}; see also the discussion in
  Appendix~A of \cite{2017A&A...599A...4K}. Both approaches reduce the
  convective velocities and lead to a higher (lower) Coriolis (Rossby)
  number. To reach a solar-like regime, $\Co > 1$ ($\Ro < 1$) is needed
  \citep{KMB11,GYMRW14,2017ApJ...836..192B,2017arXiv171010222V}. The
  Coriolis numbers achieved in the present study
  ($2.5\lesssim\Co\lesssim4.6$, see
  \Tablel{tab:runs}) compare well with, for example, those of
  \cite{NBBMT13} who also used $\Omega_0=3\Omega_\odot$ \citep[see
  Appendix~A of][]{2017A&A...599A...4K}.

\begin{figure}
\begin{center}
\includegraphics[width=0.65\textwidth]{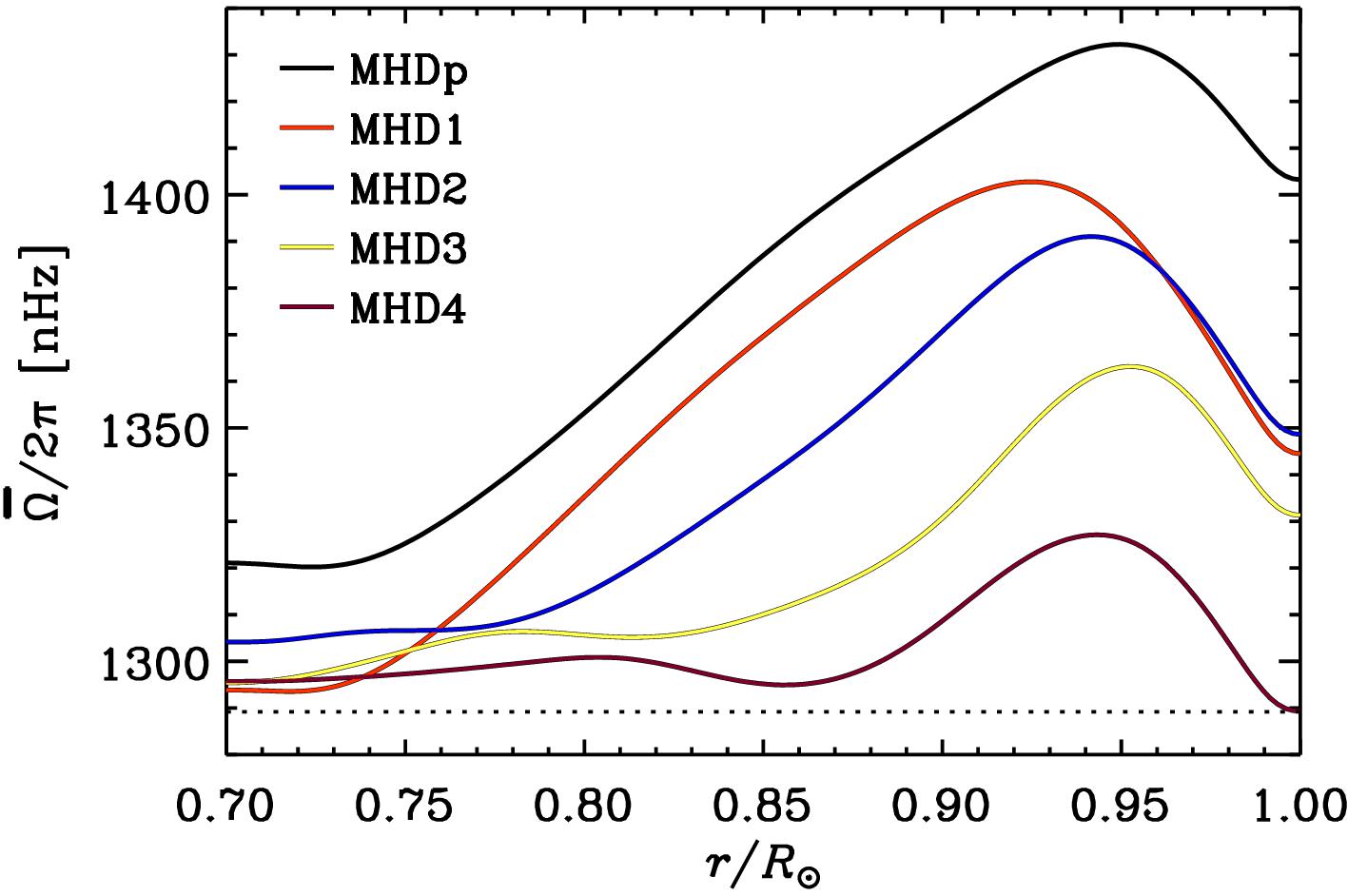}
\caption{The mean angular velocity at the equator as a function
    of radius from the runs in the MHD set (colour online).}
\label{fig:pOm_line_nd}
\end{center}
\end{figure}

\begin{figure}
\begin{center}
    \includegraphics[width=\textwidth]{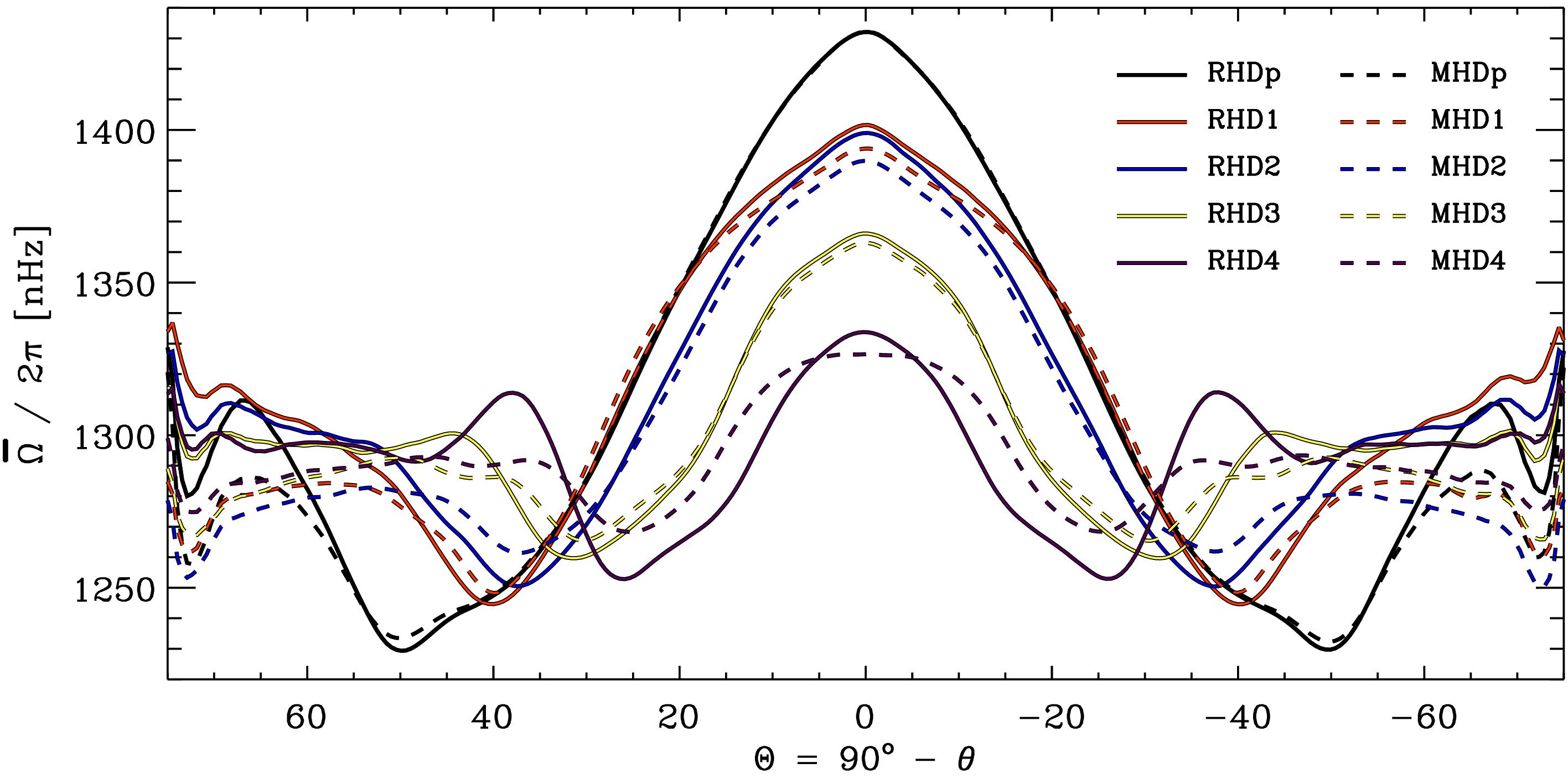}
    \caption{Mean angular velocity $\mean{\Omega}$ from $r=0.95R_\odot$
        from all runs in Sets RHD (solid lines) and MHD (dashed) (colour online).}
\label{fig:pOmega_lat}
\end{center}
\end{figure}

We find that the isocontours of $\mean{\Omega}=\Omega_0+\mean{U}_\phi/r\sin\theta$
are significantly
tilted even in the run with a fixed heat conduction profile
(MHDp). Furthermore, a
mid-latitude minimum is visible, but it is shallower, and occupies
a wider latitude range than in previous simulations with similar
rotation rates (see, e.g., Figure~4 of \citealt{ABMT15} and Figure~1 of
\citealt{2018A&A...616A..72W}). The mid-latitude minimum is most likely
responsible for the equatorward migrating activity seen in the
aforementioned studies \citep{WKKB14}. A small NSSL,
  confined at low latitudes is also
visible in the current runs; see \figu{fig:pOm_line_nd}. The
most likely reason for its
appearance is that the density stratification in the
current runs is higher ($\Delta \rho \approx 60$) than in earlier
studies with otherwise similar parameters
\citep[e.g.][]{KMB12,WKKB14,2017A&A...599A...4K} where $\Delta \rho \approx 20$.
This has also been found in recent simulations of \cite{2018arXiv181000115M}.
In theory, this allows the development of a NSSL, where the
rotational influence on the flow is weak, quantified by $\Co<1$
\citep{RKT14}.
In such a parameter regime, the non-diffusive Reynolds stress, or
$\Lambda$ effect, responsible for the generation of differential
rotation \citep[e.g.][]{R89} reduces to a single term
\citep[e.g.][]{2017arXiv171208045K} that drives a latitude-independent
radial shear, as in \cite{2014A&A...570L..12B} and \cite{Ki16}.
However, the current numerical results seem to confirm the results of
\cite{RC01} in that extreme density stratification \citep{HRY15a} is
not required for
the appearance of a NSSL.
However, the detailed reproduction of, e.g., the solar NSSL
does require high stratification and resolution to capture the
small-scale surface convection and the shorter time scales near the surface.

In Runs~MHD1 and MHD2, the rotation profile is qualitatively similar to that
in MHDp. The clearest difference is the enhanced radial gradient of
$\mean{\Omega}$ near the surface at low latitudes. Also, the region of
negative radial shear in mid-latitudes is enhanced---especially in the
Deardorff layer in Run~MHD2 (see the dashed and solid white lines in
\figu{fig:pOm}). In the remaining runs (MHD3 and MHD4), the layer of
negative radial shear is even more pronounced, but appears
predominantly within the BZ. It is also apparent that the differential
rotation creeps into the radiative interior due to the relatively high
diffusivities used in the current simulations. The local minimum of
$\mean{\Omega}$ at mid-latitudes, coinciding with the location of the
tangent cylinder of the BZ to DZ transition near the equator, becomes
more pronounced in Runs~MHD3
and MHD4.

We show the rotation profile of a hydrodynamic run RHD2 in
  \figu{fig:pOm}(f). The differences to the corresponding MHD run (MHD2)
  are most clearly visible at high latitudes, where the angular
  velocity is clearly reduced in the MHD run. Furthermore, the
  mid-latitude minimum of $\mean{\Omega}$ is somewhat shallower in the
  MHD case. A similar trend is found in all runs in the RHD and MHD
  sets; see \figu{fig:pOmega_lat} for a comparison of $\mean{\Omega}$
  at $r=0.95R_\odot$. The relatively weak influence of magnetic fields
  on differential rotation appears to differ from the results of
  \cite{2017A&A...599A...4K}, who found a strong quenching of mean
  flows in high resolution simulations. However, at comparable
  magnetic Reynolds numbers, as in the current simulations (roughly 30),
  the results of \cite{2017A&A...599A...4K} also indicate only weak
  quenching; see their Figure~2 and Table~2.

\begin{figure}
\begin{center}
\begin{minipage}{150mm}
\subfigure[MHDp]{
\resizebox*{5cm}{!}{\includegraphics{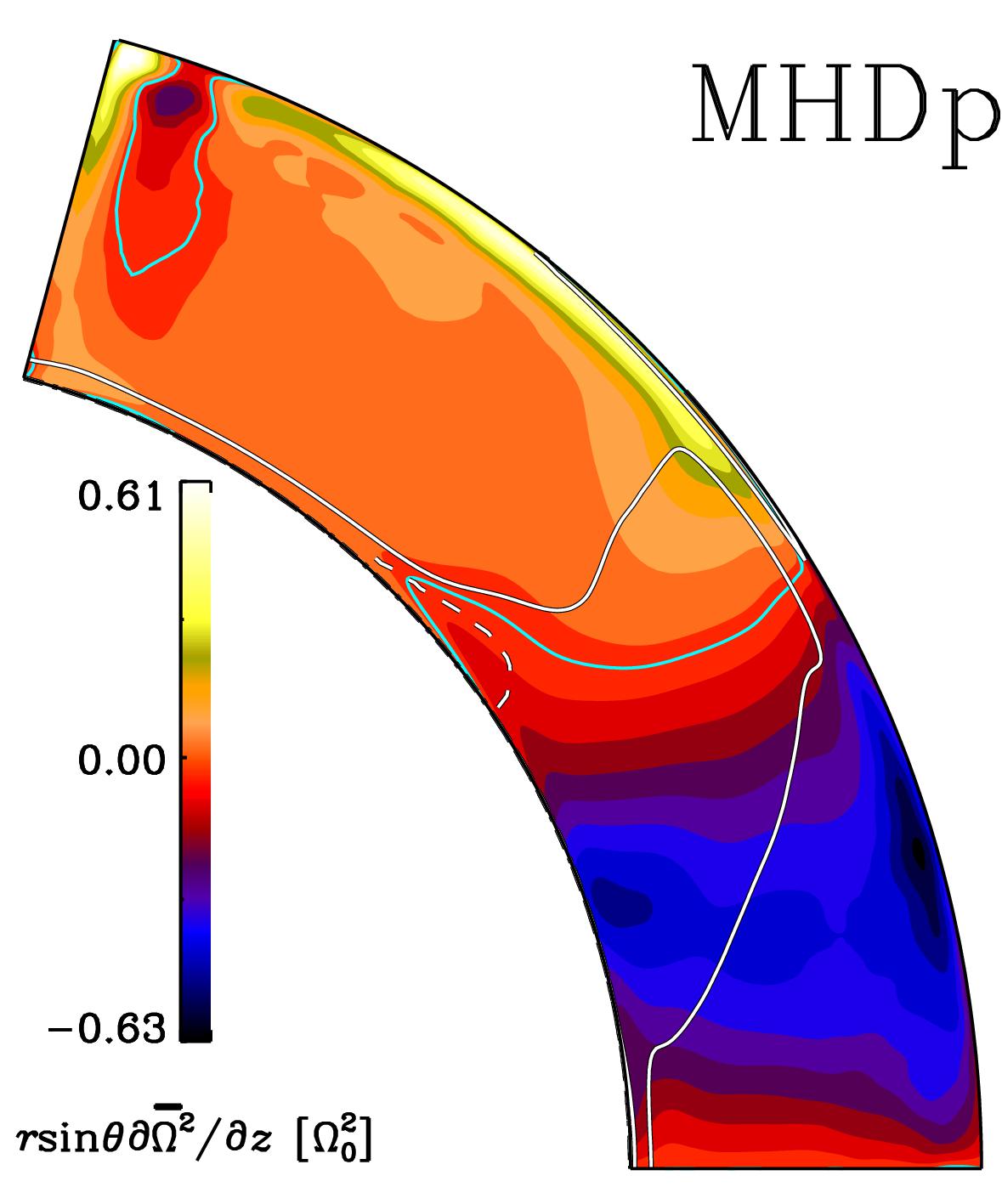}}}%
\subfigure[MHD1]{
\resizebox*{5cm}{!}{\includegraphics{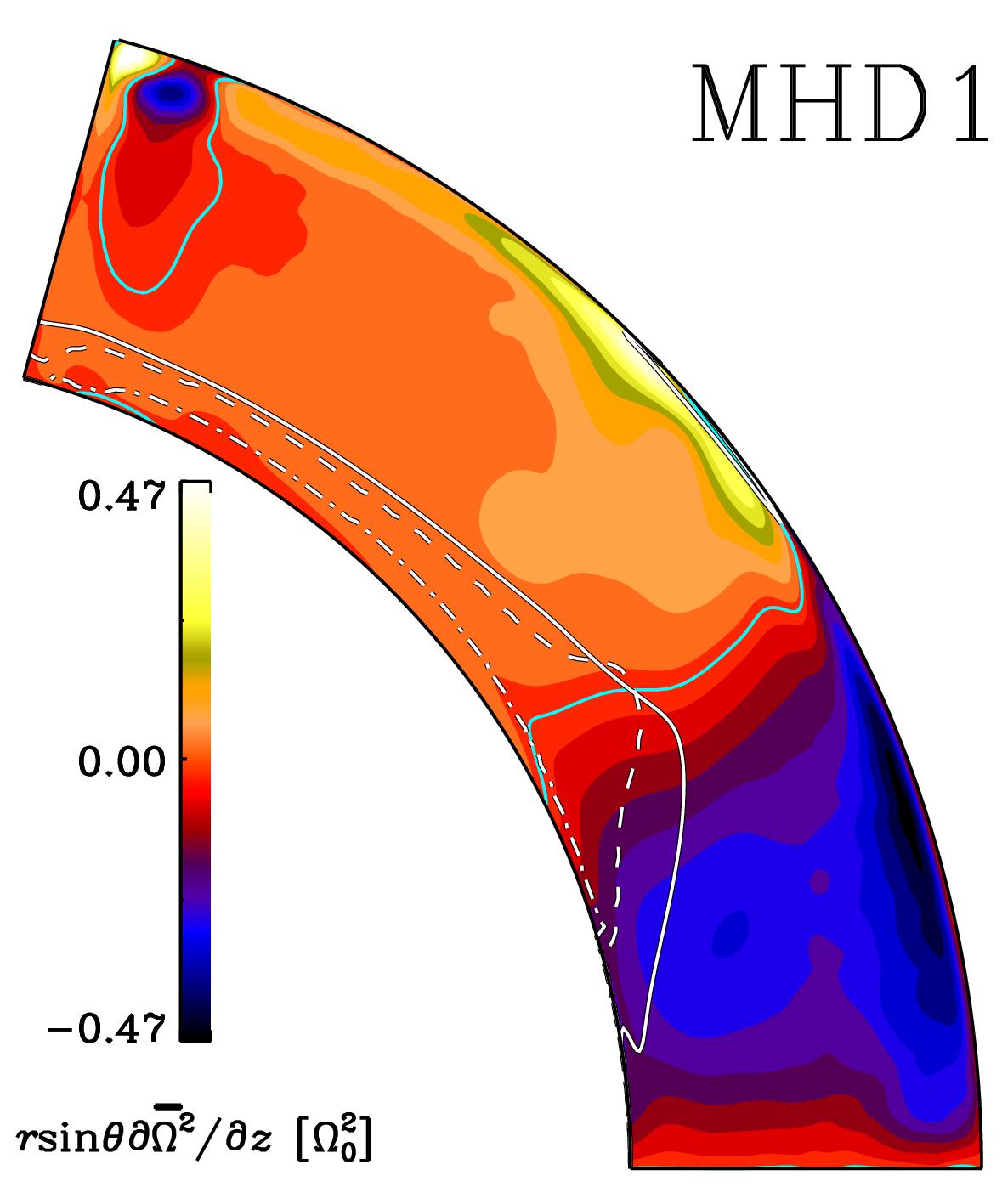}}}%
\subfigure[MHD4]{
\resizebox*{5cm}{!}{\includegraphics{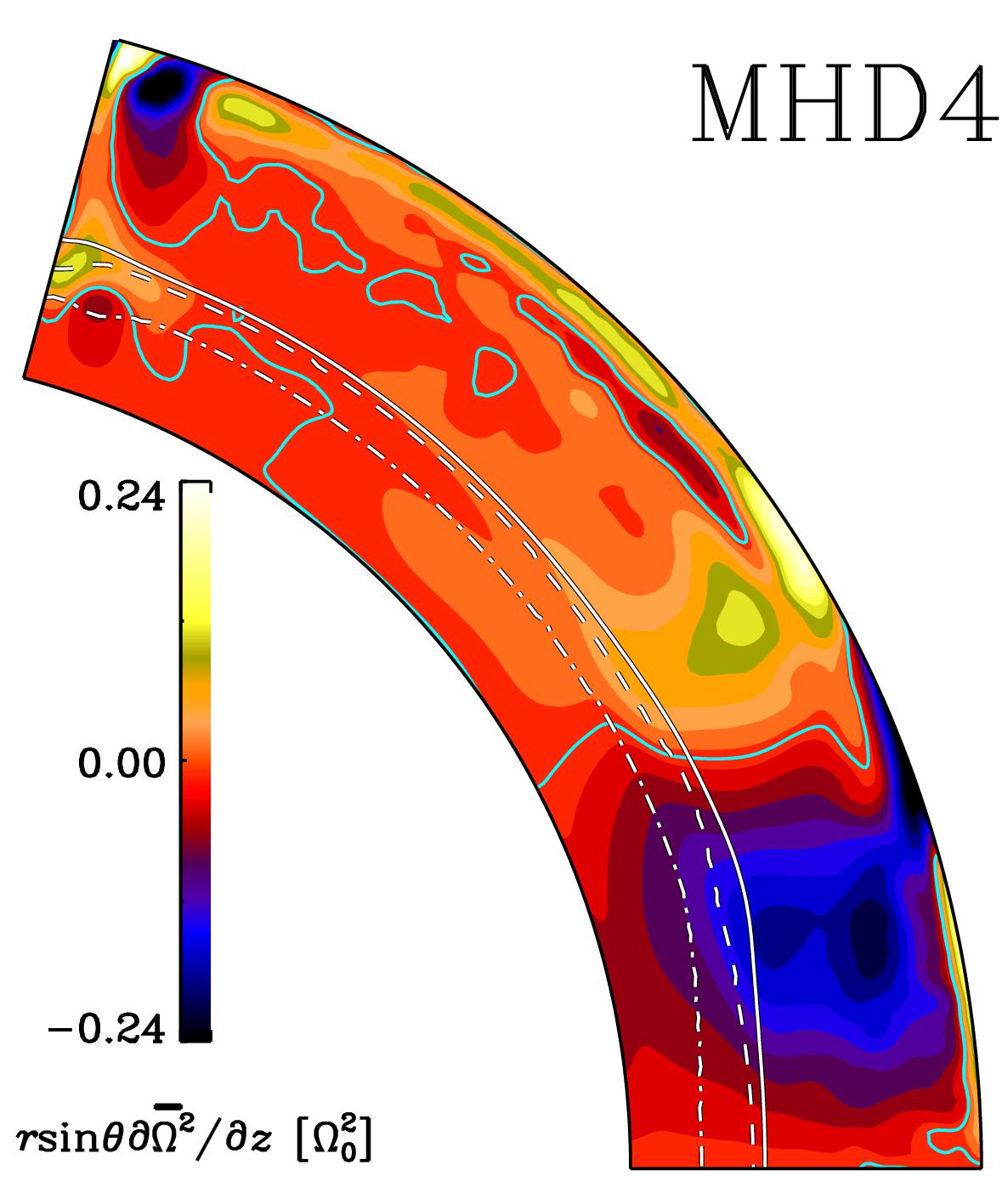}}}%
\\
\subfigure[MHDp]{
\resizebox*{5cm}{!}{\includegraphics{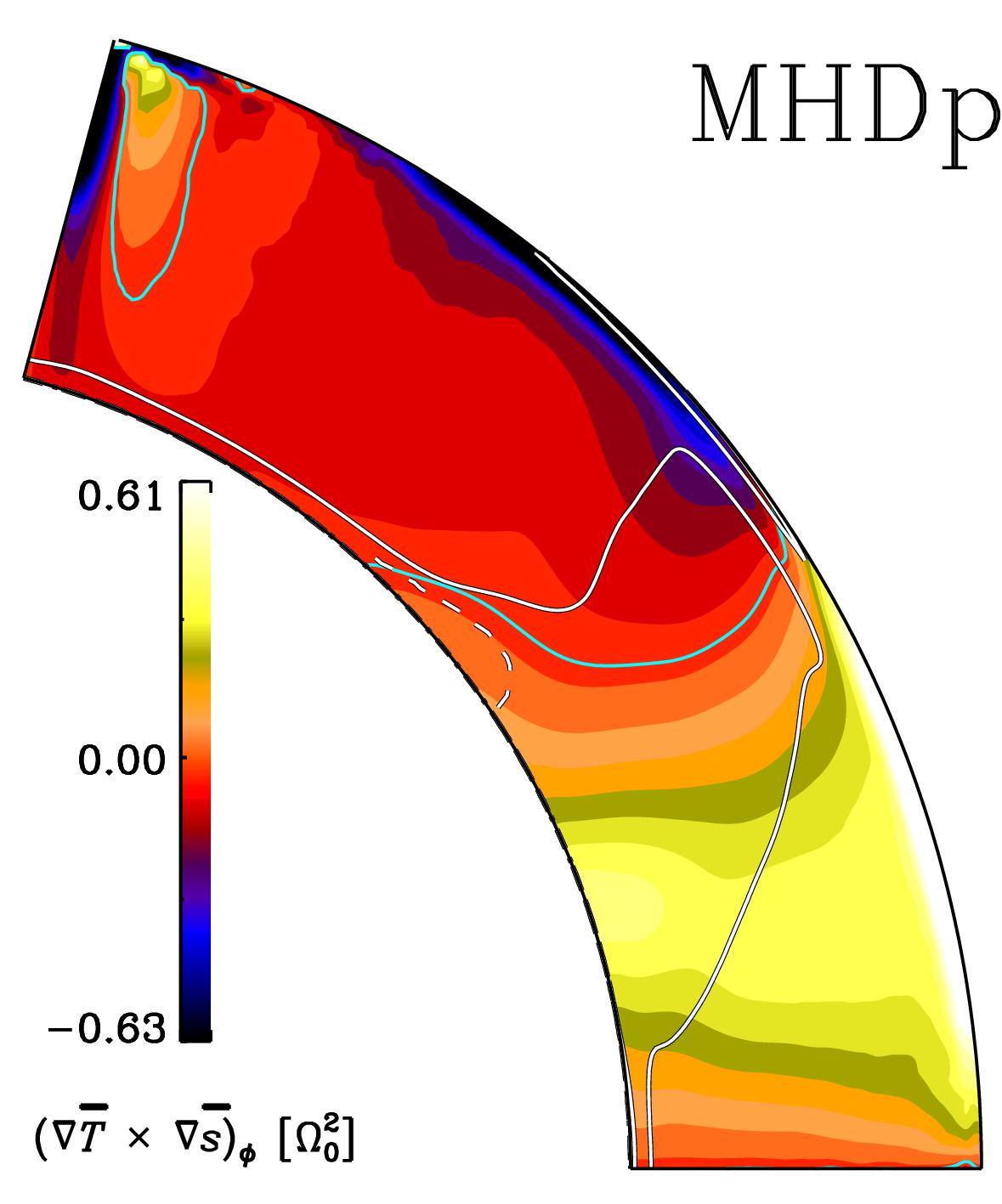}}}%
\subfigure[MHD1]{
\resizebox*{5cm}{!}{\includegraphics{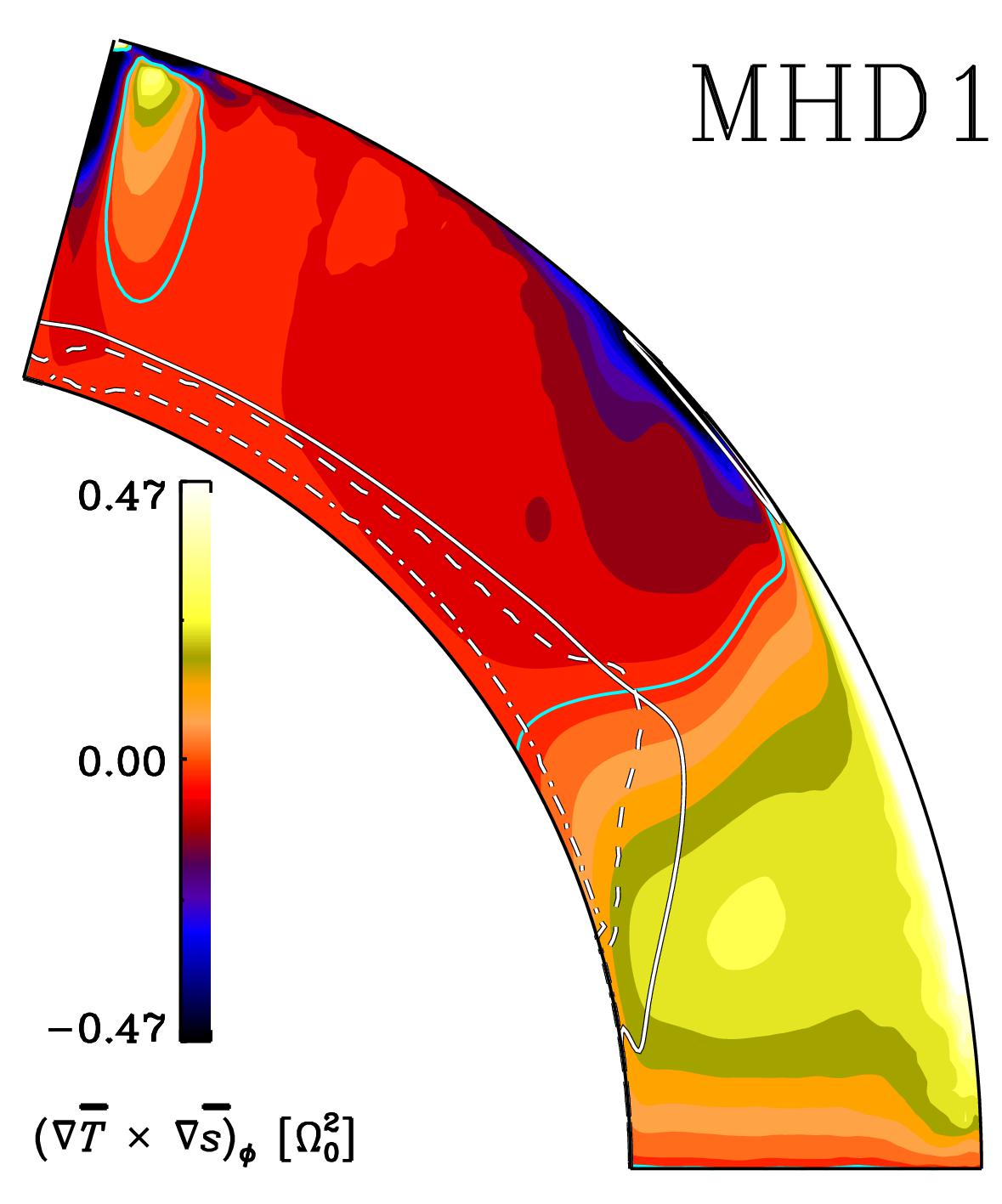}}}%
\subfigure[MHD4]{
\resizebox*{5cm}{!}{\includegraphics{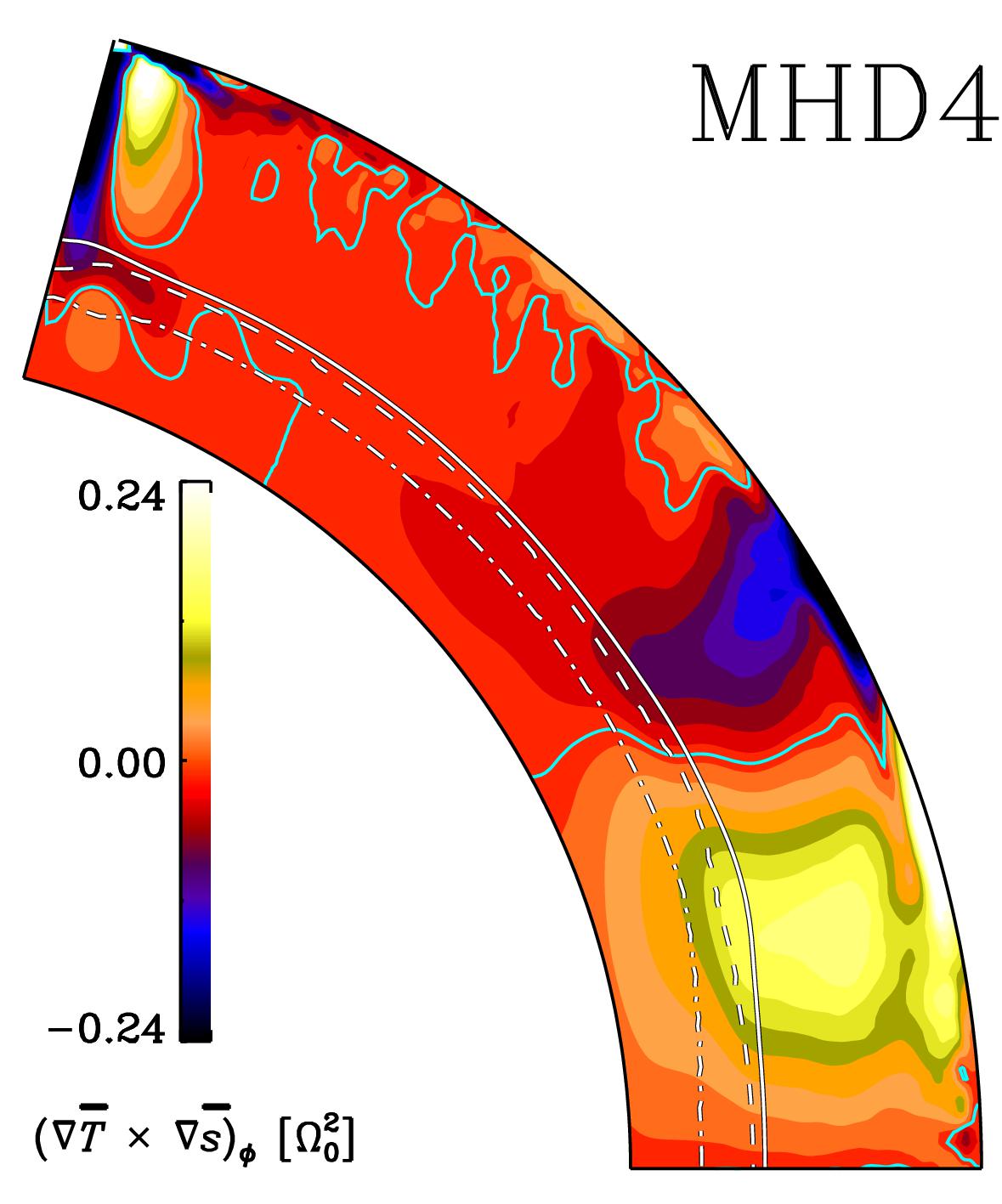}}}%
\caption{Dominant terms in the equation of mean azimuthal vorticity:
    Coriolis force $r\sin\theta\pd\mean{\Omega}^2/\pd z$ (upper row)
    and the baroclinic term $(\bm\nabla\mean{T} \times
    \bm\nabla\mean{s})_\phi$ (lower row) in
    units of $\Omega_0^2$ for
    Runs~MHDp (left panel), MHD1 (middle), and MHD4 (right). The cyan
    line indicates the zero level whereas the white solid, dashed, and
    dash-dotted lines indicate the bottoms of the BZ, DZ, and OZ,
    respectively (colour online).}
\label{fig:pbaro}
\end{minipage}
\end{center}
\end{figure}

The rotation profiles of Runs~MHDp and MHD1 in particular clearly
deviate from the Taylor-Proudman balance. To study this, we consider
the mean vorticity equation:
\begin{eqnarray}
\frac{\pd}{\pd t} (\bm\nabla \times \mUUU)_\phi\,
=\, r \sin\theta \frac{\pd \mean{\Omega}^2}{\pd z} + (\bm\nabla\mean{T} \times \bm\nabla\mean{s})_\phi + \cdots\,, \label{equ:vorti}
\end{eqnarray}
where $\pd/\pd z = \cos\theta \pd/\pd r - r^{-1} \sin\theta \pd/\pd
\theta$ is the derivative along the rotation axis, and the dots
indicate terms due to Reynolds stress and molecular
viscosity. We find that the first and second terms of the right-hand side,
corresponding to the Coriolis force and the baroclinic effect,
respectively, balance each other at all latitudes in Runs~MHDp and MHD1;
figures~\ref{fig:pbaro}(a)--(b) and (d)--(e). The extended DZ
of MHDp does not appear to lead to a significant enhancement of the
baroclinic effect. In Run~MHD4, the balance between Coriolis and
baroclinic terms is realized only at low latitudes $-30^\circ \gtrsim
\Theta \gtrsim 30^\circ$. Furthermore, the magnitudes of both terms
are clearly reduced in comparison to Runs~MHDp and MHD1. This is
likely to explain the more cylindrical isocontours of $\mean{\Omega}$
in this run.

{In a recent study, \cite{2018arXiv180709309K} showed that the
  large-scale properties of flows in hydrodynamic convection
  simulations, with the same model as here, are sensitive to changes
  in the thermal boundary conditions and the treatment of the
  unresolved photospheric layers. In particular, the setup used in the
  present study, with cooling near the surface and an isothermal top
  boundary condition, leads to larger deviations from the
  Taylor-Proudman balance than a corresponding setup where the energy
  in the near-surface layer is transported by SGS diffusion and where the
  upper boundary obeys a black body boundary condition. The latter
  was also used, for example, by \cite{2017A&A...599A...4K}. This
  suggests that the subadiabatic layer is not the main reason why the
  rotation profiles are more conical here than in previous studies.

\begin{figure}
\begin{center}
\begin{minipage}{150mm}
\subfigure[HDp]{
\resizebox*{7.5cm}{!}{\includegraphics{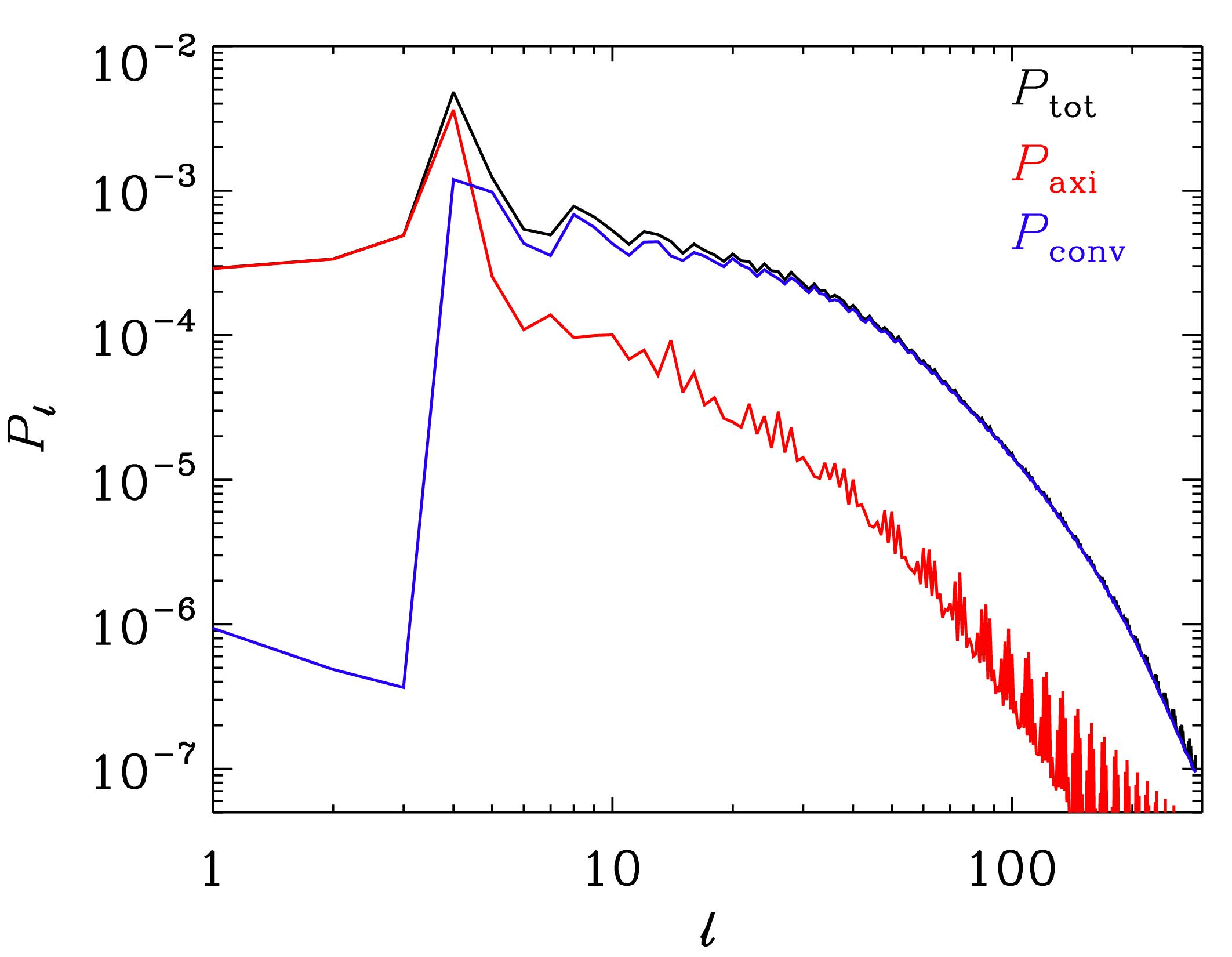}}}%
\subfigure[HD4]{
\resizebox*{7.5cm}{!}{\includegraphics{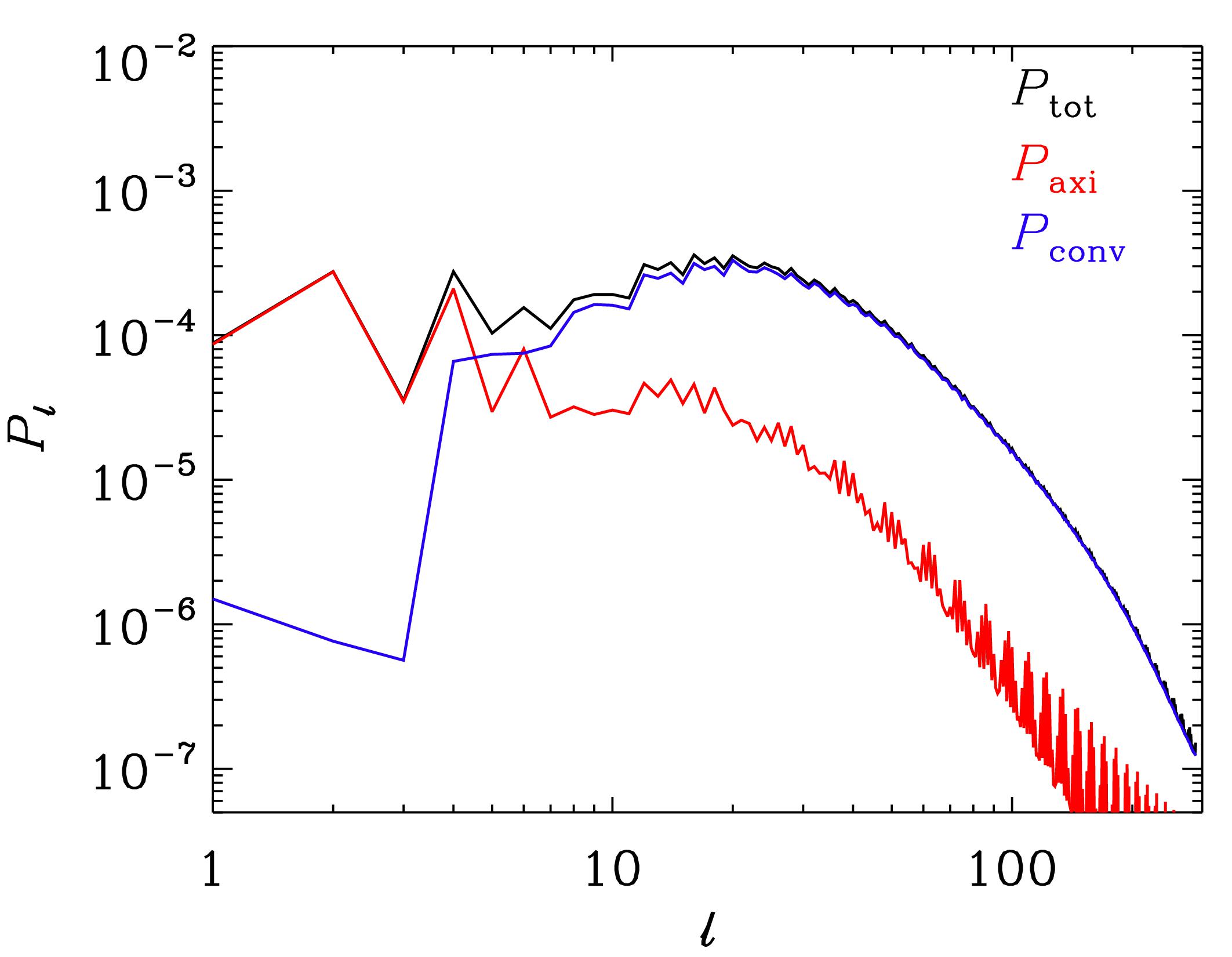}}}%
%\\
\begin{center}
\subfigure[MHD1]{
\resizebox*{7.5cm}{!}{\includegraphics{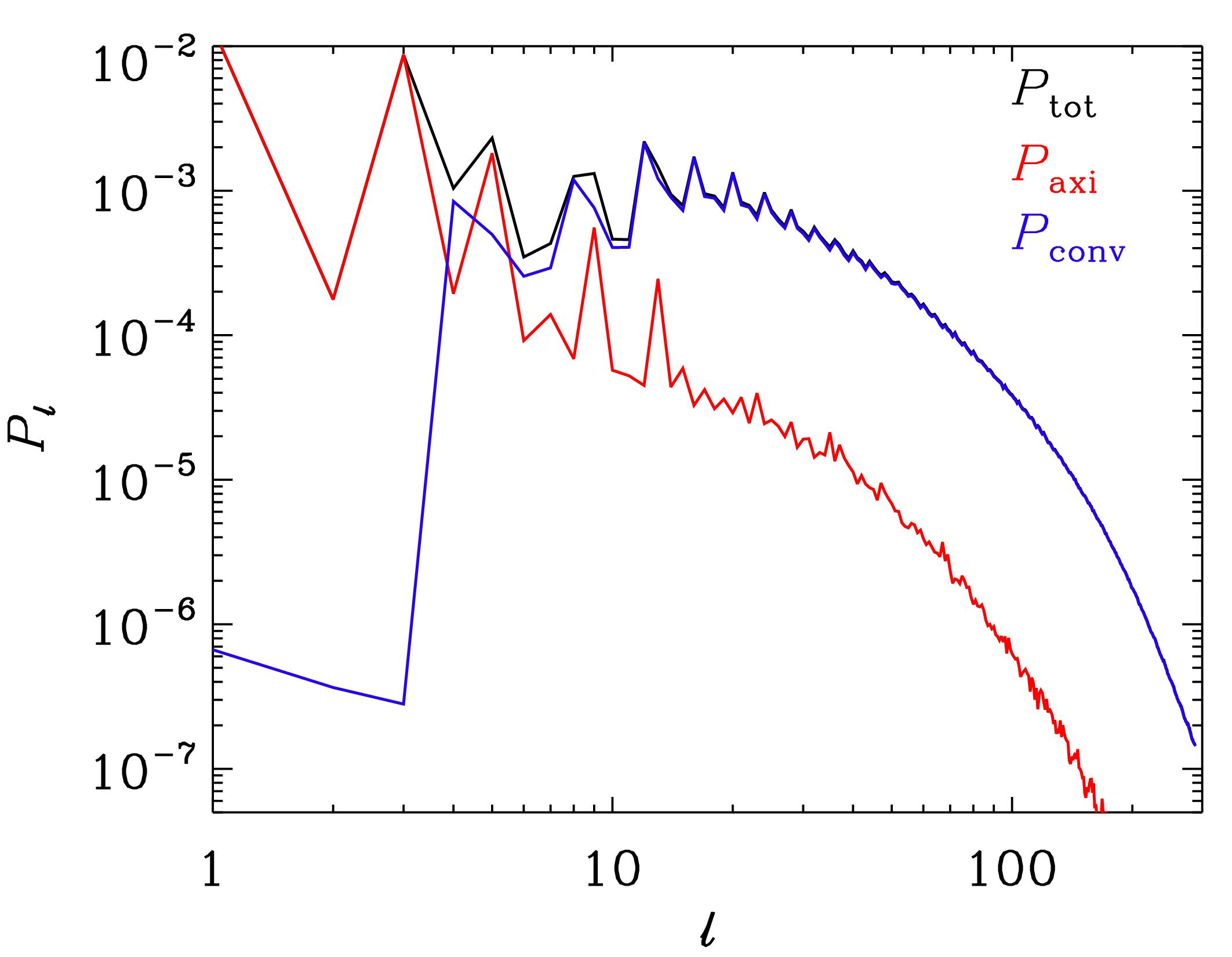}}}%
\subfigure[MHD4]{
\resizebox*{7.5cm}{!}{\includegraphics{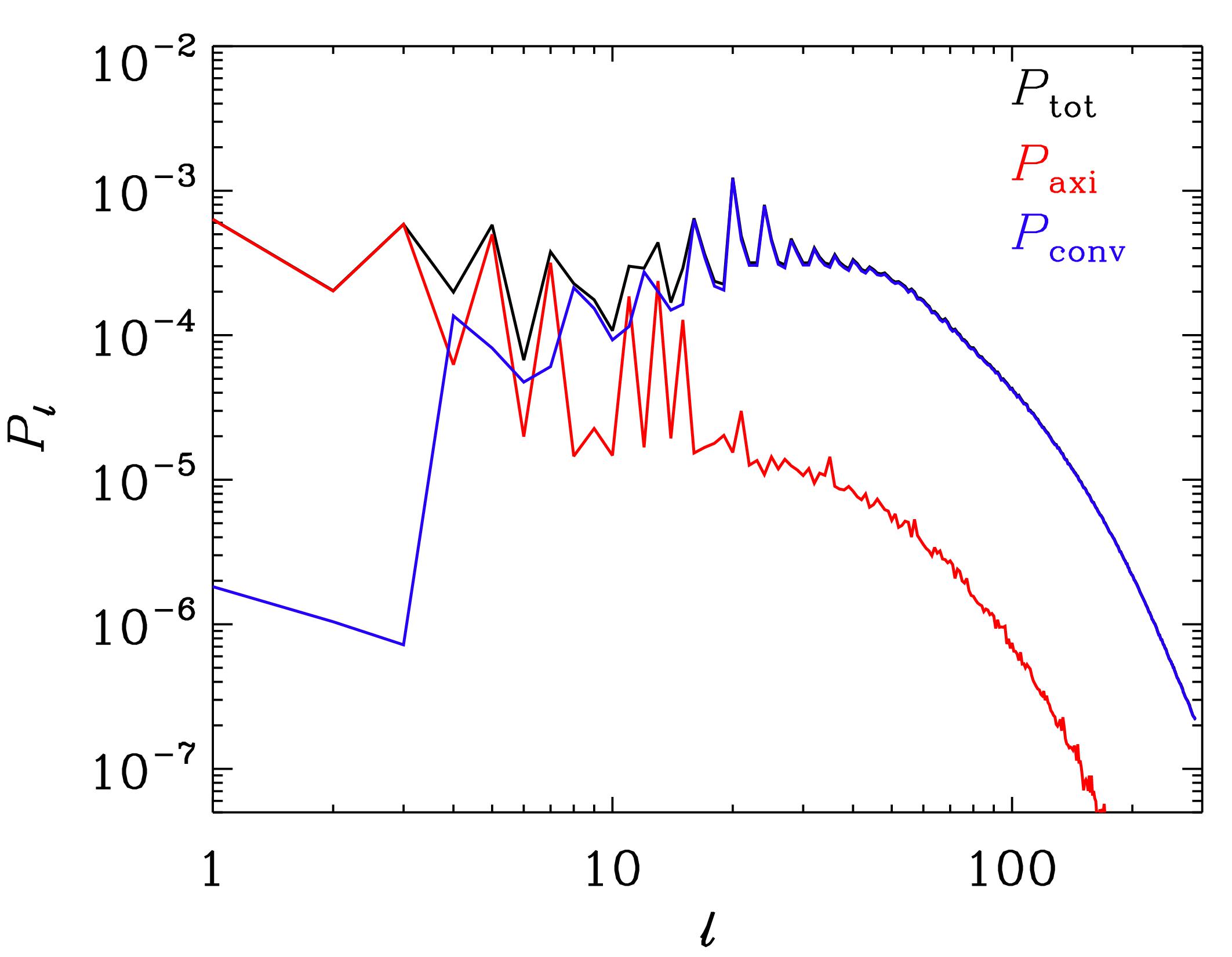}}}%
\end{center}
\caption{Power spectra of the total (black), axisymmetric (red), and
  non-axisymmetric (blue) parts of the velocity from runs (a) HDp,
  (b) HD1, (c) MHD1, and (d) MHD4 (colour online).}
\label{fig:pwrspectra}
\end{minipage}
\end{center}
\end{figure}

\begin{figure}
\begin{center}
\begin{minipage}{150mm}
\subfigure[Set 1]{
\resizebox*{7.5cm}{!}{\includegraphics{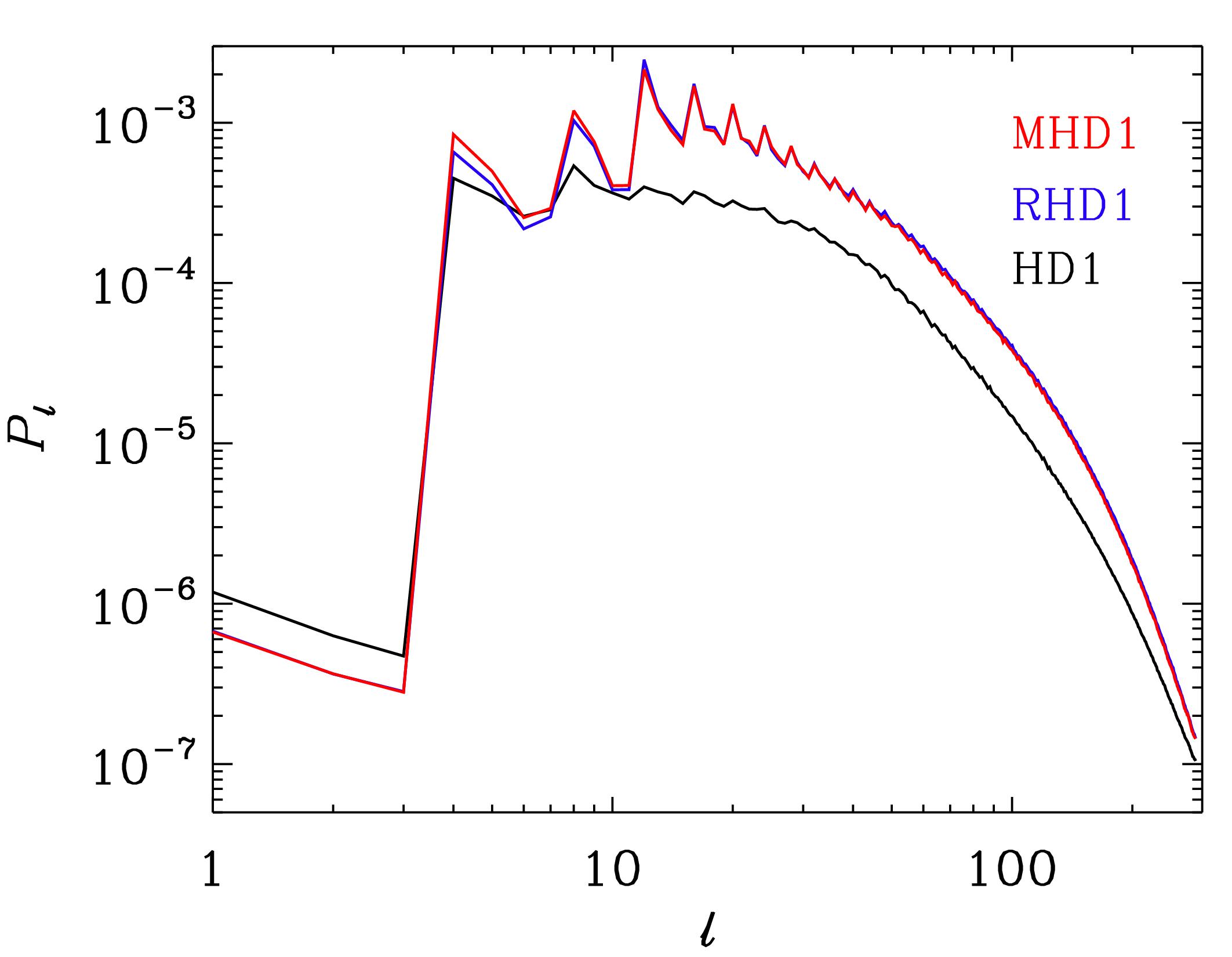}}}%
\subfigure[HD set]{
\resizebox*{7.5cm}{!}{\includegraphics{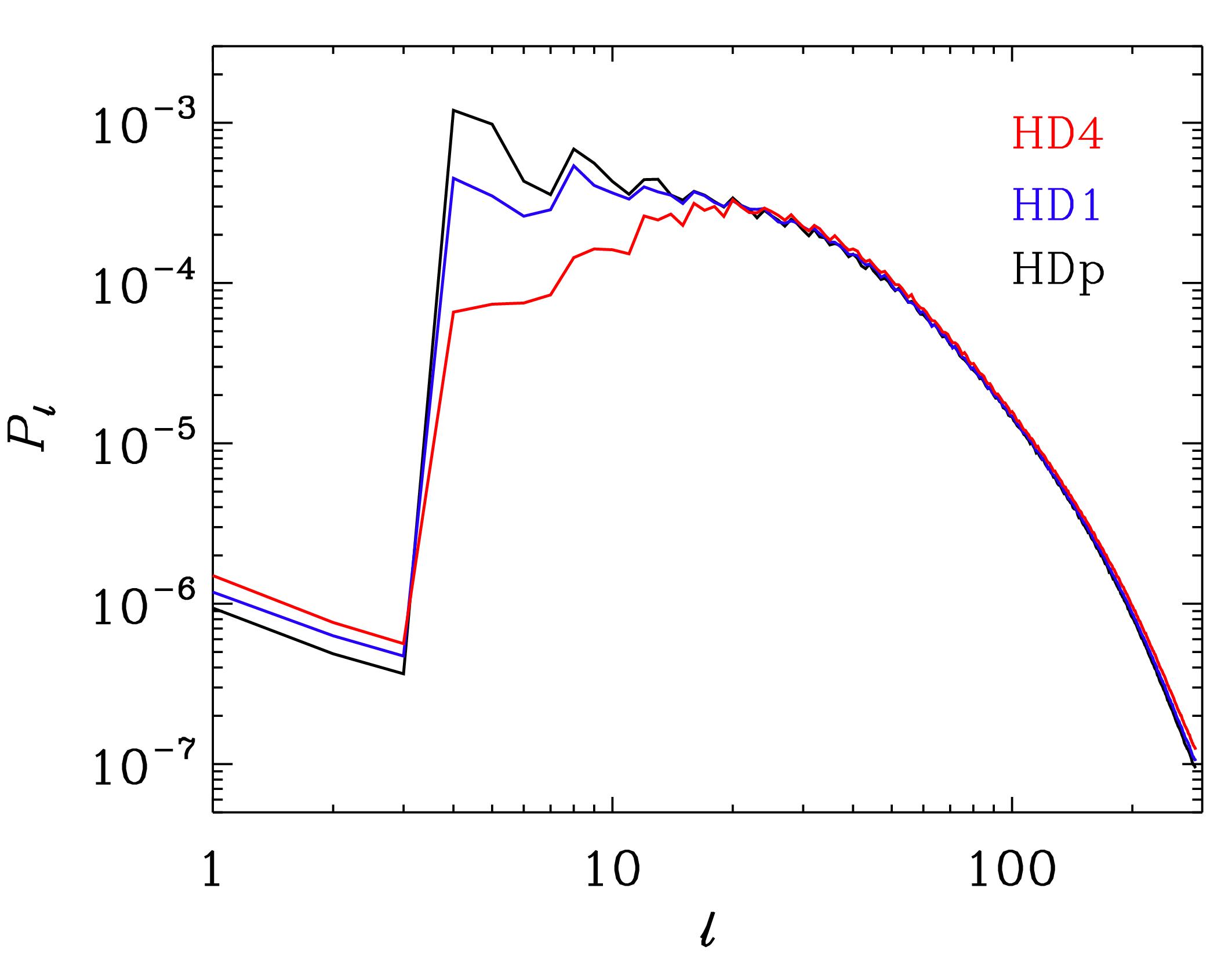}}}%
%\\
\begin{center}
\subfigure[RHD set]{
\resizebox*{7.5cm}{!}{\includegraphics{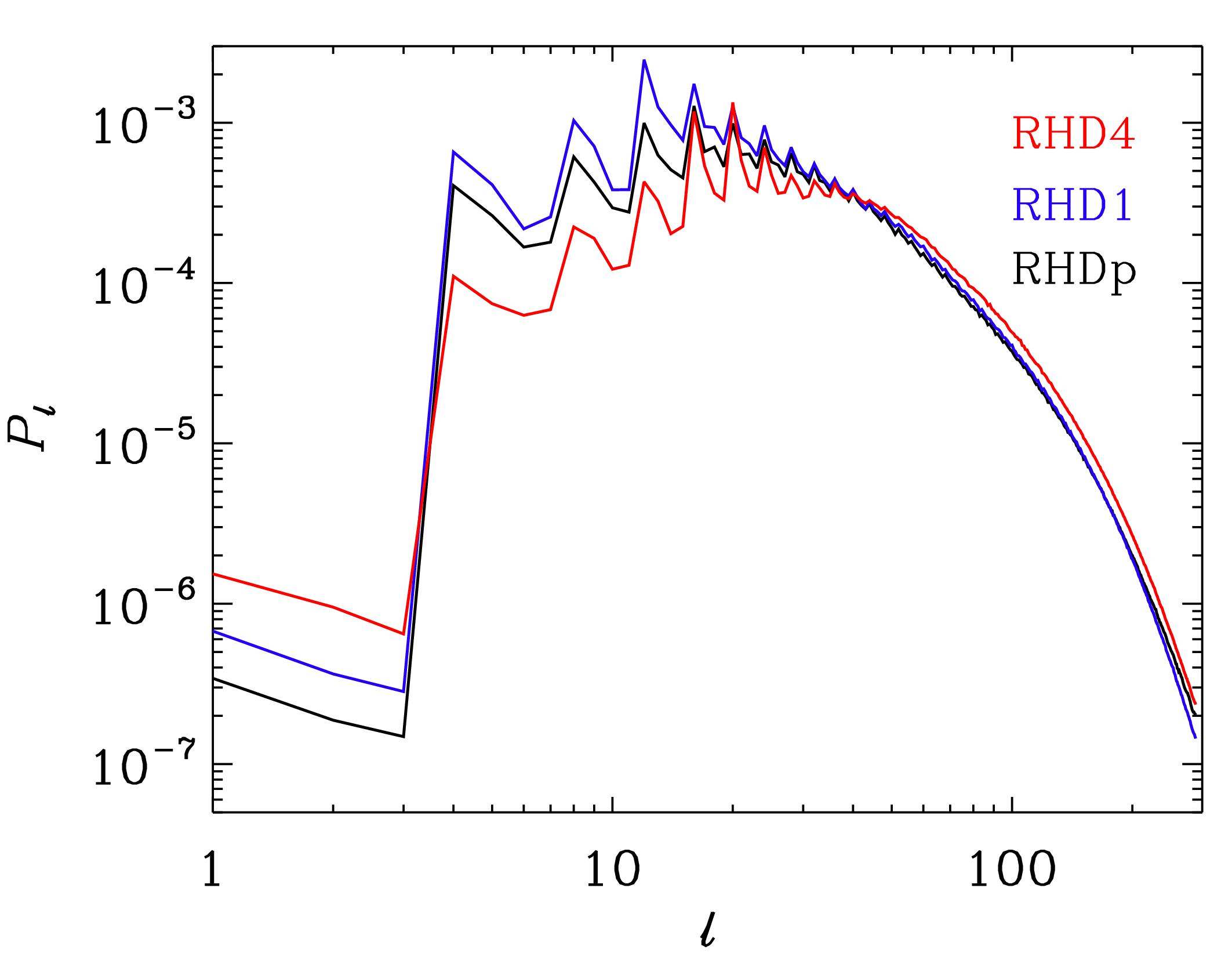}}}%
\subfigure[MHD set]{
\resizebox*{7.5cm}{!}{\includegraphics{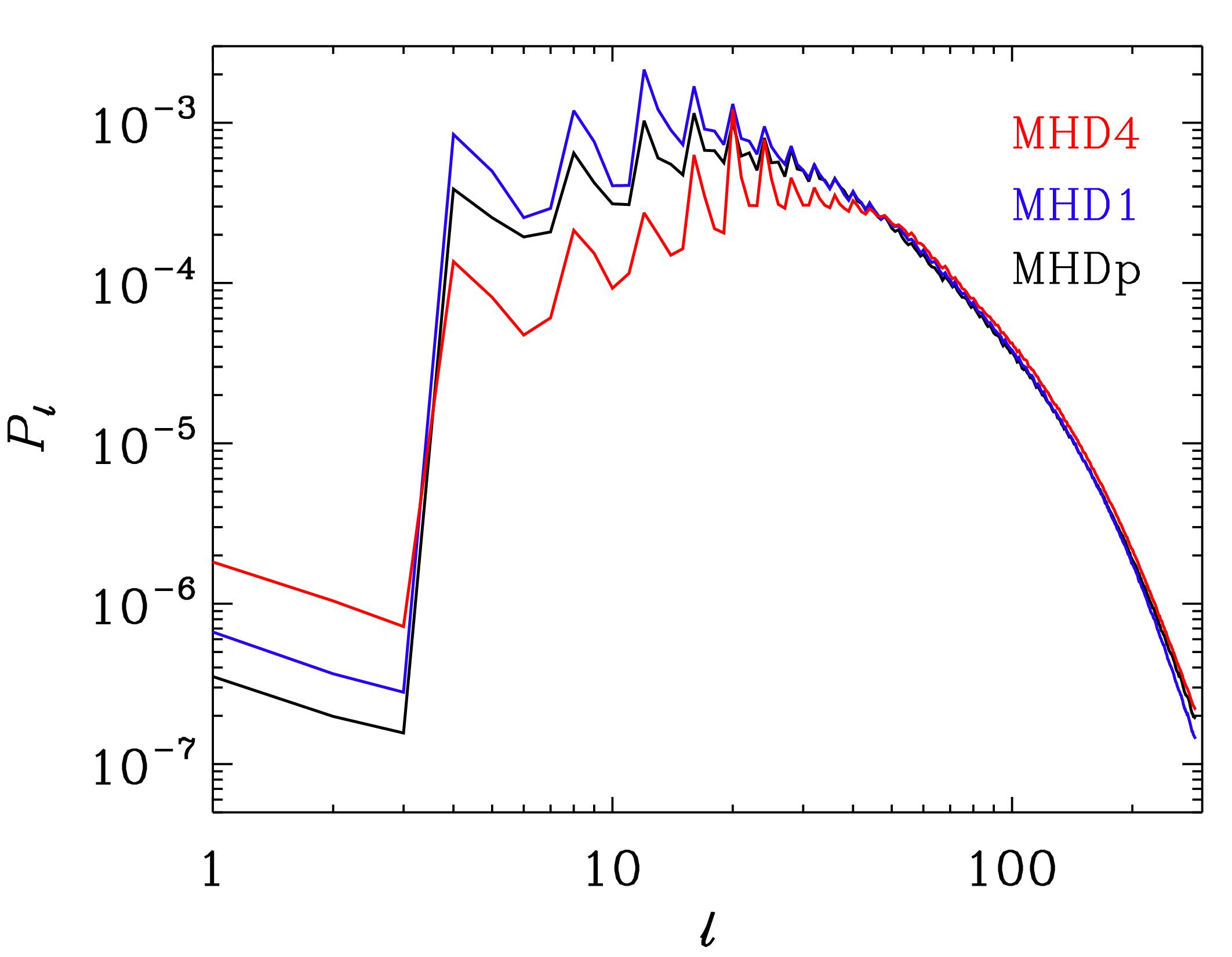}}}%
\end{center}
\caption{Comparison of the convective spectra $P_{\rm conv}$ from
    different runs:
Runs~HD1, RHD1, MHD1 with the same $K_0$ (a), and for runs `p', `1',
and `4' in Sets~HD (b), RHD (c), MHD (d), respectively (colour online).}
\label{fig:spectra_comparison}
\end{minipage}
\end{center}
\end{figure}

\subsection{Horizontal velocity spectra}

In recent studies, \cite{FH16b,FH16} investigated the effects of
increasing supercriticality of convection and the rotational influence
on the spectral energy distribution of convective flows in an effort
to find clues to solve the convective conundrum. Furthermore,
\cite{2017ApJ...845L..23K} found evidence that the structure of
convective flows changes qualitatively when a smoothly varying heat
conduction profile is used.
Hence, next we study the effects of rotation, magnetic fields, and stably stratified
layers on the spectral energy distribution in spherical domains.
To calculate the power spectra for the horizontal velocity,
we follow the same procedure as in \cite{FH16b}.
For each run, we consider a near-surface layer at $r=0.98R_\odot$
  and calculate the normalised power spectrum:
\begin{equation}
P_\ell\,=\,\left.\sum \limits_{m=-\ell}^{\ell}|u_{\ell,m}|^2\right/\sum  \limits_{\ell}\sum \limits_{m=-\ell}^{\ell}|u_{\ell,m}|^2.
\end{equation}
We separate the axisymmetric contribution, $P_{\rm axi}$, given by the $m=0$ mode
to obtain the convective velocity spectra, $P_{\rm conv}$, as the sum of the
higher $m$ modes.

Figure~\ref{fig:pwrspectra} shows the results for four representative runs.
Comparing panels (a) and (b), it is clear that a prescribed profile for
the heat conductivity (Run~HDp, panel a) leads to higher energy in the large
scales than with Kramers profiles (Run~HD4, panel b), when substantial OZ and RZ occur.
The
fact that the power at large scales is dominated by the axisymmetric
component is due to a strong coherent meridional flow that develops in
the system. Similar large-scale convective modes have been reported in
non-rotating and slowly rotating simulations in the past
\citep[e.g.][]{BP09,2018AN....339..127K}.
In Run~HD4, the meridional circulation becomes weaker, which explains the
difference in the spectra.
In addition, by increasing the radiative diffusivity in
rotating MHD Runs~MHD1 and MHD4, see \figu{fig:pwrspectra}(c) and (d),
the total energy at large scales decreases by an order of magnitude and
the peak in the convective spectra moves toward smaller scales, and
the power at large scales in the axisymmetric velocity field becomes
reduced. This
is a consequence of weaker differential rotation; see \figu{fig:pOm}.
The reason for the
changing distribution of the convective power is not so easily
distinguishable. The rotational influence on the flow is changing by
roughly 40 per cent between Runs~MHD1 and MHD4 (see the
sixth column of
\Tablel{tab:runs}) and it is unlikely that this could have caused
such a large effect.
Another possibility is that some of the large-scale convective
modes excited in Run~MHD1 are absent in the shallower CZ of Run~MHD4.

In \figu{fig:spectra_comparison} we compare the horizontal
convective velocity spectra $P_{\rm conv}$ for different runs.
In panel (a) we compare Kramers cases with the same value of $K_0$, but
adding rotation and magnetic fields. Adding rotation has a marked
effect in that the convective power is boosted at practically all
scales. While a change from vertically to horizontally dominated
turbulence as a
function of rotation has been reported earlier from spherical convection
simulations \citep[e.g.][]{KKB14}, the increase of the absolute
magnitude of the horizontal flows is a new result. We find that the
these flows are enhanced especially in the upper parts of the CZ in
the rotating runs. This could be because of shear-produced turbulence
due to the strong differential rotation in these cases.
Adding magnetic fields does not produce further visible difference.
In the non-rotating Kramers runs, the energy at large scales decreases
with increasing heat conductivity (panel (b)),
while in the rotating cases in figures~\ref{fig:spectra_comparison}(c) and (d),
the runs with lower heat conductivity
have higher energy than the run with a prescribed profile.
This is in accordance with \figu{fig:purms}.
In the rotating cases, there seems to be a threshold: at low values of heat conductivity
the energy at large scales is enhanced with respect to the same run with a prescribed profile,
while increasing the value of $K_0$ has the effect of decreasing the energy at small values of $\ell$.
This is also visible in figures~\ref{fig:pFenth}(g), (h), and (i),
where the luminosity is at first enhanced around the equator (Run~MHD1),
and then the transport of energy becomes almost isotropic throughout the CZ
for Run~MHD4.
A possible explanation is that, while the depth of the convectively
unstable layer diminishes, the horizontal extent of the largest
excited convective modes is also reduced, thus lowering the power at
the largest values of $\ell$.
Allowing a self-consistent evolution of the heat conduction profile
helps to reduce the energy at large scales,
but does not affect the small scales.
Moreover, the difference between a prescribed profile and Kramers runs is more marked
in the non-rotating cases.
Adding rotation reduces the effect of a Kramers-like opacity law.

In conclusion, we find that the effects of the Kramers-based heat
conductivity on the velocity amplitudes are rather weak and they are
not enough to resolve the problem of too high convective power in
simulations in comparison to the Sun.

\subsection{Dynamo solutions}

We find that all of the current MHD simulations show large-scale dynamo
action. Time-latitude diagrams of the mean azimuthal field are shown
in \figu{fig:pbutter}. The solution in Run~MHDp shows a cyclic
large-scale field which, however, is relatively weak and stronger
magnetic fields are mostly concentrated toward high latitudes. In Run~MHD1,
the solution does not show clear polarity reversals, although a
quasi-periodic component clearly appears; see
\figu{fig:pbutter}(b). This behaviour is qualitatively similar to that
reported in \cite{KKKBOP15} and in Run~E2 of
\cite{2017A&A...599A...4K}, which, apart from the lower density
stratification, has otherwise similar parameters as the current Run~MHD1.

\begin{figure}
\begin{center}
\begin{minipage}{150mm}
\begin{center}
\subfigure[MHDp]{
\resizebox*{7.5cm}{!}{\includegraphics{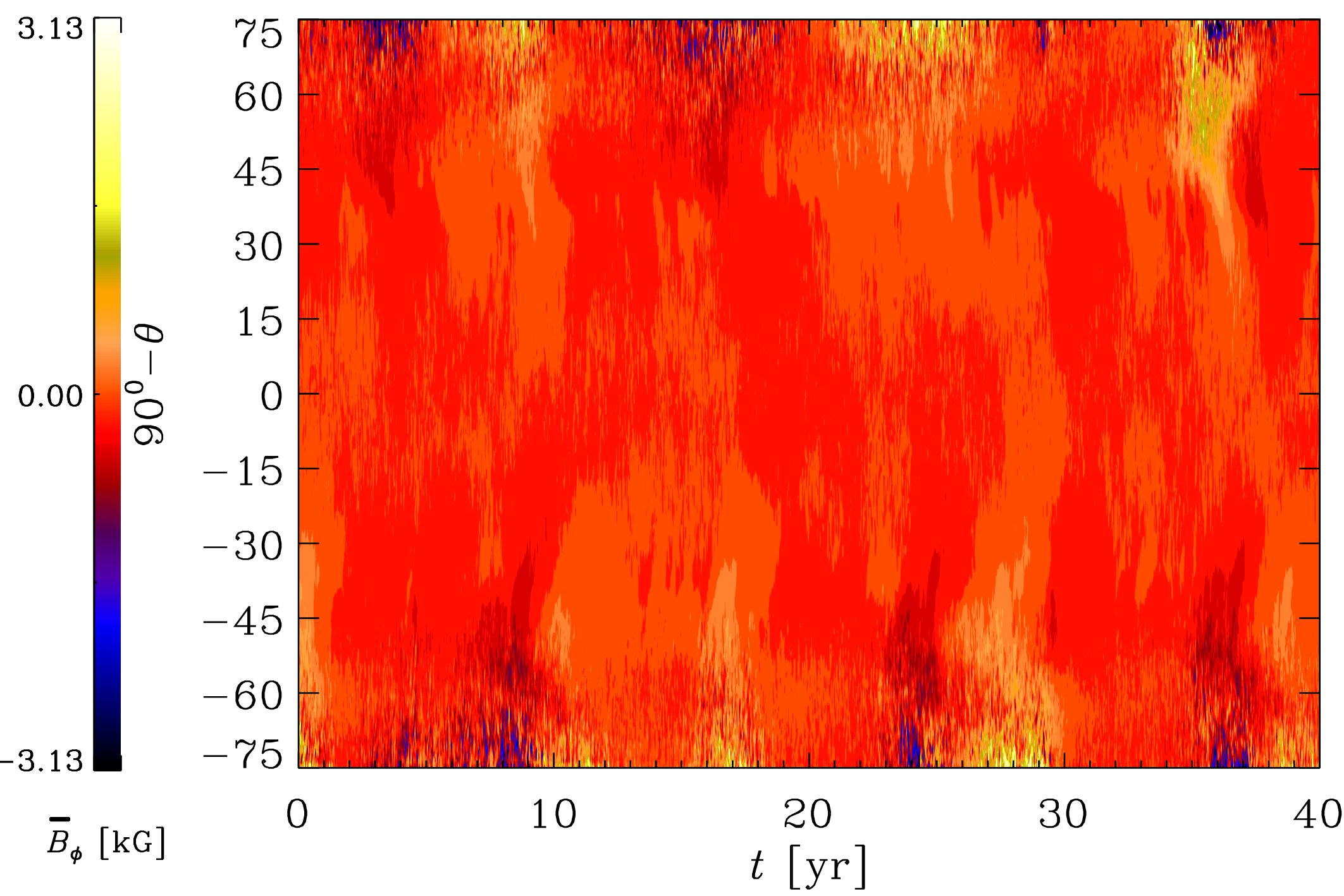}}}%
\subfigure[MHD1]{
\resizebox*{7.5cm}{!}{\includegraphics{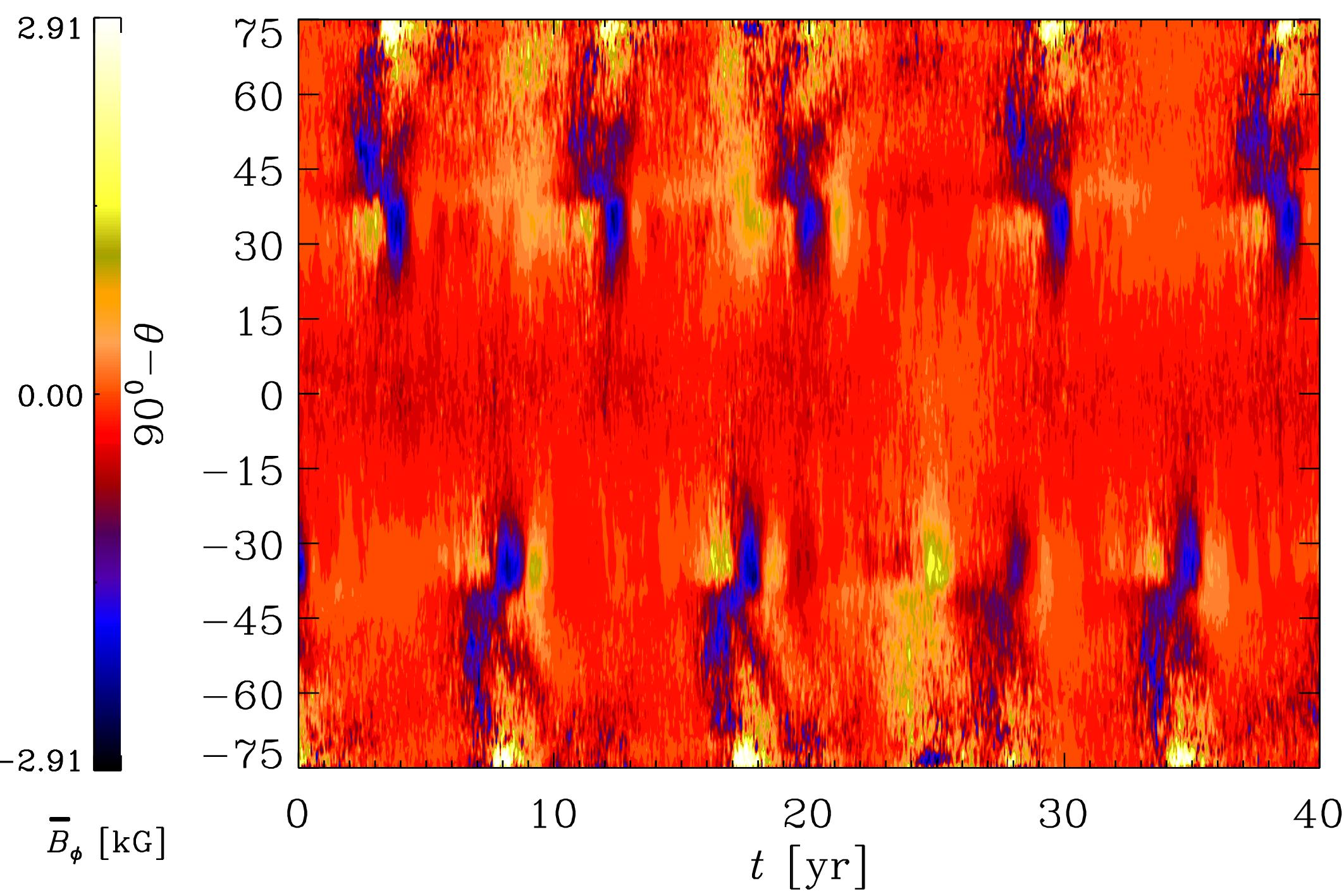}}}%
\\
\subfigure[MHD2]{
\resizebox*{7.5cm}{!}{\includegraphics{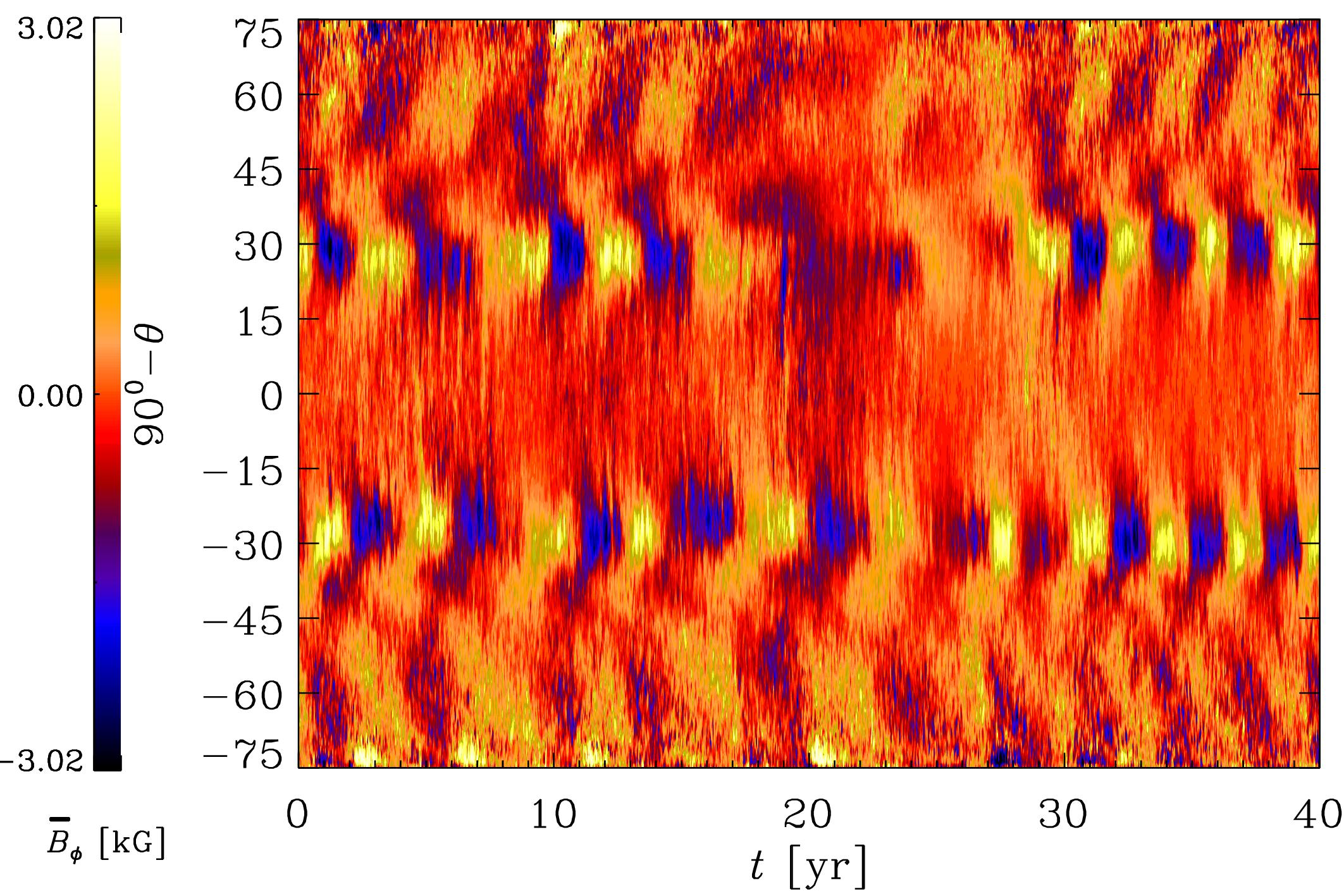}}}%
\subfigure[MHD3]{
\resizebox*{7.5cm}{!}{\includegraphics{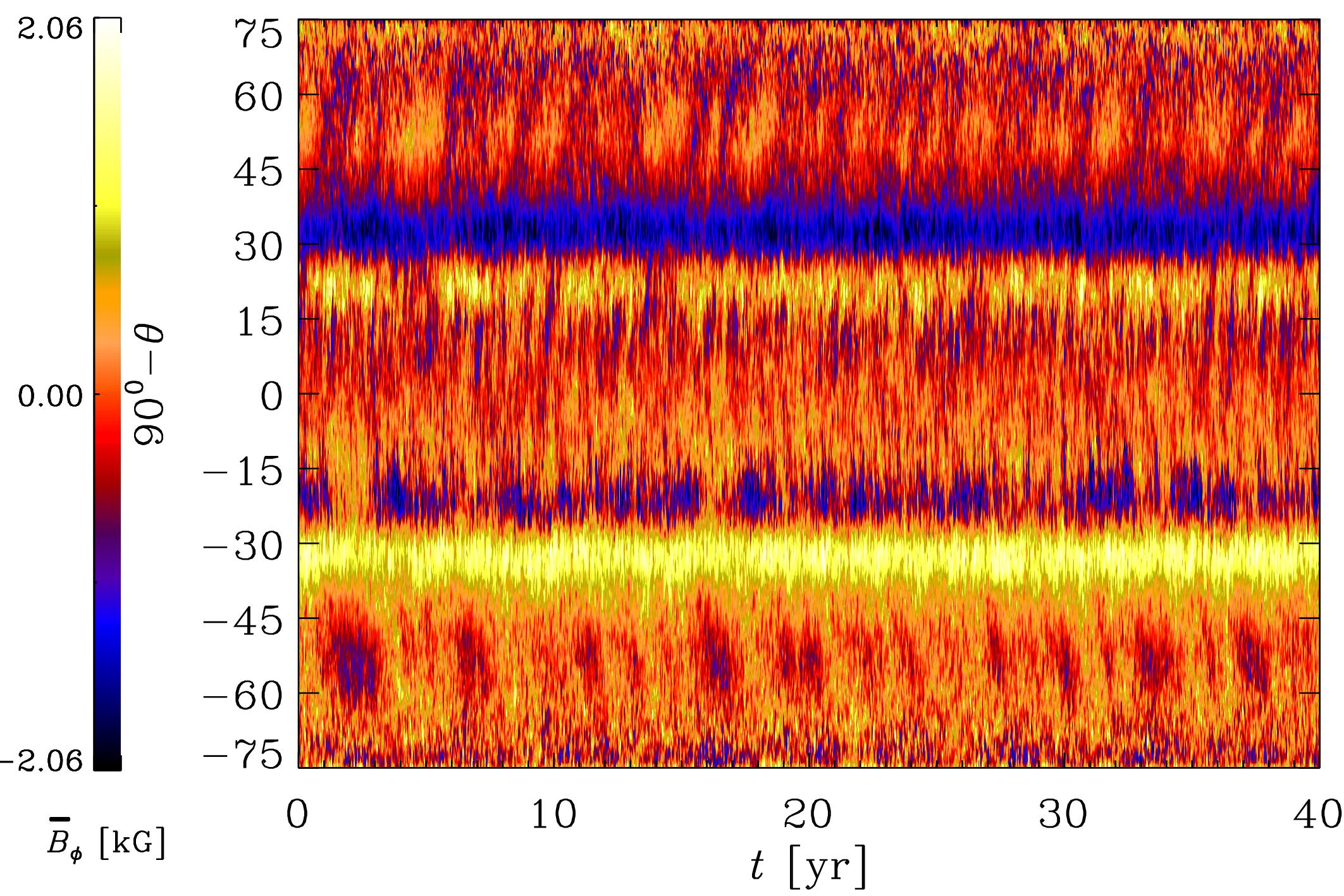}}}%
\\
\subfigure[MHD4]{
\resizebox*{7.5cm}{!}{\includegraphics{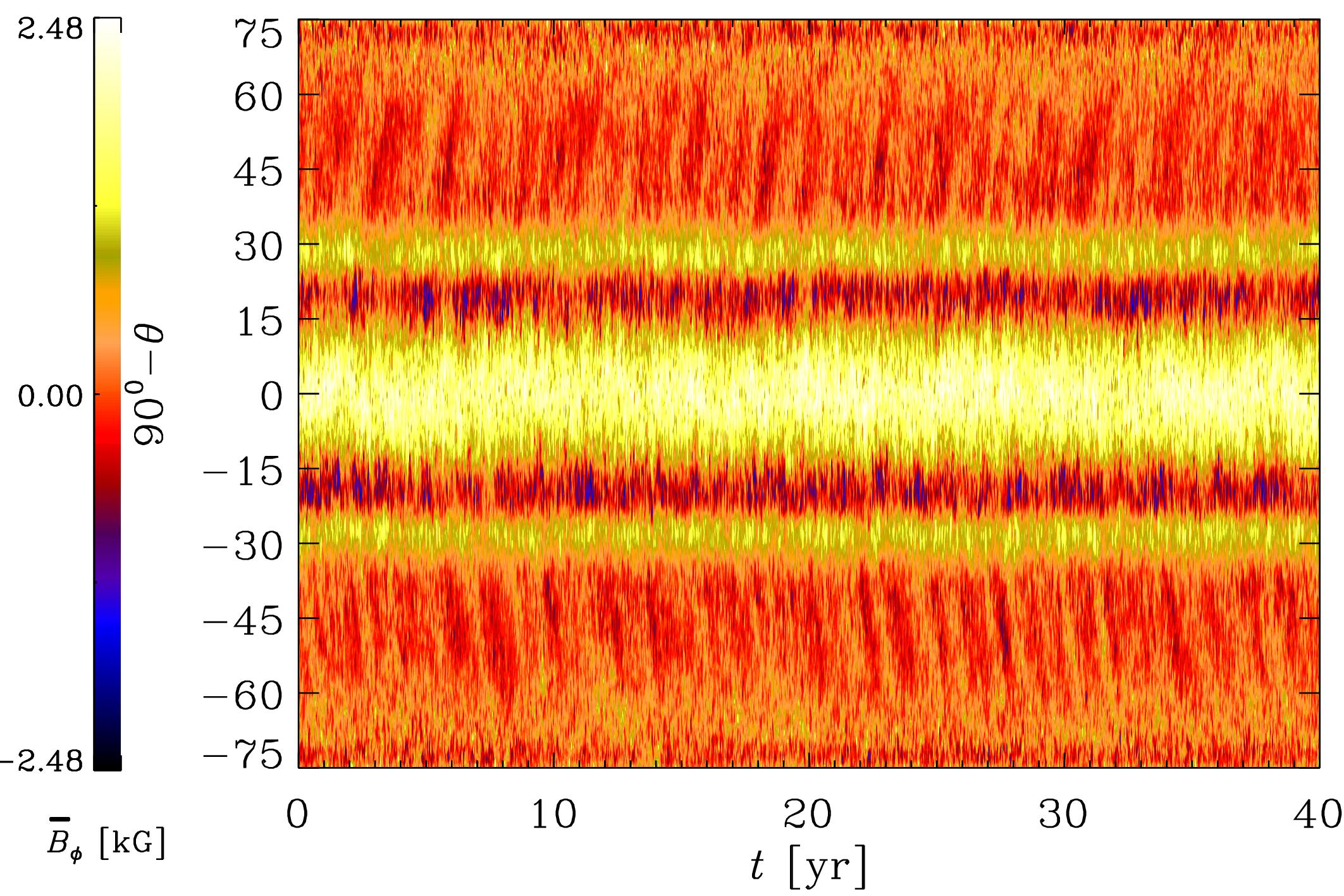}}}%
\subfigure[MHD2 deep]{
\resizebox*{7.5cm}{!}{\includegraphics{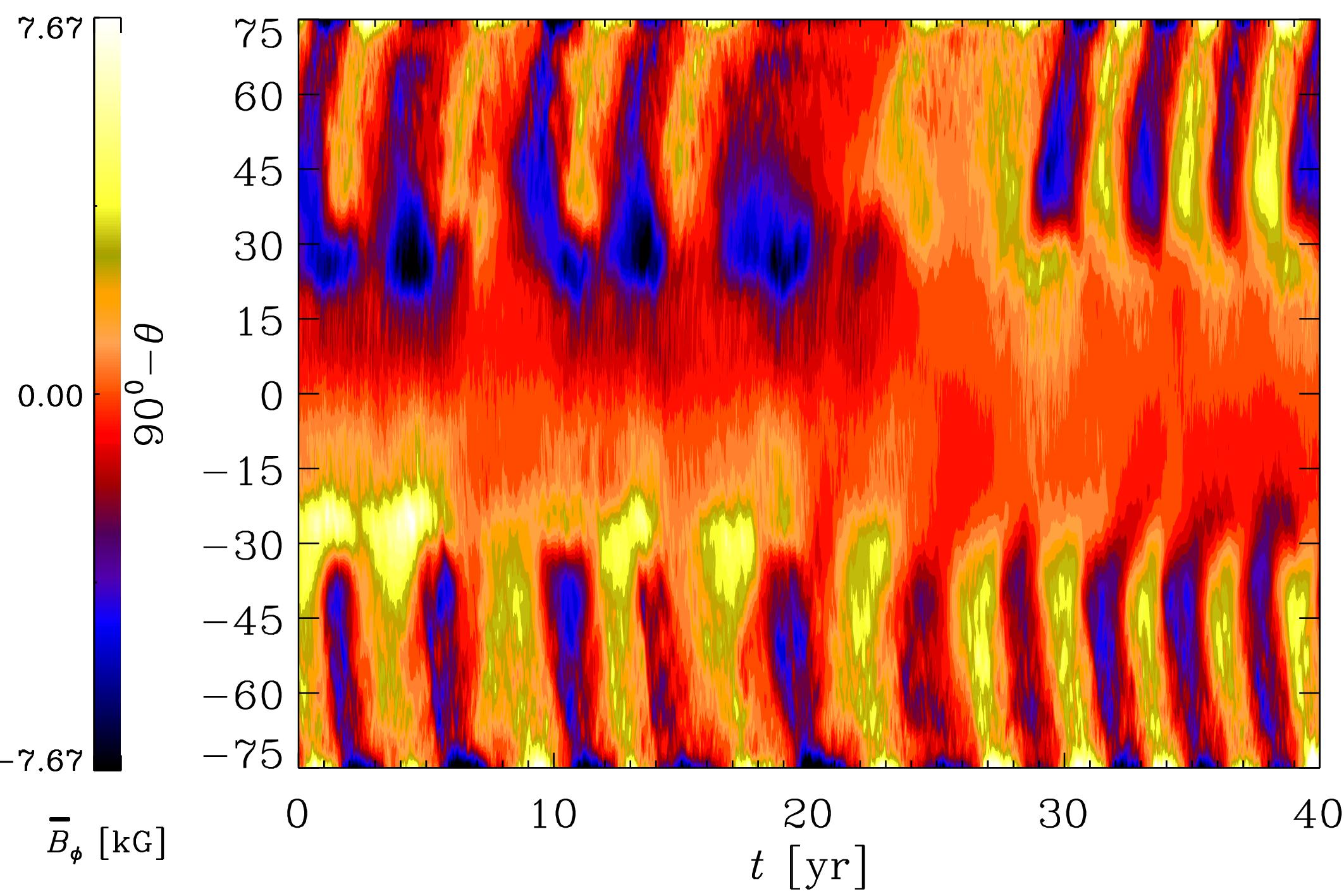}}}%
\end{center}
\caption{Panels (a)--(e): azimuthally averaged azimuthal magnetic field
  $\mean{B}_\phi$
  near the surface at $r/R_\odot=0.98$ as a function of time from a 40
  year time span from the runs in the MHD set. Panel (f) shows the
    azimuthally averaged $\mean{B}_\phi$ from $r/R_\odot=0.75$ from Run~MHD2
    from the same time span as in panel (c) (colour online).}
\label{fig:pbutter}
\end{minipage}
\end{center}
\end{figure}

\begin{figure}
\begin{center}
%PJK: for arXiv-submission
    \includegraphics[width=\textwidth]{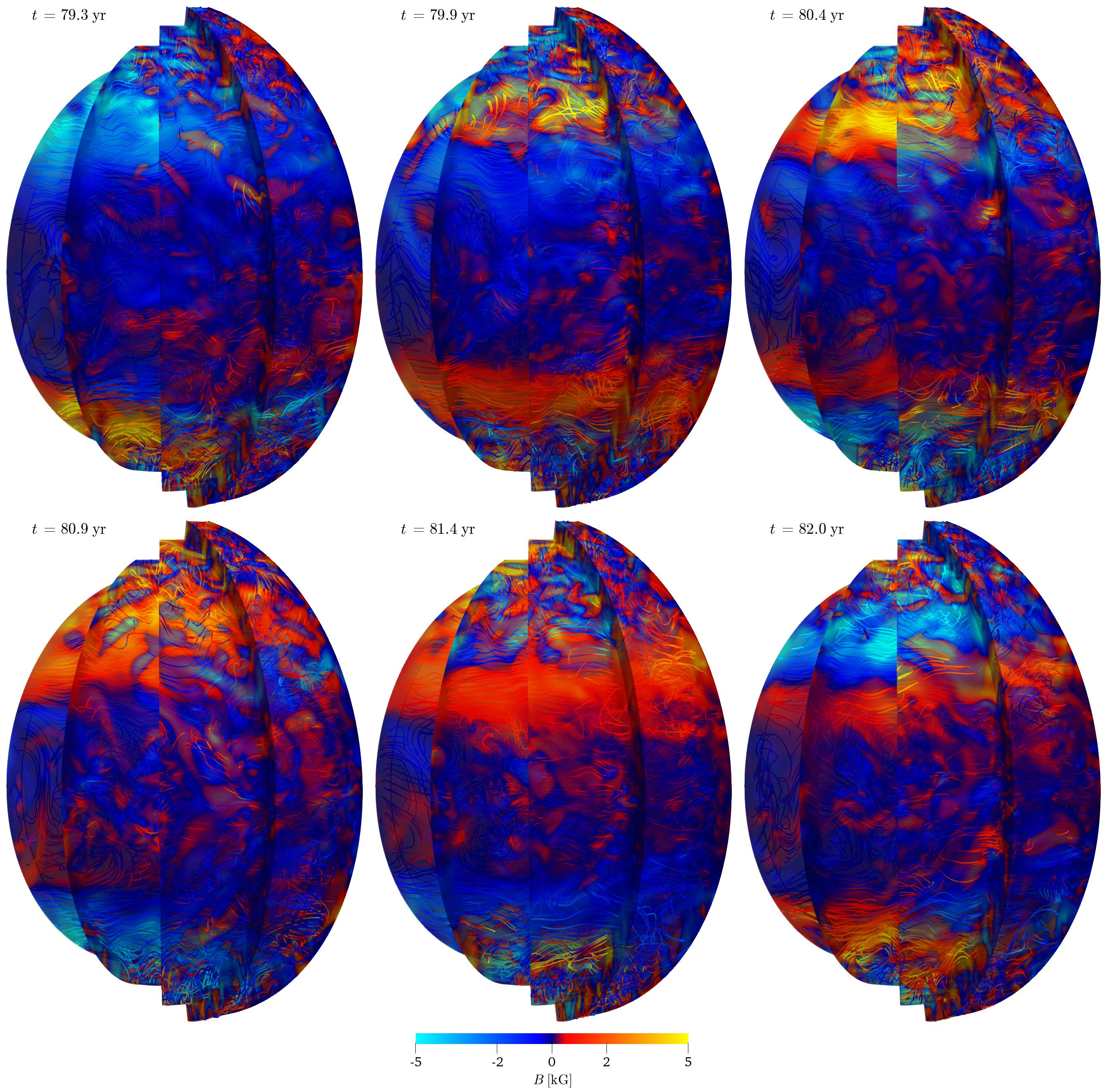}
    \caption{Magnetic fields lines and azimuthal magnetic field
      (colours) from six snapshots covering roughly one magnetic cycle
      from Run~MHD2. The panels are separated by $0.5$--$0.6$ years, as
      indicated by the legends. The colour bar indicates the field
      strength in kilogauss. An animated visualisation of the
        evolution of the magnetic field for this run is available in the online
        material (colour online).}
\label{fig:wedges_Bp}
\end{center}
\end{figure}

In Run~MHD2, a clearly oscillatory mode is excited, which is reminiscent
of earlier results
\citep{KMB12,KMCWB13,ABMT15,KKOBWKP16,SBCBN17,2018A&A...616A..72W}.
The main cycle period of a very similar run parameter-wise \citep{KKOBWKP16}
was reported to be roughly 5 years, whereas here the cycle is shorter
nearly by a factor of two.
Representations
of the magnetic fields in different phases of the cycle are shown in
\figu{fig:wedges_Bp}. Quantitative differences to Run~MHD1 are
relatively minor; the Reynolds and Coriolis numbers differ by roughly
15 per cent, see the fifth and sixth columns of \Tablel{tab:runs}. However,
the Deardorff layer is thicker at low latitudes and the region of
negative radial shear is wider at mid-latitudes in Run~MHD2 in
comparison to Run~MHD1. The dynamo solution thus
appears to be sensitive to relatively small changes in the flow
properties.

We also observe a quiescent period roughly between 15 and 25 years
in physical time that can be interpreted as a Maunder minimum-type
event \citep[see also][]{ABMT15,KKOBWKP16}. During this event, also the
dominant dynamo mode at the surface appears to change to a shorter
one at late stages. The minimum event and the changing dynamo
mode are due to a change of magnetic
field structure in the deeper layers. This is illustrated in
  \figu{fig:pbutter}(f), where $\mean{B}_\phi$ near the
  bottom of the CZ is shown. The period of the oscillatory mode at
  latitudes $|\Theta|\gtrsim 30^\circ$ decreases after a readjustment
  around $t=15$--25 years. This is also where the polarity of
  the near--equator field changes, which could suggest a dynamo mode
  in the deep parts with a much longer period. The time series
  is too short to determine the dynamo period of its statistics
  reliably, however. These results are
in agreement with the conclusions of \cite{KKOBWKP16}.

\begin{figure}
\begin{center}
\includegraphics[width=.5\textwidth]{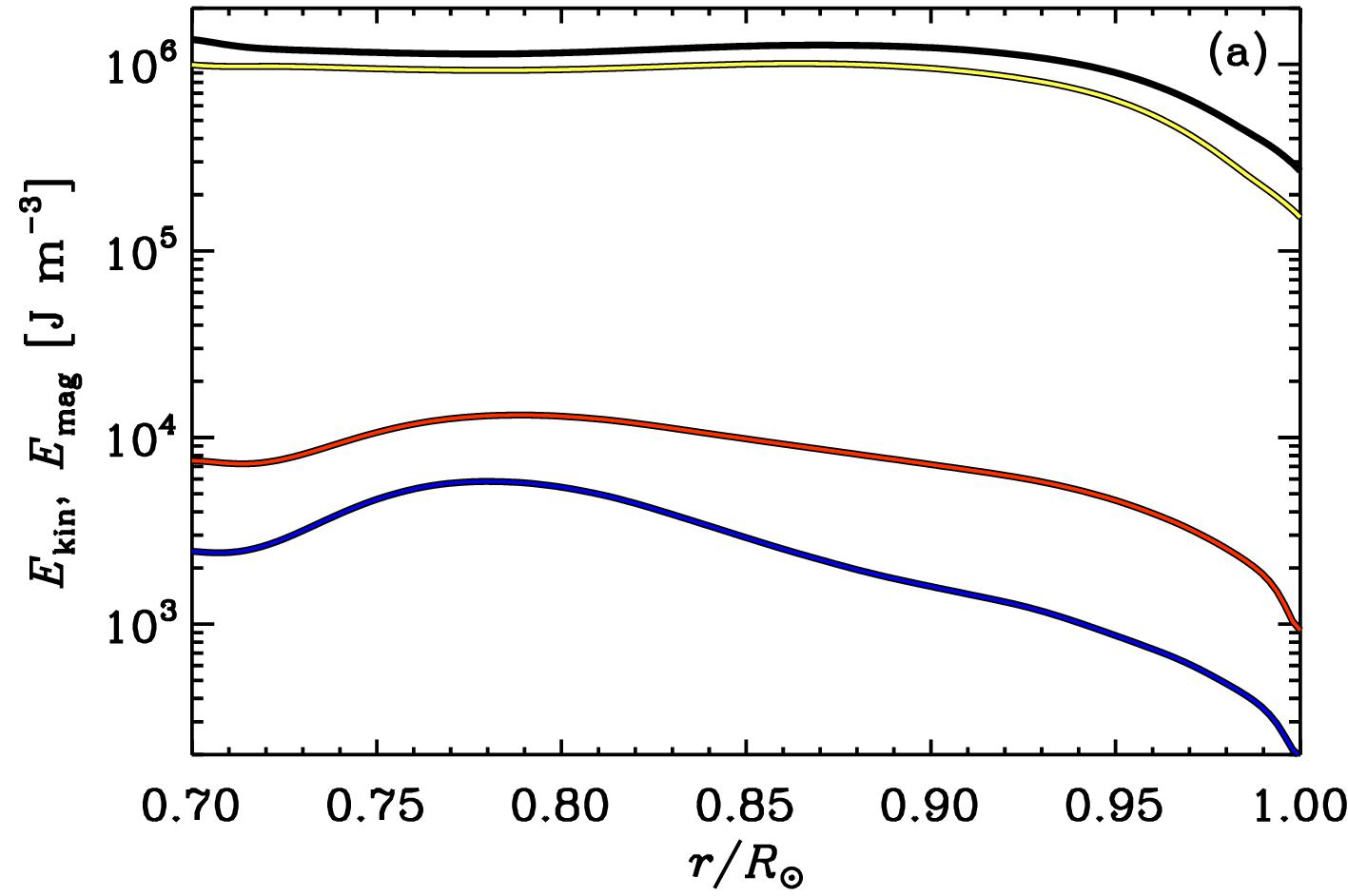}\includegraphics[width=.5\textwidth]{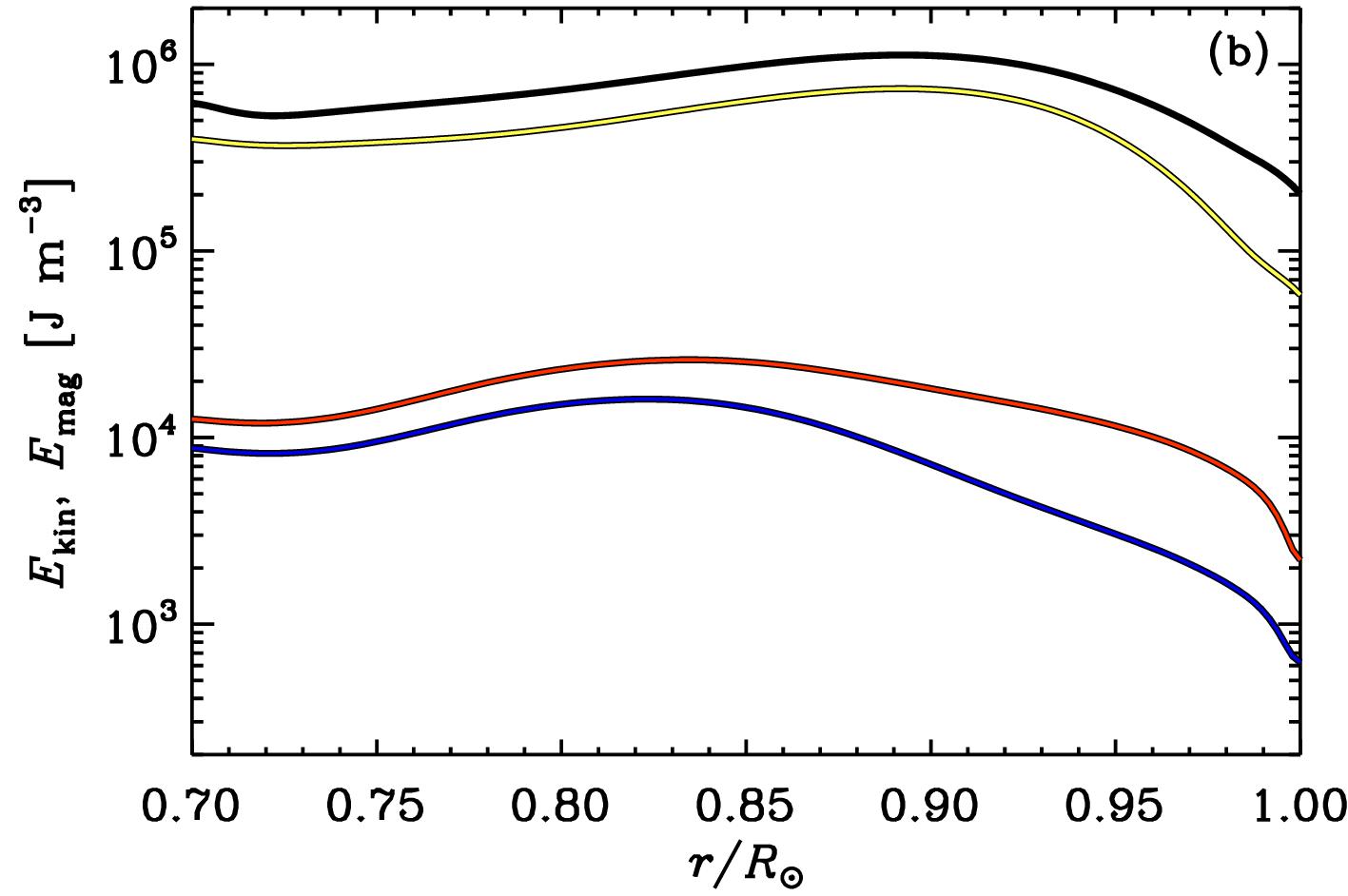}\\
\includegraphics[width=.5\textwidth]{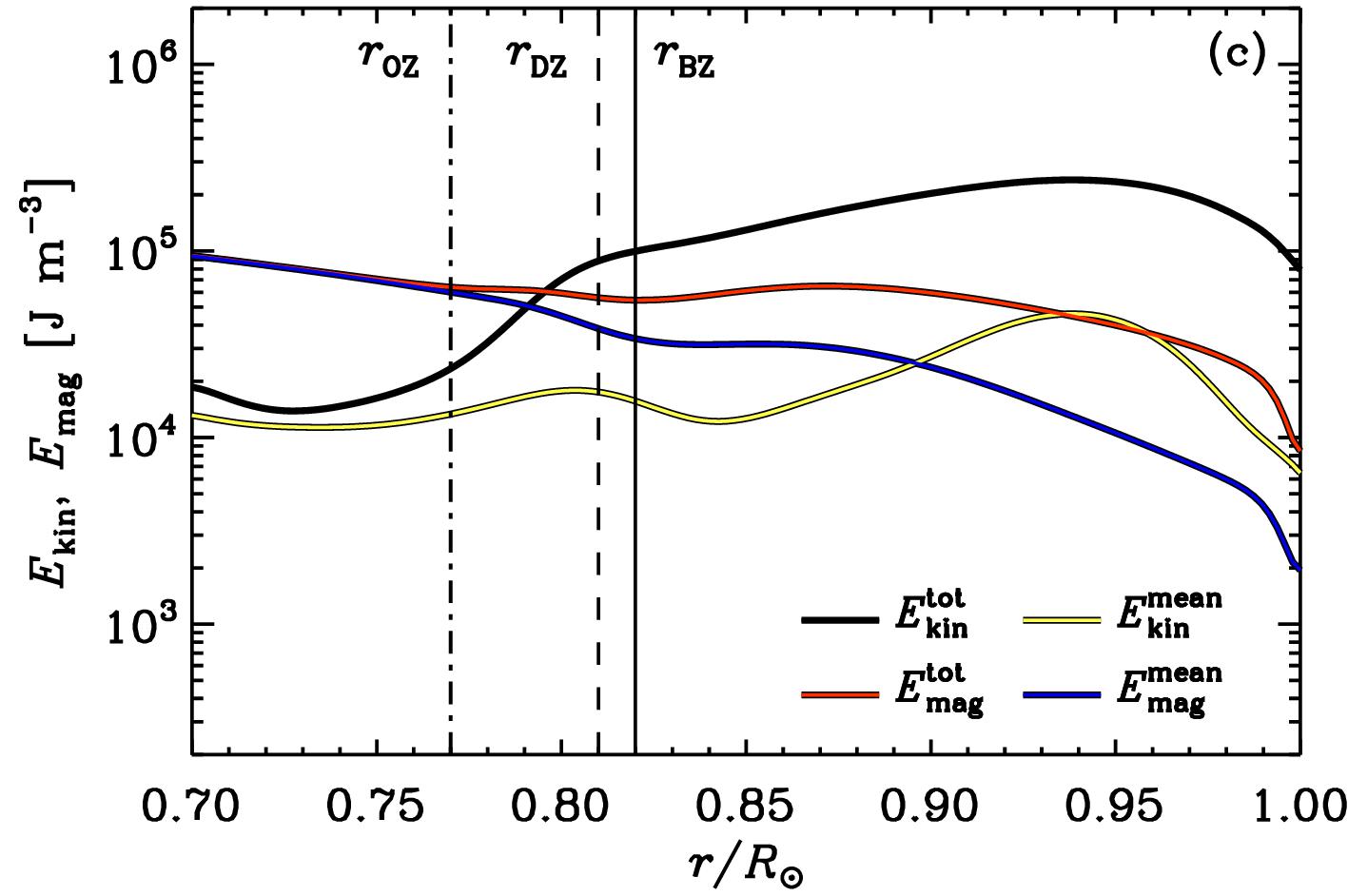}\includegraphics[width=.5\textwidth]{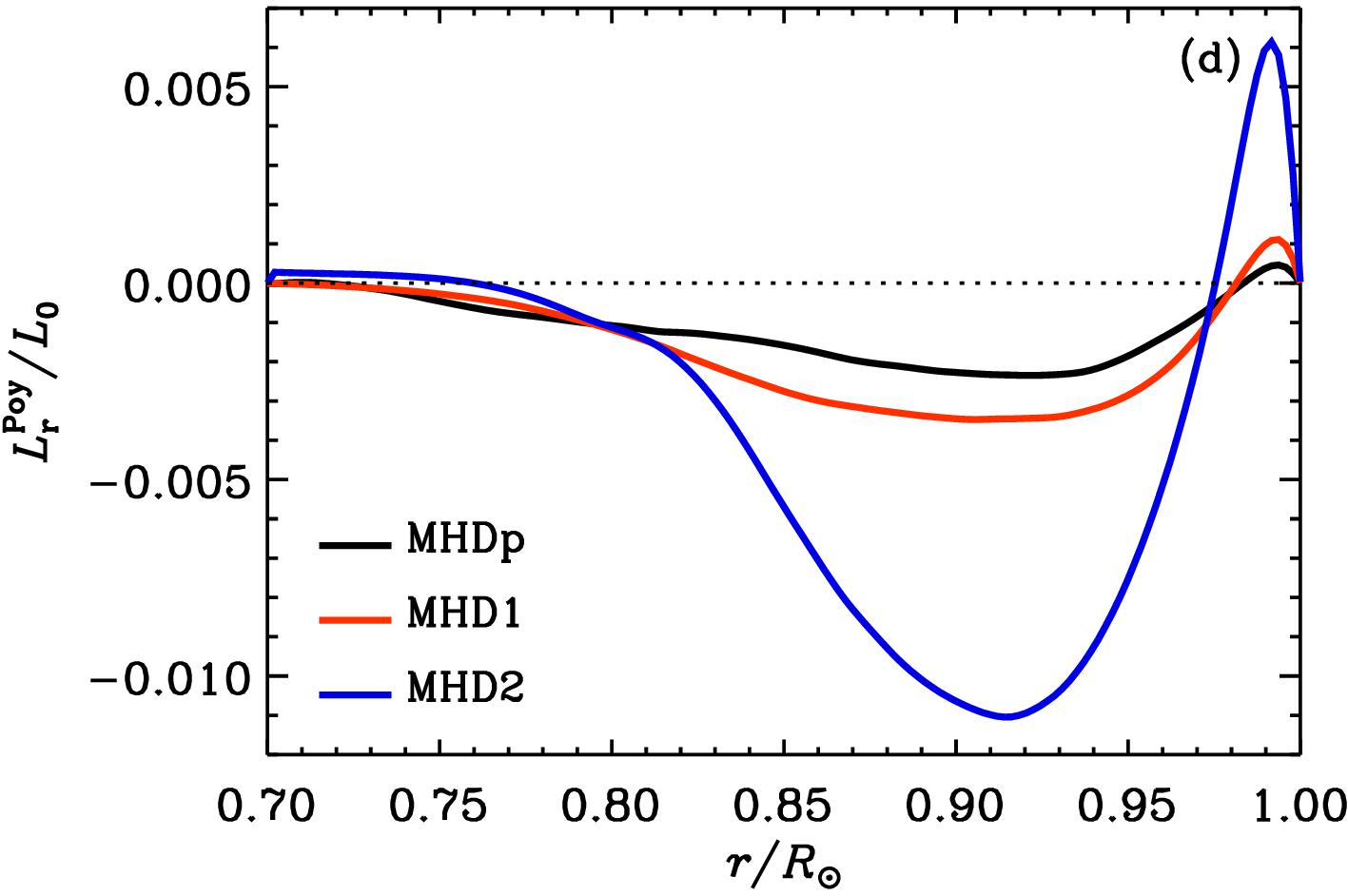}\\
\caption{Horizontally and temporally averaged energy densities of
    the total flow (black), mean flow (yellow), total magnetic field
    (red), and mean magnetic field (blue) for Runs~MHDp (a), MHD1 (b),
    and MHD4 (c). In panel (c) the bottoms of the BZ,
    DZ, and OZ
    are indicated with solid, dashed, and dash-dotted lines,
    respectively. Panel (d) shows the horizontally
    averaged luminosity corresponding to the radial Poynting flux for
    the same runs (colour online).}
\label{fig:magene}
\end{center}
\end{figure}

The total magnetic energy $E_{\rm mag}^{\rm
    tot}=\langle\bm{B}^2/2\mu_0\rangle_{\theta\phi}$ in Runs~MHDp and
  MHD1 is on the order of $10^4$~J~m$^{-3}$; see
  \figu{fig:magene}(a)--(b). The energy of the mean field $E_{\rm
    mag}^{\rm
    mean}=\langle\mean{\bm{B}}^2/2\mu_0\rangle_{\theta\phi}$, on the
  other hand, is roughly 20--30 percent of $E_{\rm mag}^{\rm tot}$ in
  MHDp and somewhat more in MHD1. The total kinetic energy $E_{\rm
    kin}^{\rm tot}=\langle\rho \bm{U}^2/2\rangle_{\theta\phi}$ is of
  the order of $10^6$~J~m$^{-3}$ in MHDp and somewhat less in
  MHD1. The contribution of the mean flows (differential rotation and
  meridional circulation), $E_{\rm kin}^{\rm mean}=\langle\rho
  \mean{\bm{U}}^2/2\rangle_{\theta\phi}$, is higher in MHDp in
  comparison to MHD1. This could reflect the effect of higher mean
  magnetic fields in the latter. The overall magnetic energy in these
  two runs is roughly an order of magnitude less than in runs with
  similar Coriolis number and Reynolds numbers in
  \cite{2017A&A...599A...4K} (their run E2). The current runs differ
  from those of \cite{2017A&A...599A...4K} in that the density
  stratification is roughly three times higher. Furthermore, the
  radial thermal boundary conditions and the treatment of the
  near-surface cooling differ from those used in
  \cite{2017A&A...599A...4K} as explained in \sect{sec:diffrot}. The
  dynamics of the
  simulations are sensitive to the thermal boundary conditions as well
  as the parameterisation of the near-surface layers
  \citep{2018arXiv180709309K} which can also influence the dominant
  dynamo mode via the flow. Lastly, the SGS flux of entropy is here
  applied to the fluctuations as opposed to the total
  entropy. However, the current data is insufficient to track down the
  cause of the differences in the magnetic energy levels.

In Runs~MHD3 and MHD4, the dynamo switches
to a non-oscillatory mode. These solutions are superficially
  similar to quasi-stationary magnetic `wreaths' at mid-latitudes
  found in several studies \citep[e.g.][]{BBBMT10,NBBMT11}. These runs
  did not, however, include a radiative layer below the convection
  zone, although they operate in a similar Coriolis number regime. One
  possible explanation is that the rotation profiles of these runs do
  not have a local minimum at mid-latitudes that is seen in later
oscillatory solutions. In Runs~MHD3 and MHD4, the surface appearance
of the toroidal magnetic field
reflects the occurrence of a large-scale field stored beneath the CZ in the stably
stratified layers; see panel (c) of \figu{fig:magene}. The
  overall magnitude of the magnetic field is significantly higher than
  in the cases without an RZ. We note that,
  for example, \cite{GSdGDPKM15}, \cite{2018A&A...616A..72W}, and
  \cite{2018ApJ...863...35S} found changing dynamo modes as a function
of the rotational influence on the flow. The Coriolis number based on the
depth of the revised CZ varies by roughly 40 per cent in the runs of
the MHD set; see the sixth column of \Tablel{tab:runs}, which is a
  possible explanation for the change of the dominant dynamo mode.

\begin{figure}
\begin{center}
\begin{minipage}{150mm}
\subfigure[MHDp]{
\resizebox*{5cm}{!}{\includegraphics{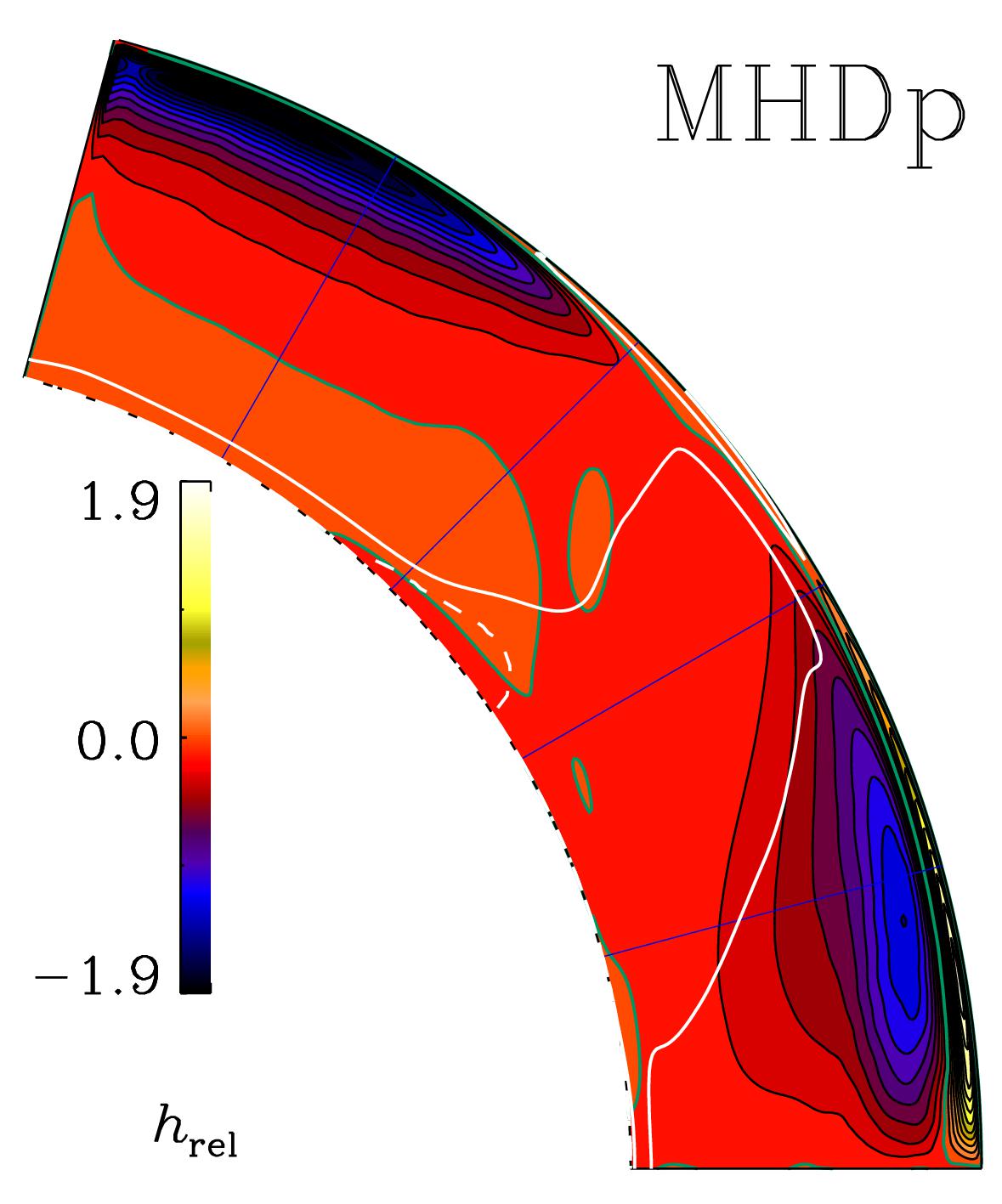}}}%
\subfigure[MHD1]{
\resizebox*{5cm}{!}{\includegraphics{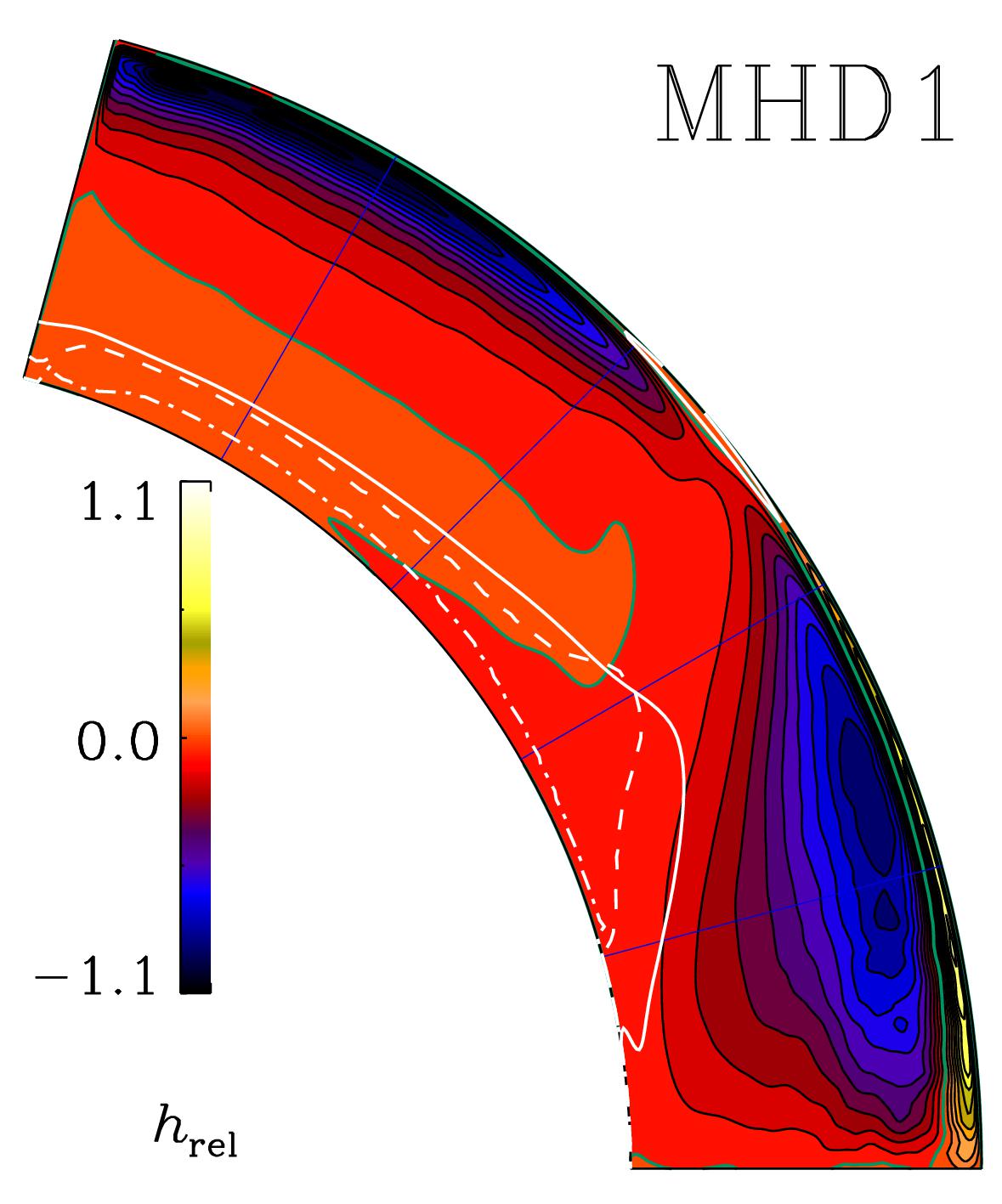}}}%
\subfigure[MHD2]{
\resizebox*{5cm}{!}{\includegraphics{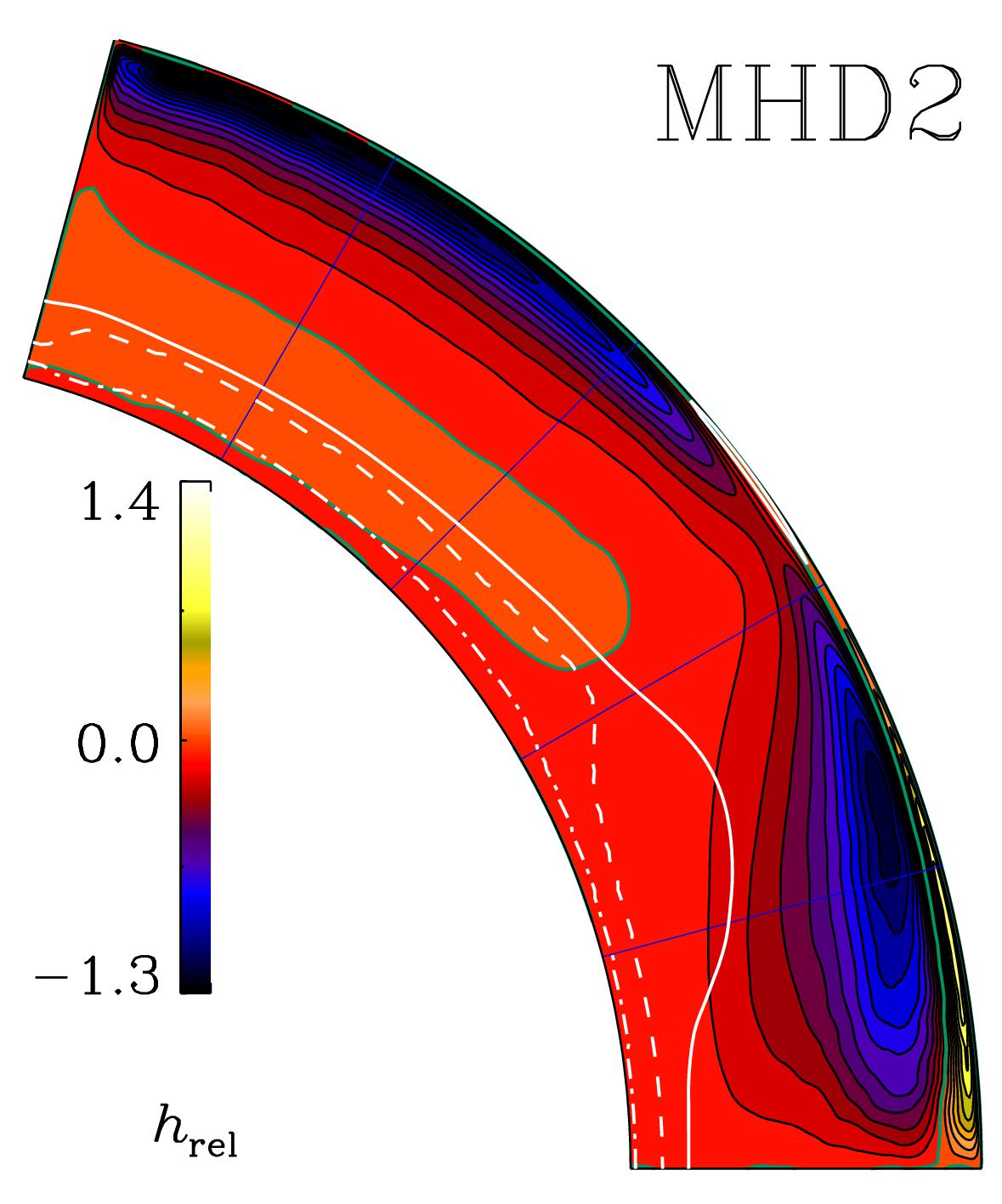}}}%
\\
\begin{center}
\subfigure[MHD3]{
\resizebox*{5cm}{!}{\includegraphics{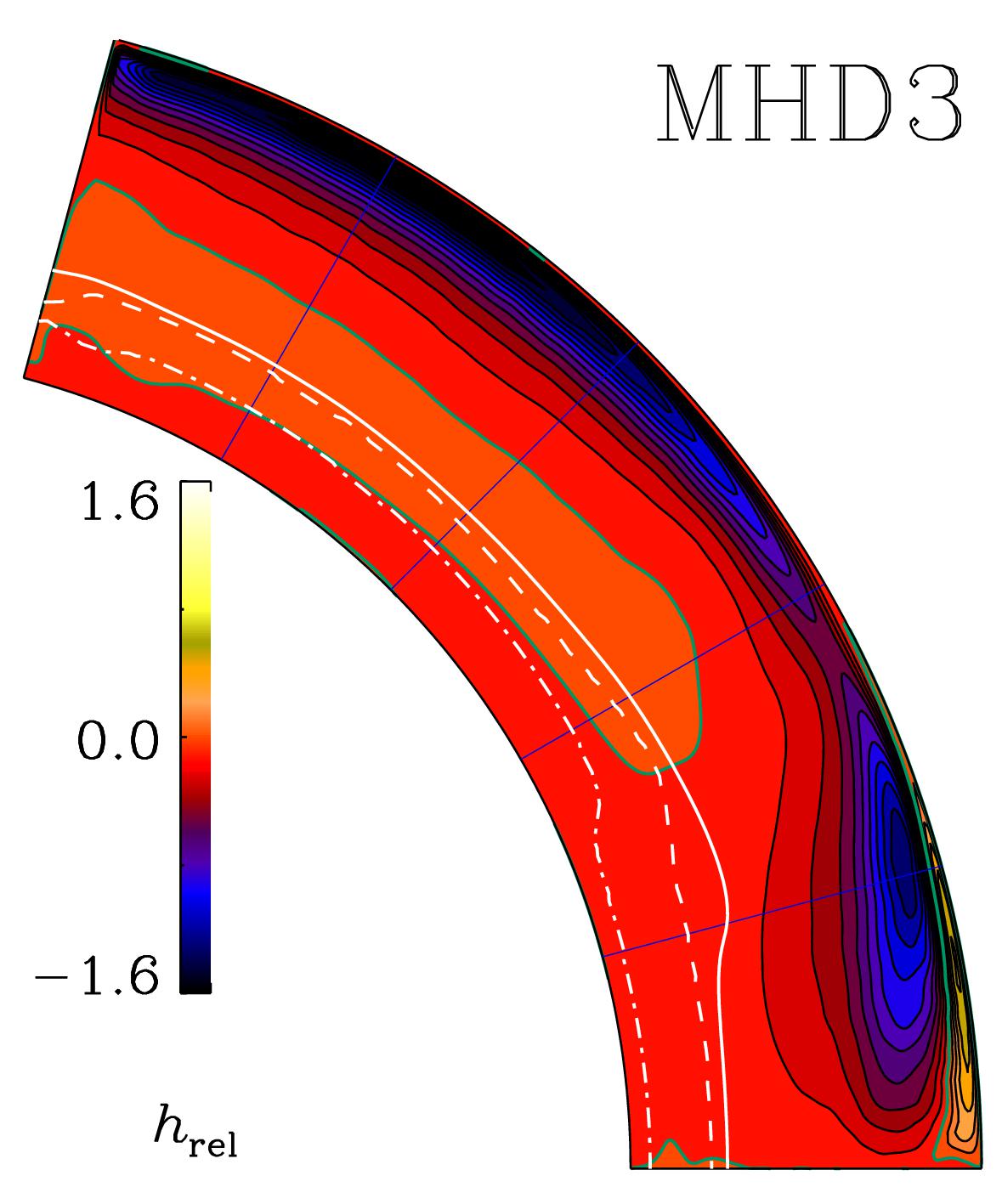}}}%
\subfigure[MHD4]{
\resizebox*{5cm}{!}{\includegraphics{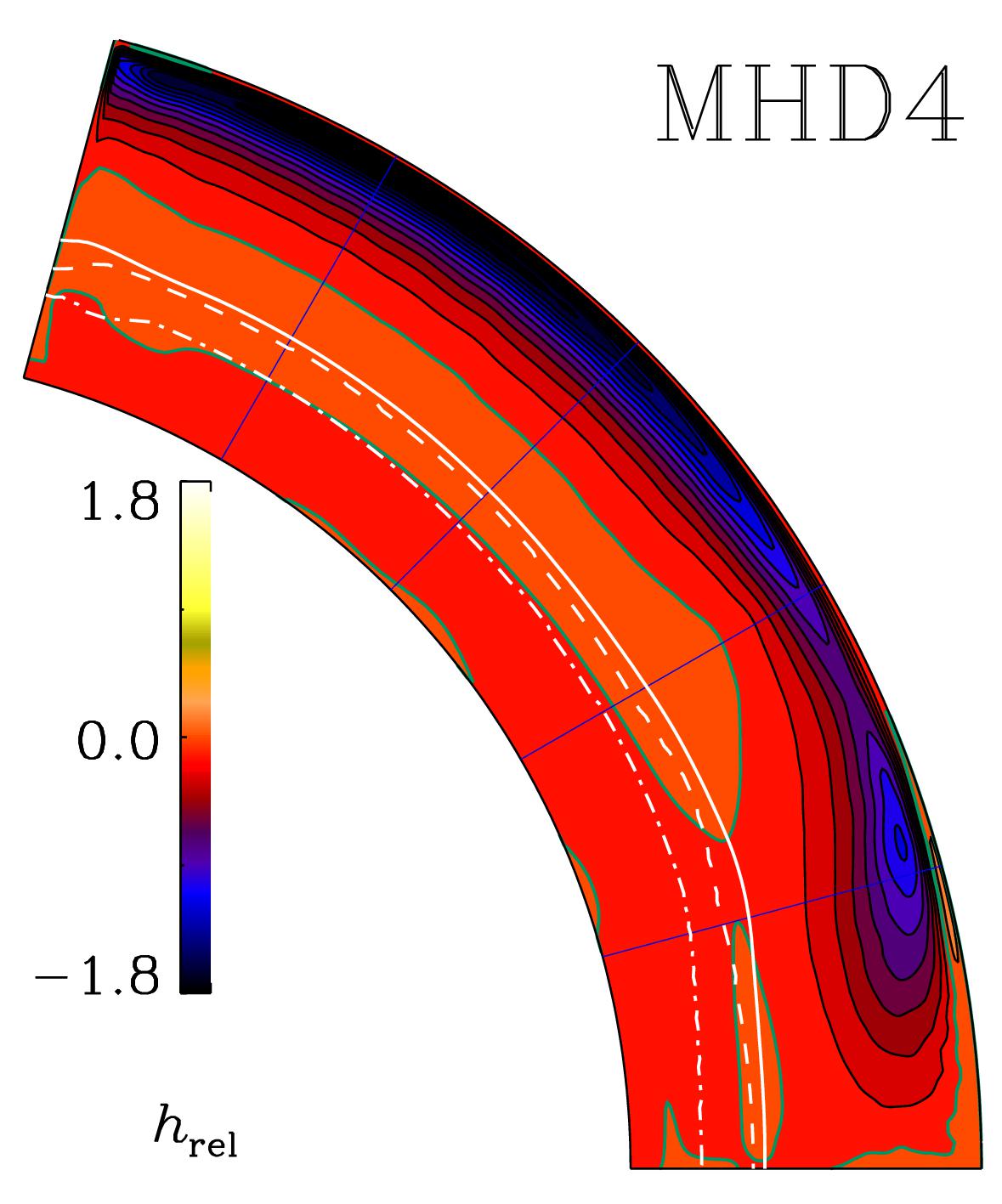}}}%
\end{center}
\caption{Time-averaged relative kinetic helicity from the MHD runs.
As in \figu{fig:pFenth}, the black and
white solid, dashed, and dash-dotted lines indicate the bottoms of
the buoyancy, Deardorff, and overshoot zones, respectively. The green
line indicates the zero level (colour online).}
\label{fig:poum}
\end{minipage}
\end{center}
\end{figure}

We further study the magnetic energy transport due to the Poynting flux,
\begin{eqnarray}
\bm{F}^{\rm Poy} \,= \,(\bm{E} \times \bm{B})\big/\mu_0\,,
\end{eqnarray}
where $\bm{E}=-\bm{U}\times \bm{B} + \eta \mu_0 \bm{J}$ is the
electric field. We consider the luminosity of the horizontally
averaged radial component of $\FFF^{\rm Poy}$, $\langle L_r^{\rm Poy}
\rangle_{\theta\phi} = 4\pi r^2 \langle F_r^{\rm Poy}
\rangle_{\theta\phi}$ in \figu{fig:magene}(d).  The magnitude of
$\langle L_r^{\rm Poy} \rangle_{\theta\phi}$ is
at most on the order of one per cent of the total flux in Run~MHD4 and
between 0.2 and 0.3 per cent in Runs~MHDp and MHD1. This suggest that
the Poynting flux has an almost negligible effect on the total energy
transport. Furthermore, the flux always points downwards, which
agrees with previous results in Cartesian \citep{NBJRRST92} and
spherical \citep{BMT04} geometries.

\begin{figure}
\begin{center}
\includegraphics[width=.33\textwidth]{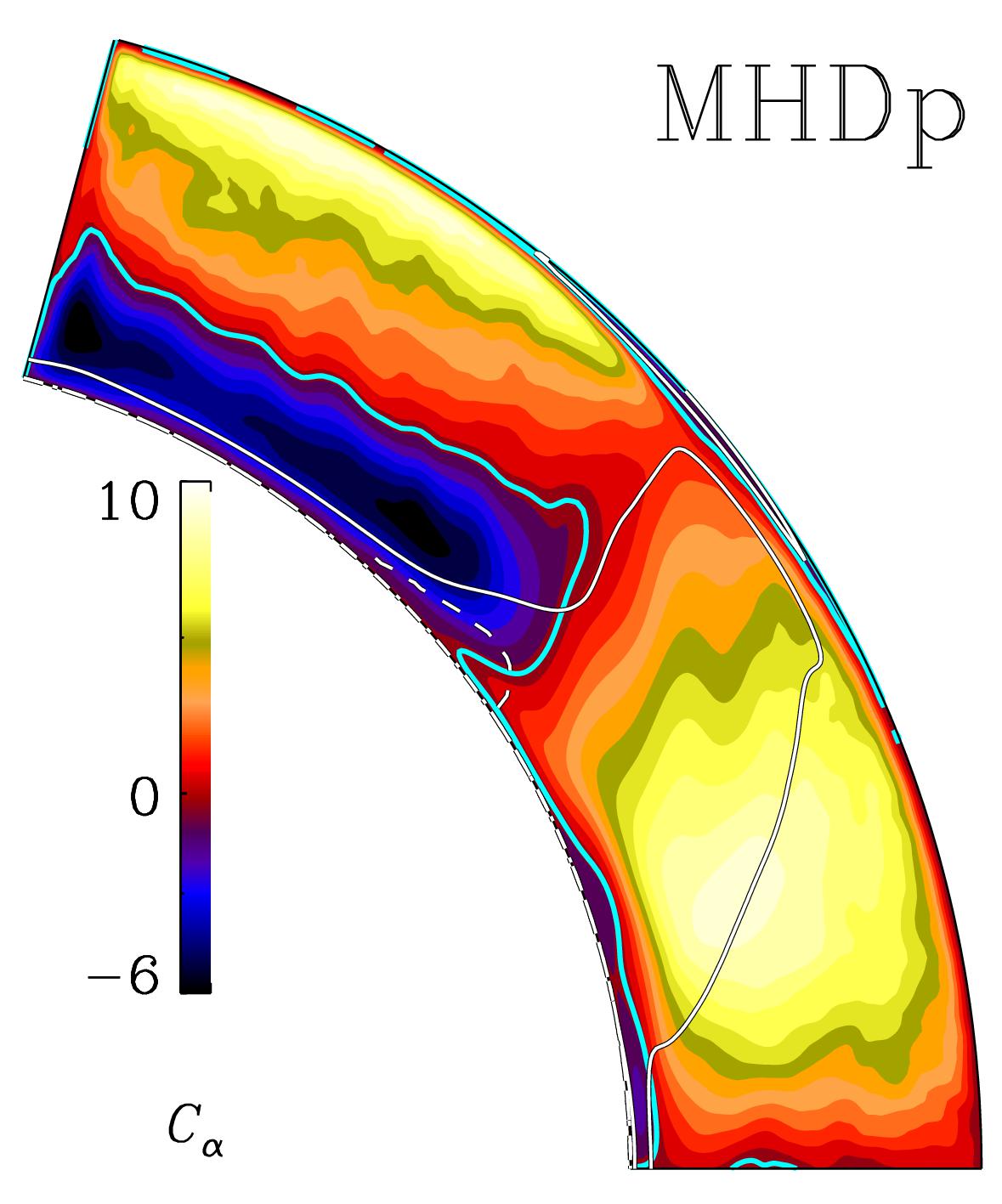}\includegraphics[width=.33\textwidth]{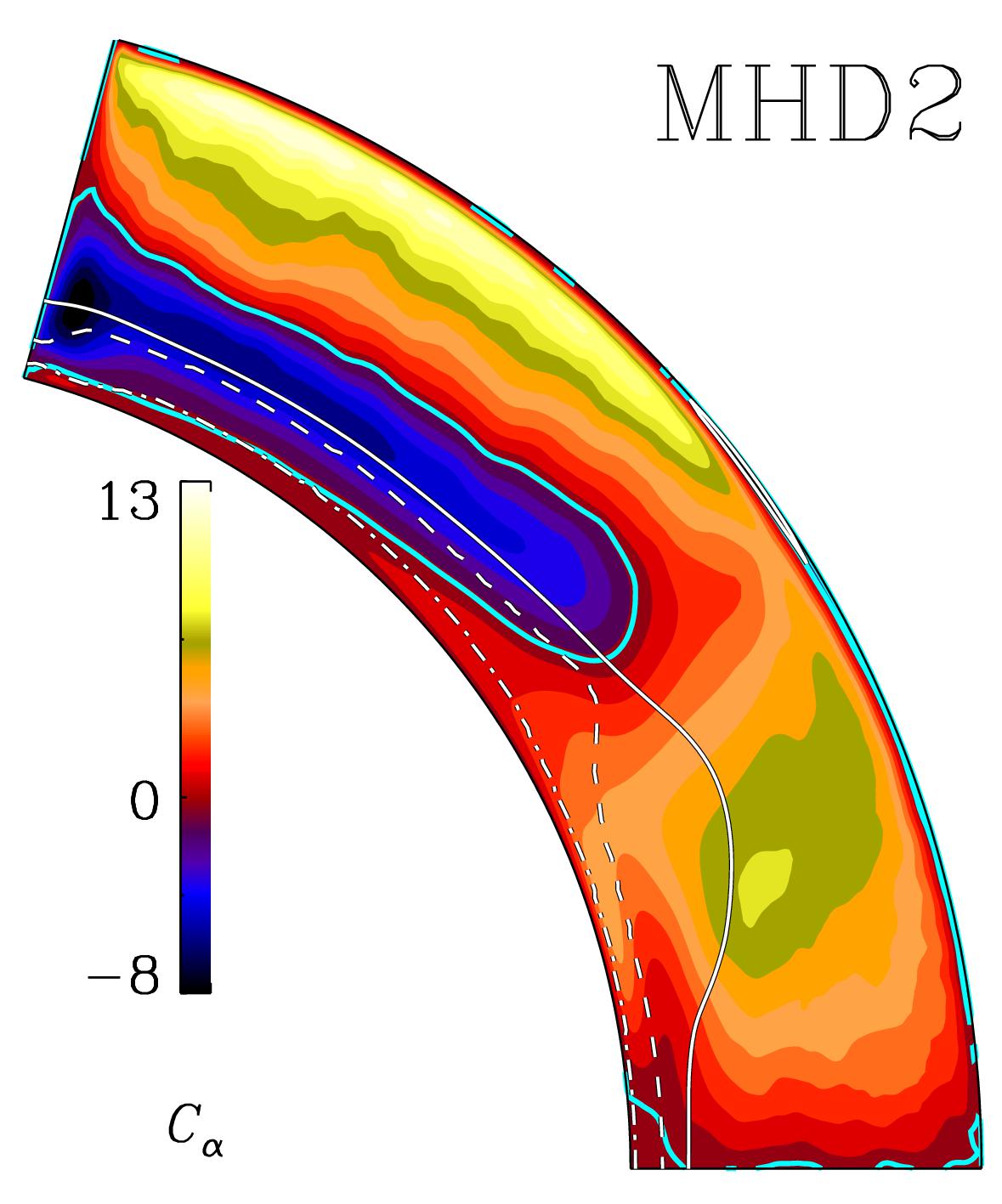}\includegraphics[width=.33\textwidth]{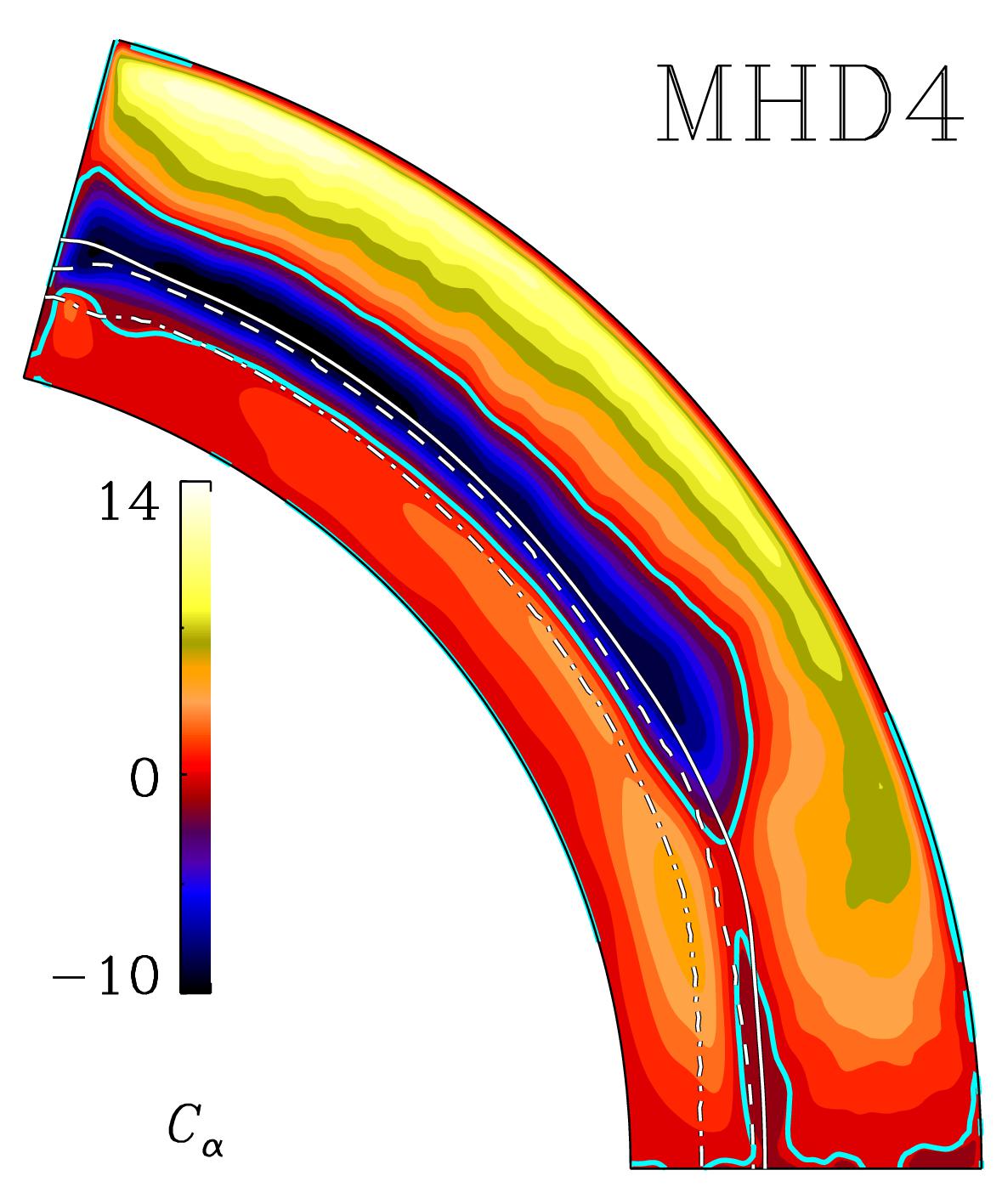}\\
\includegraphics[width=.33\textwidth]{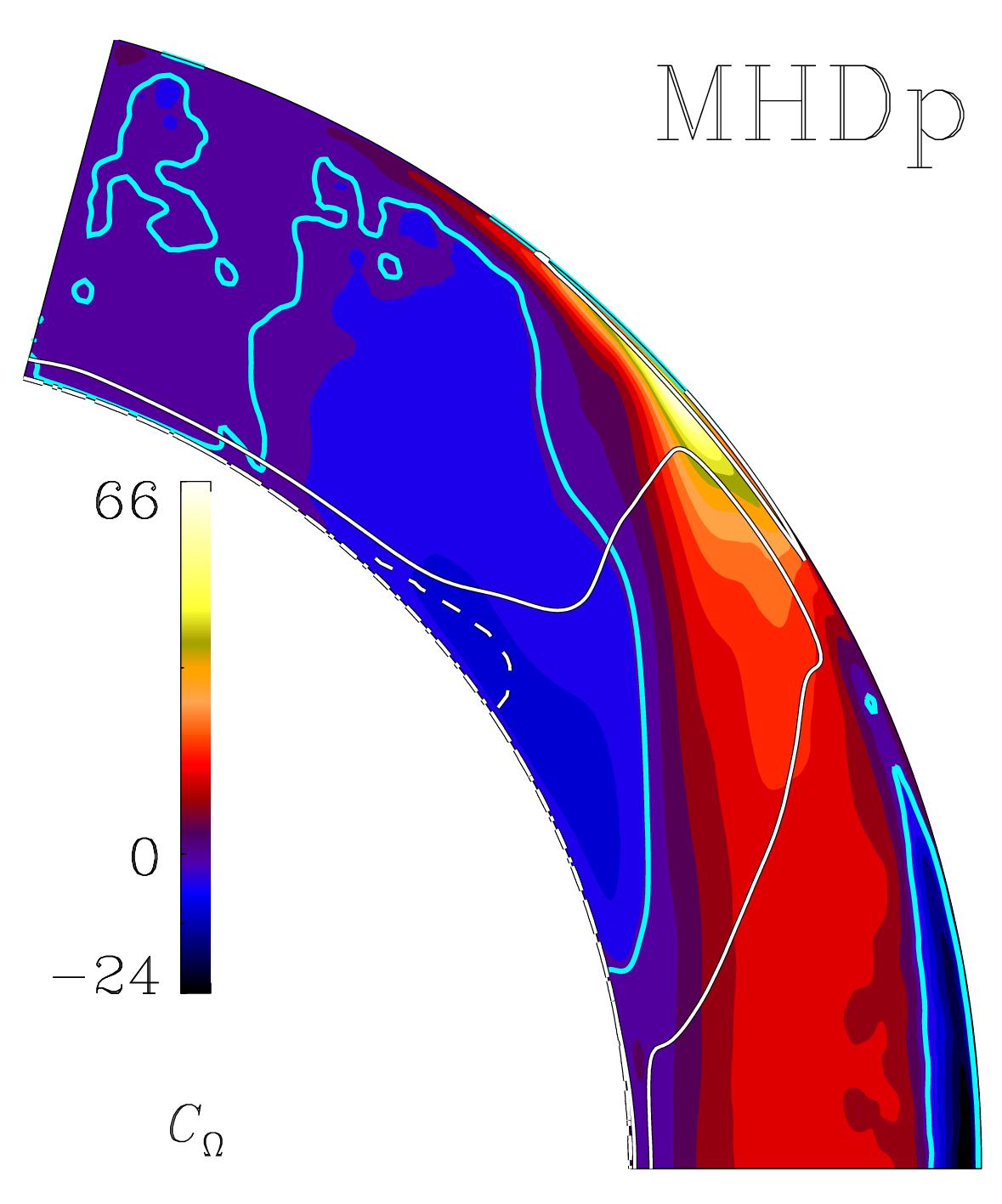}\includegraphics[width=.33\textwidth]{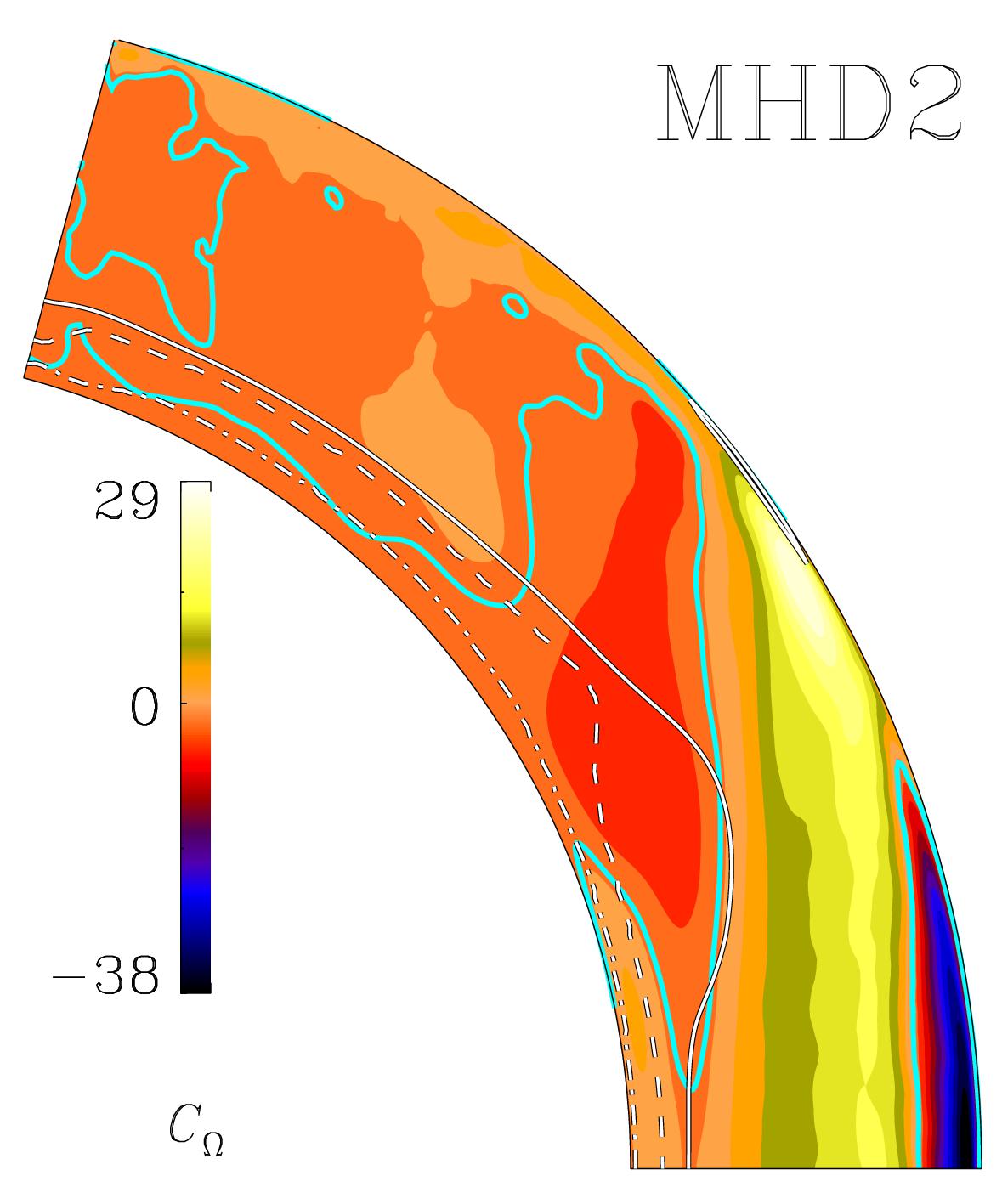}\includegraphics[width=.33\textwidth]{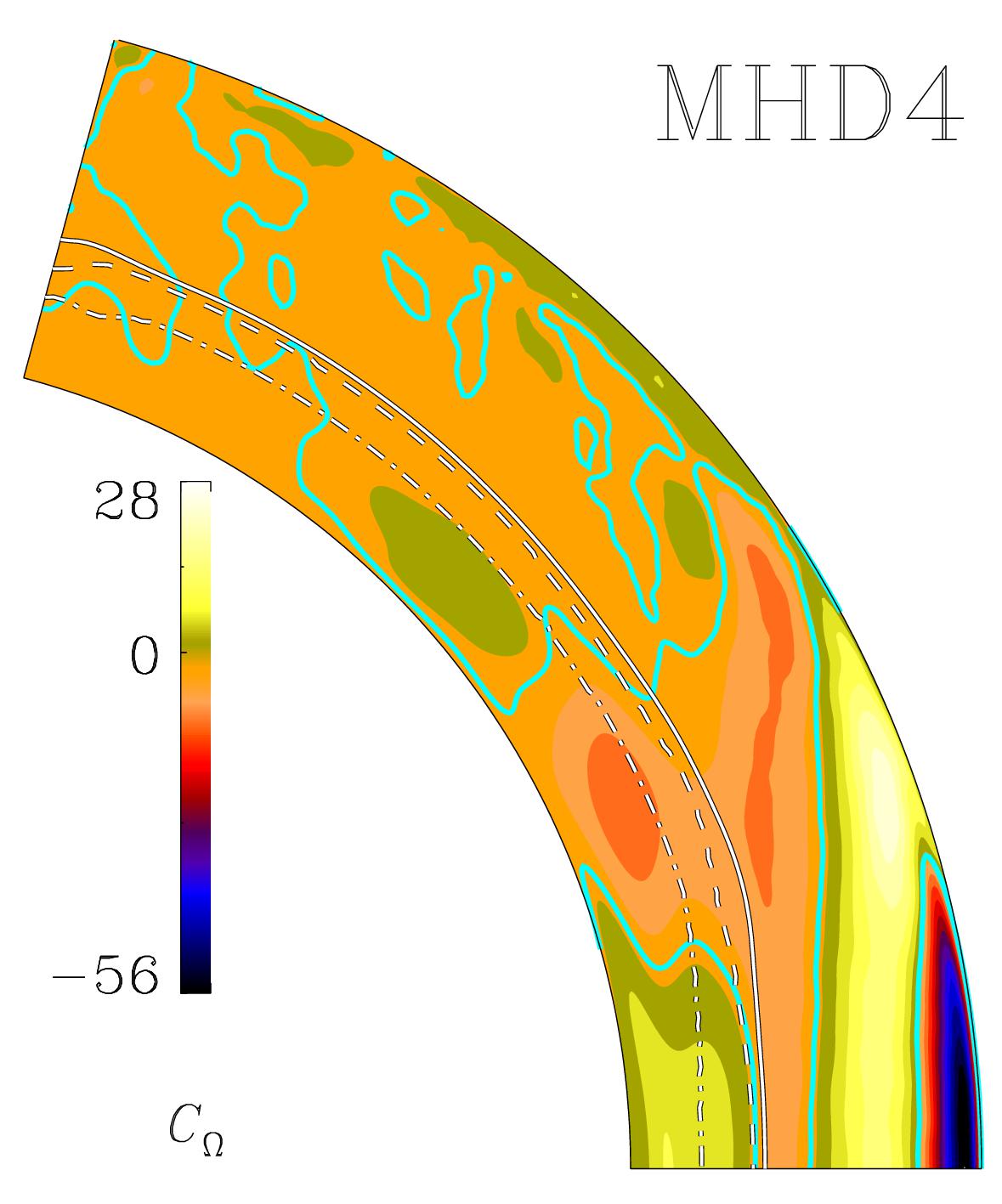}\\
\caption{Dynamo numbers $C_\alpha$ (upper row) and $C_\Omega$ (lower row)
  for Runs~MHDp (left panels), MHD2 (middle), and MHD4 (right). The cyan
    contours indicate the zero levels whereas the bottoms of the BZ,
    DZ, and OZ are indicated with solid, dashed, and dash-dotted
    lines, respectively (colour online).}
\label{fig:pdynnum}
\end{center}
\end{figure}

Given that the dominant dynamo mode changes as a function of depth
of the stably stratified layers below the CZ, it is of interest to
study the diagnostics that are commonly held responsible for the
generation of large-scale magnetic fields and cycles. One such
diagnostics is the kinetic helicity of the flow, which can, for
high conductivity, be associated with the $\alpha$ effect of
mean-field electrodynamics
\citep[][]{1966ZNatA..21..369S,KR80}. Furthermore, it has been shown
by numerical simulations that the sign of the kinetic helicity can
change under certain conditions in the deep parts of the convection
zone and lead to a change of the propagation direction of the dynamo
wave \citep{DWBG16}.
Similar reversals have routinely been seen in simulations of
stratified convection \citep{BNPST90,KKB09a}, but the change in the
migration direction occurs only for sufficiently deep helicity
reversals, which is what was demonstrated by \cite{DWBG16}.
Such reversals of the resulting $\alpha$ effect have been utilised in
mean-field dynamo theory starting with the work of \cite{Yo72}.
The relevance of the kinetic helicity reversal for the Sun is
  that mean-field theory \citep{KR80} and typical simulations
  \citep[e.g.][]{OSB01,KKOS06} predict a positive $\alpha$ effect in
  the northern hemisphere. Furthermore, a predominantly positive
  radial gradient of angular velocity is present in the solar CZ. In
  $\alpha\Omega$ dynamos, this
  combination leads to poleward migration of dynamo waves due to the
  Parker--Yoshimura rule \citep{Pa55b,Yo75}. Thus, reversing the sign
  of $\alpha$ would resolve this issue.

Figure~\ref{fig:poum} shows time-averaged relative kinetic helicity $h_{\rm
  rel}=\overline{\uuu\cdot\ooo}/\urms\omega_{\rm rms}$ from the MHD runs.
Here, $\ooo=\nab\times\uuu$ is the vorticity of the fluctuating velocity.
We do find a region of inverted helicity at the base of the CZ
in all runs. However, this region is not very pronounced and is concentrated at high latitudes in
Runs~MHDp and MHD1--2. Although the region of positive helicity
extends to lower latitudes in Runs~MHD3 and MHD4, it is still confined
within the tangent cylinder with respect to the bottom of the BZ. Only
in Run~MHD4, a clear inversion is seen at low latitudes near the
equator.

The origin of the large-scale magnetic fields in the current
  simulations cannot be determined without a detailed analysis
  involving the computation of turbulent transport coefficients with,
  for example, the test-field method
  \citep[e.g.][]{SRSRC05,SRSRC07,2018A&A...609A..51W}. However,
  earlier studies \citep{WKKB14,2018A&A...609A..51W} have indicated
  that, at least, the appearance of latitudinal dynamo waves can be
  fairly accurately predicted by the Parker--Yoshimura rule.
  This rule essentially states that the sign of
  the product of the $\alpha$ effect and radial gradient of $\Omega$
  determines the propagation direction of the latitudinal dynamo
  wave. For an equatorward wave, $\alpha \pd\mean{\Omega}/\pd r <0$ in
  the northern hemisphere of the Sun. We proceed to study this by means
  of local dynamo numbers \citep[as was also done in][]{KMCWB13}
\begin{eqnarray}
C_\alpha\,=\,\frac{\alpha \Delta r}{\eta_{\rm t0}}, \hspace{2cm} C_\Omega\,=\,\frac{\pd\mean{\Omega}/\pd r (\Delta r)^3}{\eta_{\rm t0}}\,,
\end{eqnarray}
where $\alpha = -\onethird \tau (\mean{\uuu\cdot\ooo} -
\mean{\jjj\cdot\bbb}/\mean{\rho})$ is a proxy of the $\alpha$ effect,
including the contributions from kinetic and current
helicities \citep{PFL76}. Furthermore, $\eta_{\rm t0}=\onethird \tau
\urms^2(r,\theta)$ is an estimate of the turbulent magnetic
diffusivity, where $\tau=\alpha_{\rm MLT} \Hp/\urms(r,\theta)$ is
the convective turnover time, $\alpha_{\rm MLT}=1.7$ is
the mixing length parameter, and $\Hp=-(\pd \ln p/\pd r)^{-1}$ is the
pressure scale height.

We show representative results of $C_\alpha$ and $C_\Omega$ in
  \figu{fig:pdynnum}. In Run~MHDp, the product $C_\alpha C_\Omega$ is
  negative in the upper parts of the CZ at high latitudes and in a
  shallow layer near the equator. However, in the former (latter),
  $C_\Omega$ ($C_\alpha$) is small, which could explain the
  absence of cycles. In Run~MHD2, a sizeable mid-latitude region
  shows a negative $C_\alpha C_\Omega$. This occurs at the same
  location where the equatorward migrating fields are seen in
  \figu{fig:pbutter}(c). A similar but somewhat smaller area appears
  also in Runs~MHD3 and MHD4; see the rightmost panels of
  \figu{fig:pdynnum}, although these runs do not support migrating
  solutions. A possible explanation is that the strong magnetic field
  developing in the radiative, strongly subadiabatic, layer forming in
  these runs is inhibiting dynamo migration.
The behaviour seen in the current simulations with radiative layers is not likely to occur to
the same extent in real stars where the small magnetic diffusivity
will not allow substantial magnetic fields to penetrate into the radiative
layers below the CZ.

\section{Conclusions}

In the current study we have presented the first simulations
of convection in rotating spherical coordinates with
a heat conduction prescription based on the Kramers opacity
law. In such models the radiative flux adapts to the
  thermodynamic state in a dynamical fashion such that the depth of
the CZ is not fixed a priori.
We have demonstrated that in such setups, the depth of the CZ is controlled
by the overall efficiency of convective energy transport.
Enhancing the radiative energy transport reduces the fraction of
energy transported by convection in the deep parts, and is associated
with the appearance of stably stratified
Deardorff, overshoot, and radiative layers below the Schwarzschild-unstable layer. The enhanced
luminosity in the current simulations implies a moderate
Kelvin--Helmholtz time and allows the models to be evolved
to a thermally saturated state in a reasonable time. Thus we do not have
to resort to artificially enhancing the heat conductivity in the convectively stable
layers immediately below the CZ
  \citep[e.g.][]{2017ApJ...836..192B,2017ApJ...843...52H}. Such
  procedure leads to a more abrupt transition to the RZ and increased
  stiffness of the upper part of the CZ. This is likely to have
  repercussions for the interaction of the dynamics of the RZ and CZ.
We have shown
that the
presence of such a stable layer has several interesting implications
for the dynamics of convection.

Although the up- and
downflows contribute roughly equally to the energy transport in all of the cases studied here,
the presence of stably stratified overshoot and radiative layers are
reflected in the force balance. This suggests fundamentally different
dynamics in systems with and without such layers. In the rotating cases with
$\Omega_0=3\Omega_\odot$ and without significant stably stratified
layers, the convective energy transport is highly anisotropic with
mid-latitude regions producing an almost negligible contribution to the
overall luminosity. If, on the other hand, stably stratified layers
are present, the latitudinal dependence of convective energy transport is
much weaker.
Although the spectral power in convective motions is slightly
  reduced in cases with Kramers opacity, this effect is too small to
  account for the discrepancy between solar observations and simulations,
  most notably the problem of anti-solar differential
  rotation at solar parameters \citep[e.g.][]{KKB14,HRY15a}.

The changes in the rotation profiles and large-scale magnetism are
more subtle and the interpretation is less straightforward. However, the
current simulations show clearly a NSSL at low
latitudes---irrespective of the prescription of radiative
diffusion. This is possibly
due to the somewhat higher density stratification in the current
simulations in comparison to several previous studies. The appearance
of stably stratified layers at the bottom of the domain tends to
produce a layer of negative radial shear at the base of the
CZ. However, this leads to clearly equatorward migrating large-scale
magnetic fields only in a single case. Although an inversion of the
kinetic helicity is observed in the OZ and the lower parts of the CZ in
our cases with the shallowest convection zone, they exhibit
quasi-stationary large-scale magnetic fields.

\section*{Acknowledgement}{
The referees are acknowledged for their helpful comments on
the manuscript. The authors wish to thank CSC -- IT Center for
Science Ltd.\ in Espoo, Finland, and the Gauss Center for
Supercomputing through the Large-Scale computing project ``Cracking the
Convective Conundrum'' in the Leibniz Supercomputing Centre's SuperMUC
supercomputer in Garching, Germany for computational resources. This
work was supported in part by
the Deutsche Forschungsgemeinschaft Heisenberg programme (grant
No.\ KA 4825/1-1; PJK), the Academy of Finland ReSoLVE Centre of
Excellence (grant No.\ 272157; MJK, PJK), the National Science Foundation Astronomy and
Astrophysics Grants Program (grant 1615100), and the University of
Colorado through its support of the George Ellery Hale visiting
faculty appointment.
MV acknowledges a postgraduate fellowship from the SOLSTAR
Max Planck Research Group and the enrolment in the framework
of the International Max Planck Research School for Solar
System Science at the University of G\"ottingen (IMPRS) in Germany.
FS was supported by the Max Planck Society grant ``Science Projects in
Preparation for the PLATO Mission.''}

\bibliographystyle{gGAF}
%\bibliography{bib}
\markboth{\rm P.J.\ K\"APYL\"A et al.}{\rm GEOPHYSICAL AND ASTROPHYSICAL FLUID DYNAMICS}

\vspace{36pt}

\end{document}